\def\draft{1}
\def\llncs{0}
\def\anon{0}

\def\showtableofcontents{1}
\def\cameraready{0}

\makeatletter
\newsavebox\myboxA
\newsavebox\myboxB
\newlength\mylenA

\newcommand*\xoverline[2][0.75]{%
    \sbox{\myboxA}{$\m@th#2$}%
    \setbox\myboxB\null
    \ht\myboxB=\ht\myboxA%
    \dp\myboxB=\dp\myboxA%
    \wd\myboxB=#1\wd\myboxA
    \sbox\myboxB{$\m@th\overline{\copy\myboxB}$}
    \setlength\mylenA{\the\wd\myboxA}
    \addtolength\mylenA{-\the\wd\myboxB}%
    \ifdim\wd\myboxB<\wd\myboxA%
       \rlap{\hskip 0.5\mylenA\usebox\myboxB}{\usebox\myboxA}%
    \else
        \hskip -0.5\mylenA\rlap{\usebox\myboxA}{\hskip 0.5\mylenA\usebox\myboxB}%
    \fi}
\makeatother

\ifnum\llncs=0
\documentclass[11pt]{article}
\usepackage{fullpage}
\usepackage[dvipsnames]{xcolor}
\usepackage{cite}
\else
\documentclass[orivec,runningheads,envcountsect]{llncs}
\usepackage[T1]{fontenc}
\usepackage{lmodern}
\fi

\title{Incrementally Verifiable Oracle Aided Computation}
\ifnum\anon=1
\author{Submission to STOC 2024}
\date{\today}
\fi

\ifnum\anon=0
\author{Shafi Goldwasser\thanks{MIT, Berkeley and RESI. \texttt{shafi@csail.mit.edu.}} 
\and Yael Tauman Kalai\thanks{MIT and RESI. \texttt{tauman@mit.edu}.} }

\fi

\usepackage{hyperref}
\definecolor{DarkBlue}{RGB}{0,0,150}
\definecolor{DarkRed}{RGB}{150,0,0}
\definecolor{DarkGreen}{RGB}{0,150,0}
\hypersetup{colorlinks,breaklinks,urlcolor=[rgb]{0,0,1},linkcolor=DarkRed,citecolor=DarkRed}
\usepackage[hyperpageref]{backref}
\usepackage{xspace}

\ifnum\cameraready=0

\usepackage{fullpage}
\usepackage{multicol}
\usepackage{amsmath}
\usepackage{amssymb,amsfonts}
\usepackage{float}
\usepackage{graphicx}
\usepackage{enumitem}
\usepackage{color}
\usepackage{comment}
\usepackage{appendix}
\usepackage{mathtools}
\usepackage{bm,bbm}
\usepackage{dashbox}
\usepackage{framed}
\usepackage{stmaryrd}

\usepackage[utf8]{inputenc}
\usepackage{balance}
\usepackage{csquotes}
\usepackage{microtype}
\usepackage{breqn}
\usepackage{titlesec}

\usepackage{booktabs}
\usepackage{nicefrac}
\PassOptionsToPackage{hyphens}{url}
\usepackage[hyphens]{url}
\usepackage{tikz}
\usepackage{tikz-qtree}

\fi 

\usepackage[lambda,operators,logic,asymptotics,keys]{cryptocode}

\newcommand{\ynote}[1]{\ifnum\draft=1 $\ll$\textsf{\color{red} Yael: { #1}}$\gg$ \fi}
\newcommand{\snote}[1]{\ifnum\draft=1 $\ll$\textsf{\color{blue} Shafi: {#1}}$\gg$ \fi}

\newcommand{\commentout}[1]{}

\ifnum\llncs=0

\usepackage{amsthm}
\usepackage[capitalise,nameinlink]{cleveref}
\crefformat{footnote}{#2\footnotemark[#1]#3}

\newtheorem{theorem}{Theorem}

\newtheorem{definition}[theorem]{Definition}
\newtheorem{claim}[theorem]{Claim}

\newtheorem{remark}[theorem]{Remark}

\newtheorem{construction}[theorem]{Construction}

\theoremstyle{definition}

\Crefname{claim}{Claim}{Claims}

\else

\spnewtheorem{prop}{Property}{\bfseries}{\itshape}
\Crefname{prop}{Property}{Properties}
\spnewtheorem{informaltheorem}{Informal Theorem}{\bfseries}{\itshape}
\Crefname{informaltheorem}{Informal Theorem}{Informal Theorems}
\spnewtheorem{fact}{Fact}{\bfseries}{\itshape}
\Crefname{fact}{Fact}{Facts}
\Crefname{conjecture}{Conjecture}{Conjecture}

\spnewtheorem{claim}{Claim}{\bfseries}{\itshape}
\Crefname{claim}{Claim}{Claims}
\fi

\usepackage[nointegrals]{wasysym}

\usepackage{array,arydshln,multirow,booktabs,adjustbox,graphicx}

\newcommand{\probcond}[2]{\Pr \left[
\begin{array}{l}#1\end{array} \; : \;
\begin{array}{l}#2\end{array} \right]
}

\def\ddefloop#1{\ifx\ddefloop#1\else\ddef{#1}\expandafter\ddefloop\fi}

\def\ddef#1{\expandafter\def\csname bb#1\endcsname{\ensuremath{\mathbb{#1}}\xspace}}
\ddefloop ABCDEFGHIJKLMNOPQRSTUVWXYZ\ddefloop
\def\ddef#1{\expandafter\def\csname c#1\endcsname{\ensuremath{\mathcal{#1}}\xspace}}
\ddefloop ABCDEFGHIJKLMNOPQRSTUVWXYZ\ddefloop

\def\ddef#1{\expandafter\def\csname m#1\endcsname{\ensuremath{\mathbf{#1}}\xspace}}
\ddefloop ABCDEFGHIJKLMNOPQRSTUVWXYZabcdefghijklmnopqrstvwxyz\ddefloop

\def\ddef#1{\expandafter\def\csname v#1\endcsname{\ensuremath{\boldsymbol{#1}}\xspace}}
\ddefloop ABCDEFGHIJKLMNOPQRSTUVWXYZabcdefghijklmnopqrstuvwxyz\ddefloop

\def\ddef#1{\expandafter\def\csname v#1\endcsname{\ensuremath{\boldsymbol{\csname #1\endcsname}}\xspace}}
\ddefloop {alpha}{beta}{gamma}{delta}{epsilon}{varepsilon}{zeta}{eta}{theta}{vartheta}{iota}{kappa}{lambda}{mu}{nu}{xi}{pi}{varpi}{rho}{varrho}{sigma}{varsigma}{tau}{upsilon}{phi}{varphi}{chi}{psi}{omega}{Gamma}{Delta}{Theta}{Lambda}{Xi}{Pi}{Sigma}{varSigma}{Upsilon}{Phi}{Psi}{Omega}{ell}\ddefloop

\newcommand{\BARG}{\ensuremath{\mathsf{BARG}}\xspace}

\newcommand{\seBARG}{\ensuremath{\mathsf{seBARG}}\xspace}
\newcommand{\SNARG}{\ensuremath{\mathsf{SNARG}}\xspace}

\newcommand{\RAM}{\ensuremath{\mathsf{RAM}}\xspace}

\newcommand{\Gen}{\ensuremath{\mathsf{Gen}}\xspace}

\newcommand{\Verify}{\ensuremath{\mathsf{Verify}}\xspace}

\newcommand{\Extract}{\ensuremath{\mathsf{Extract}}\xspace}

\newcommand{\Open}{\ensuremath{\mathsf{Open}}\xspace}

\newcommand{\Hash}{\ensuremath{\mathsf{Hash}}\xspace}

\newcommand{\rt}{\ensuremath{\mathsf{rt}}\xspace}

\newcommand{\crs}{\ensuremath{\mathsf{crs}}\xspace}

\providecommand{\sk}{\ensuremath{\mathsf{sk}}\xspace}

\newcommand{\td}{\ensuremath{\mathsf{td}}\xspace}

\ifnum\cameraready=0

\fi

\newcommand{\LWE}{\ensuremath{\mathsf{LWE}}\xspace}

\newcommand{\DLIN}{\ensuremath{\mathsf{DLIN}}\xspace}
\newcommand{\DDH}{\ensuremath{\mathsf{DDH}}\xspace}
\newcommand{\QR}{\ensuremath{\mathsf{QR}}\xspace}

\newcommand{\NP}{\ensuremath{\mathsf{NP}}\xspace}

\renewcommand{\poly}{\mathsf{poly}}

\newcommand{\ppt}{\ensuremath{\mathsf{PPT}}\xspace}
\newcommand{\PPT}{\ensuremath{\mathsf{PPT}}\xspace}
\newcommand{\secp}{\lambda}

\newcommand{\msf}[1]{\ensuremath{\mathsf{#1}}}

\providecommand{\vk}{\ensuremath{\mathsf{vk}}}
\providecommand{\star}{^*}

\newcommand{\Sig}{\mathsf{Sig}}
\newcommand{\sig}{\mathsf{sig}}

\providecommand{\hk}{\msf{hk}}

\newcommand{\HT}{{\sf HT}}

\newcommand{\IVC}{\mathsf{IVC}}
\newcommand{\calO}{\mathcal{O}}

\renewcommand{\negl}{\msf{negl}}

\mathchardef\mhyphen="2D
\newcommand{\bit}{\{0, 1\}}

\newlength{\protowidth}

\newlength{\cprotowidth}

\providecommand{\hk}{\mathsf{hk}}
\providecommand{\secparam}{1^\lambda}
\providecommand{\secpar}{\lambda}

\newcommand{\ignore}[1]{}

\makeatletter
\newcommand{\ostar}{\mathbin{\mathpalette\make@circled\star}}
\newcommand{\make@circled}[2]{%
  \ooalign{$\m@th#1\smallbigcirc{#1}$\cr\hidewidth$\m@th#1#2$\hidewidth\cr}%
}
\newcommand{\smallbigcirc}[1]{%
  \vcenter{\hbox{\scalebox{0.77778}{$\m@th#1\bigcirc$}}}%
}
\makeatother

\newcommand{\loc}{\mathrm{loc}}
\newcommand{\Adv}{\mathsf{Adv}}
\newcommand{\err}{\mathsf{err}}
\newcommand{\val}{\mathsf{v}}

\begin{document}
\maketitle

\begin{abstract}

Incrementally verifiable computation ($\IVC$) allows a prover
to maintain a short proof that can be efficiently updated
after each computation step, without retaining or recomputing
the preceding execution.  In the context of AI safety,  the need for incremental verification is particularly important, 
as AI systems interact with users over time and perform further
computation after each interaction. 
However, standard $\IVC$ schemes do not directly address a crucial feature of 
AI tasks: their dependence on external information and expertise.
Indeed, the UK AISI Alignment Team~\cite{IrvingM25} emphasizes that 
verification methods for AI safety should accommodate access
to external information and human judgment. This is the topic of this paper.

We construct $\IVC$ schemes for {\em oracle-aided} computations under the Learning with Errors ($\LWE$) assumption by first constructing succinct non-interactive arguments ($\SNARG$s), and then extending them to the incremental setting. 
We consider the following two
settings, where below, $\lambda$ is the security parameter, $n$ is
the input length, and $T$, $S$, and $Q$ are the time, space,
and oracle-query bounds.
\begin{itemize}
    \item \textbf{Robust computations.}
    For bounded-space computations, including non-deterministic
    computations, in which every accepting purported execution
    on a false input contains at least an $\varepsilon$-fraction
    of incorrect oracle answers, we construct {\em privately
    verifiable} $\SNARG$s and $\IVC$. For $\IVC$, the
    corresponding robustness condition is imposed on the
    claimed configuration. Public parameters, private
    verification keys, and proofs have length
    $\poly(\lambda,S,\log T,1/\varepsilon)$,
    and verification uses $O(\lambda/\varepsilon)$ oracle calls.
    We complement this result with a lower bound ruling out (deterministic) publicly verifiable $\SNARG$s and $\IVC$s with public parameters of length $o(n)$,
    and $o(Q)$ verifier oracle queries.

    \item \textbf{Signed oracles.}
    For oracles that digitally sign their query-answer pairs,
    we construct publicly verifiable $\SNARG$s and $\IVC$
    without a robustness assumption, and with oracle-free
    setup and verification. Public parameters and proofs
    have length $\poly(\lambda,S,\log T)$.
    For deterministic computations with oracle query and
    answer lengths $\poly(\lambda,\log T)$,
    we remove the dependence on $S$, obtaining public
    parameter and proof lengths
    $\poly(\lambda,\log T)$.

\end{itemize}
All our constructions have adaptive completeness and
soundness. Sub-exponential hardness of $\LWE$ suffices
for all constructions; polynomial hardness suffices in
the deterministic signed-oracle setting when $T$ is
polynomially bounded in $\lambda$. All cryptographic
assumptions are required to hold even given access to
the underlying oracle.
\end{abstract}

\ifnum\llncs=0 \thispagestyle{empty} \fi

\ifnum\showtableofcontents=1
{
\newpage
\setcounter{tocdepth}{2}
\tableofcontents
\thispagestyle{empty}
}
\fi

\newpage

\pagenumbering{arabic}

\section{Introduction}
\label{sec:intro}

Incrementally verifiable computation ($\IVC$)~\cite{Valiant08} allows a
computation to maintain a short proof of the computation  performed so far.
After each step, the prover updates a succinct proof certifying the
current configuration, without retaining or recomputing the preceding
execution. 
$\IVC$ is particularly relevant to the way we use AI systems:
we interact with them over time, providing new inputs and
receiving responses as a conversation or task unfolds. Each
new interaction prompts further computation that builds on
the system's current state. The computation thus proceeds
incrementally, and we would like verification to proceed
alongside it. 

Several $\IVC$ schemes have been proposed in the literature,
including schemes based on standard cryptographic assumptions
such as $\LWE$~\cite{PP22,DGKV22,AldemaTshuvaO26}. These results,
however, do not directly address an important feature of many
AI tasks: their dependence on external information and expertise.
As emphasized by the UK AISI Alignment Team~\cite{IrvingM25},
verification methods for AI safety should accommodate access
to external information and human judgment. Oracle-aided computation
provides a natural model for such computations: an AI system may
query an external database, consult a specialized service,
or ask for human judgment, with each consultation represented
by an oracle query.
Extending $\IVC$ to
this setting requires certifying that the computation is consistent with the underlying oracle. Moreover,
efficient verification requires the verifier to make 
substantially fewer oracle queries than made by the computations
they verify.
Combining these two goals leads to the question at
the center of this work:
\begin{quote}
\emph{Can an oracle-aided computation maintain a short,
efficiently updatable proof of its correctness that can be
verified using few oracle queries?}
\end{quote}

Oracle access introduces a fundamental obstacle to constructing
succinct proofs that can be verified using few oracle queries,
even for a single computation.  As was observed by Barbara, Chiesa, and Guan~\cite{cryptoeprint:2024/728}, 
for an arbitrary
computation, one incorrect answer can change the outcome while leaving
all ordinary transitions valid. A verifier that makes substantially
fewer oracle queries than the computation may fail to inspect precisely
that answer. Barbara et~al.\ prove that 
relativized succinct arguments, where the verifier makes a sublinear number of oracle calls, do not exist, even with
interaction and computational soundness.

 We study two approaches to overcoming this obstacle.
In both settings, our constructions satisfy \emph{adaptive
completeness and soundness}: completeness holds even when
the input and honest execution are chosen after seeing
the public parameters, and soundness holds even when a
cheating prover chooses its claim after seeing these
parameters.
\begin{itemize}
    \item The first approach follows Chen, Kalai, and
Xi~\cite{ChenKX26} and restricts soundness to
\emph{robust} computations, in which an incorrect
conclusion requires a substantial fraction of the
oracle answers to be wrong. Chen et~al.\ showed that
this robustness condition enables doubly-efficient
interactive verification.
We  construct
{\it non-interactive} succinct arguments ($\SNARG$s) and then extend them to  $\IVC$ for robust oracle-aided bounded-space
computations, including non-deterministic computations, under the sub-exponential hardness of $\LWE$. Our schemes
are {\em privately verifiable}. Namely, checking a proof requires a secret verification key
generated during setup and withheld from the prover, who
can construct and update proofs without access to this key. 
We complement this result with a lower bound: for general
robust oracle-aided computations, a $\SNARG$ or $\IVC$ scheme
with oracle-free setup and adaptive completeness and soundness
cannot have both public parameters of length sublinear in
the input length and a deterministic public verifier making
a number of oracle queries sublinear in the computation's
query count.

\item We would nevertheless like to obtain \emph{publicly verifiable}
$\SNARG$s and $\IVC$. For AI oversight, we want a proof to
serve as a certificate that can be shared and independently
checked by users, auditors, and other systems, without access
to a private verification key. Our second approach makes
this possible by requiring the oracle to authenticate its
responses. Specifically, a \emph{signed oracle} attaches a
digital signature to each query-answer pair. This is a
natural requirement when the oracle represents an external
database or service: its provider can sign responses as they
are issued, making their authenticity publicly checkable.
We show that signed oracles enable publicly verifiable
$\SNARG$s and $\IVC$ under $\LWE$ assumptions. Signing
the oracle's responses also removes the robustness
restriction altogether: our constructions apply to
arbitrary computations with access to signed oracles,
even when a single incorrect oracle answer can change
the outcome. Moreover, verification requires no oracle
queries.
\end{itemize}

Taken together, our results delineate two complementary possibilities for incrementally verifiable oracle-aided computation. Robustness enables succinct incremental verification of unauthenticated oracle computations using private verification and a small number of oracle queries; signed oracles enable public, oracle-free verification even for non-robust computations. The main technical innovations of this work lie in the robust
setting, where we develop new techniques for proving soundness
and preserving it under incremental proof updates. The
signed-oracle constructions build on established techniques;
their contribution is primarily conceptual, showing that
authenticating oracle responses enables publicly verifiable
$\SNARG$s and $\IVC$ without any robustness requirement.

\subsection{Our Results in More Detail}
\label{sec:intro:results}

Throughout this work, we denote by $\secp$ the security parameter, $n$ the input
length, and $T$, $S$, and $Q$ the time, space, and oracle-query
bounds. The underlying oracle is fixed independently of setup, and
setup makes no oracle queries. Our constructions have adaptive
completeness and soundness: in particular, an adversary may choose
its statement and proof after seeing the public parameters. All
cryptographic assumptions are required to hold even given access to
the underlying oracle.

\paragraph{Robust computations: from interaction to $\SNARG$s and $\IVC$.}

A \emph{purported execution} follows the machine's transition
rules but may use incorrect oracle answers. We say that a
computation is $\varepsilon$-robust on an input if the input
is accepting, or every accepting purported execution uses
incorrect answers on at least an $\varepsilon$-fraction of
its oracle calls. Given a complete accepting purported
execution on a false robust input, a verifier could detect
an incorrect oracle answer by checking a small random sample
of its oracle calls. Our first result obtains this guarantee
using a $\SNARG$ and preserves it under incremental proof
updates.

\begin{theorem}[Robust oracle-aided computation, informal]
Assuming sub-exponential hardness of $\LWE$, there exist a
privately verifiable $\SNARG$ and a privately verifiable
$\IVC$ scheme for every time-$T$, space-$S$ oracle-aided
computation, including non-deterministic computations.
Both schemes satisfy adaptive completeness and, for
$\varepsilon$-robust claims, adaptive soundness, with the
following guarantees:
\begin{itemize}
    \item The public parameters, private verification key,
    and proof have length
    $\operatorname{poly}(\lambda,S,\log T,1/\varepsilon)$.

    \item The verifier makes $O(\lambda/\varepsilon)$
    oracle queries.

    \item Each $\IVC$ proof update and verification takes
    time $\operatorname{poly}(n,\lambda,S,\log T,1/\varepsilon)$.

    \item The prover updates its proof after each step
    without access to the private verification key or the
    preceding execution history.\footnote{The $\SNARG$
    construction fixes the number of oracle calls in every
    purported execution. The $\IVC$ construction supports
    a growing number of calls, subject to an a priori
    upper bound.}
\end{itemize}
\end{theorem}

Here, adaptive completeness allows the input and honest
execution to be chosen after seeing the public parameters.
Adaptive soundness guarantees that no efficient adversary,
given these parameters, can produce an accepting proof of
a false $\varepsilon$-robust claim except with negligible
probability.
For $\IVC$, robustness is imposed on the configuration being certified:
every purported execution reaching an incorrect claimed configuration
must contain an $\varepsilon$-fraction of incorrect oracle answers.
Robustness of the final output alone does not imply this property for
every intermediate configuration. It does imply the required condition
for an accepting terminal configuration on a robust false input.
The formal $\SNARG$ and $\IVC$ results appear in
\Cref{thm:robust-oracle-snarg}
(\Cref{sec:robust-oracle-construction}) and
\Cref{thm} (\Cref{sec:IVC-robust}), respectively.
\Cref{sec:overview:robust-snarg} and
\Cref{sec:overview:robust-IVC} give the corresponding overviews.



\paragraph{A barrier to deterministic public verification.}

The private verification key conceals which oracle answers
will be checked. The following theorem rules out deterministic
public verification with short public parameters and few
oracle queries.

\begin{theorem}[Public-verification lower bound, informal]
There does not exist a non-interactive argument for general robust
oracle-aided computations with oracle-free setup, adaptive
completeness and soundness, public parameters of length $o(n)$,
and a deterministic public verifier making $o(Q)$ oracle queries.
This holds even for deterministic computations making nonadaptive
oracle queries, and regardless of the proof length. The same
impossibility applies to $\IVC$ with these guarantees.
\end{theorem}
The negative result already holds for the simple computation
that, on input $x$, makes the $Q$ oracle queries
$(x,1),\ldots,(x,Q)$ and outputs the majority of the resulting
bits. For every candidate $\SNARG$, our proof constructs
an oracle, fixed independently of setup, and a cheating
prover that produces an accepting proof of a false robust
claim with non-negligible probability over the public
parameters. Thus, under these requirements, robustness
alone does not suffice for deterministic public verification.
The $\IVC$ consequence follows by considering the proof at
termination. See \Cref{thm:impossibility}
(\Cref{sec:impossibility}) and the overview in
\Cref{sec:overview:impossibility}.

\paragraph{Public verification through signed oracles.}

For many applications, we would like a certificate that anyone can
verify, without a private verification key or further access to the
external sources. A result might be checked by several auditors,
passed to another computation, or verified after the original
oracle is no longer available. We also want to support computations
that depend critically on individual oracle answers and therefore
need not be robust.

Our second approach is to require the oracle to authenticate its
answers. A \emph{signed oracle}, on query $u$, returns the answer
$a=O(u)$ together with a digital signature on the pair $(u,a)$.
The oracle holds the signing key, and its verification key is public
and authenticated.
An external database or service can therefore make its responses
verifiable by signing them when they are supplied. For a versioned
database, the version identifier can be included in the query.
The answers themselves can be arbitrary; no randomness or algebraic
structure is required.

Signatures have previously been used in oracle-assisted proof
systems, beginning with Chiesa and Tromer~\cite{ChiesaT10}.
Here they authenticate the external information on which the
application's computation depends. Independently of our work,
Chiesa, Guan, and Y{\i}ld{\i}z~\cite{ChiesaGY26} also study signed
application oracles in the context of zero-knowledge verification.
Our signed-oracle results provide succinct non-interactive and
incremental verification under LWE.

\begin{theorem}
[Signed oracle-aided computation, informal]
Computations with access to signed oracles admit publicly
verifiable $\SNARG$s and $\IVC$, with adaptive completeness
and adaptive soundness, under the assumptions below.
No robustness assumption is required, and neither setup
nor verification makes oracle queries.
\begin{itemize}
    \item For time-$T$, space-$S$ non-deterministic
    computations, sub-exponential hardness of $\LWE$
    suffices. Public parameters and proofs have length
    $\operatorname{poly}(\lambda,S,\log T)$, and $\IVC$
    proof updates and verification take time
    $\operatorname{poly}(n,\lambda,S,\log T)$.

    \item For deterministic time-$T$ computations with
    oracle queries and answers of length
    $\operatorname{poly}(\lambda,\log T)$,
    sub-exponential hardness of $\LWE$ suffices.
    Public parameters and proofs have length
    $\operatorname{poly}(\lambda,\log T)$, and
    verification and $\IVC$ proof updates take time
    $n\cdot\operatorname{poly}(\lambda,\log T)$.
    When $T$ is polynomially bounded in $\lambda$,
    polynomial hardness of $\LWE$ suffices.
\end{itemize}
The hardness assumptions are understood to hold even
given access to the underlying oracle. The honest prover
obtains only the authenticated oracle responses used
by the underlying execution.
\end{theorem}

The bounded-space results appear in \Cref{thm:signed-oracle-snarg} (\Cref{sec:signed-oracles}) and \Cref{thm:IVC-signed} (\Cref{sec:IVC-signed}).
The fully succinct deterministic $\SNARG$ is \Cref{thm:short-signed-oracles} (\Cref{sec:short-signed-oracles}), and the
corresponding $\IVC$ refinement appears at the end of \Cref{sec:IVC-signed}.
For larger deterministic time bounds, the constructions require
correspondingly stronger, $\operatorname{poly}(T)$-hard security;
polynomial hardness alone is not sufficient for that extension.
\Cref{sec:overview:signed} gives an overview of both constructions.

The contribution of this approach is primarily conceptual:
authenticating external answers makes public succinct and
incremental verification possible even when an individual answer
can determine the outcome. The constructions adapt established
hash-and-batch-argument and updatable-batch-argument techniques,
with local checks for signatures. Their soundness uses local
extraction to turn an incorrect authenticated answer into a
signature forgery. In particular, the verifier checks a succinct
proof of the authenticated execution; it need not receive and
verify every signature separately.

\subsection{Related Work}
\label{sec:related}

\paragraph{Oracle-aided verification and AI oversight.}

The study of oracle-aided verification is motivated in part by
approaches to AI oversight in which a verifier has limited access to
external information or judgment. Irving, Christiano, and
Amodei~\cite{IrvingCA18} proposed AI safety via debate, and
Brown-Cohen, Irving, and Piliouras~\cite{BrownCohenIP24} developed
doubly-efficient debate protocols. These models use competing
provers. Chen, Kalai, and Xi~\cite{ChenKX26} study single-prover
verification and obtain doubly-efficient interactive proofs and
arguments for robust computations and for computations with
low-degree oracles. In particular, their robust interactive
argument uses $O(1/\varepsilon)$ verifier oracle queries for
constant soundness error. We study non-interactive arguments and
incremental proof maintenance for bounded-space computations,
and show a further limitation on deterministic public verification
even under robustness.

\paragraph{Signed oracle models.}

Chiesa and Tromer~\cite{ChiesaT10} introduced a
\emph{signed-input-and-randomness} functionality and used it to
construct short non-interactive arguments and proof-carrying data.
The functionality supplies signed randomness to support extraction
and recursive proof composition; their verifier already operates
without access to it. They also discuss realizations using signature
cards and trusted services. Maller, Mohnblatt, and
Zapico~\cite{MallerMZ25} introduce the open-and-sign random-oracle
model to obtain security guarantees for recursive proof systems,
including IVC of polynomial depth. These works concern oracle
functionalities used by the proof system itself. Our setting
authenticates arbitrary answers supplied by the application's
external sources.

Independent work of Chiesa, Guan, and
Y{\i}ld{\i}z~\cite{ChiesaGY26} also considers this use of signed
oracle answers. They study zero knowledge for oracle-aided
computations, prove an impossibility result in the random-oracle
model, and obtain zero-knowledge verification with efficient
prover and verifier for signed oracles under collision-resistant
hashing. Our work focuses on succinct non-interactive arguments
and IVC under LWE, with the time--space bounds stated above,
and also treats robust computations with unsigned oracles.

\paragraph{Succinct arguments from standard assumptions.}

Our constructions build on the development of batch arguments
and SNARGs from standard cryptographic assumptions.
Choudhuri, Jain, and Jin~\cite{FOCS:ChoJaiJin21} construct SNARGs for
deterministic polynomial-time computations from LWE.
Kalai, Vaikuntanathan, and Zhang~\cite{TCC:KalVaiZha21} obtain SNARGs
for bounded-space non-deterministic computations under
sub-exponential LWE. Our bounded-space signed-oracle result
preserves the polynomial dependence on space and polylogarithmic
dependence on time available for oracle-free
$\mathsf{NTISP}(T,S)$. The additional issue is establishing that
locally extracted oracle responses agree with the underlying
oracle, using either signature unforgeability or the quantitative
guarantee supplied by robustness.

\paragraph{Incrementally verifiable computation.}

$\IVC$ originates in the work of Valiant~\cite{Valiant08}.
Chiesa and Tromer~\cite{ChiesaT10} introduced proof-carrying data,
and Bitansky, Canetti, Chiesa, and Tromer \cite{STOC:BCCT13}
developed recursive composition and bootstrapping for SNARKs and
proof-carrying data. Paneth and Pass~\cite{PP22} and,
independently, Devadas, Goyal, Kalai, and
Vaikuntanathan~\cite{DGKV22} obtain $\IVC$ for deterministic
computations from LWE using rate-1 batch arguments.

Our incremental constructions use the updatable batch arguments
of Aldema Tshuva and Oshman~\cite{AldemaTshuvaO26}, who develop a
general framework for lifting delegation schemes to $\IVC$ across
several models of computation. Their primitive supports adding
statements and witnesses as they become available. We use this
framework to maintain local certificates of an oracle-aided
execution. In the robust setting, the additional challenge is to
preserve hidden sampling and quantitative soundness for a growing
sequence of unauthenticated oracle responses.

\subsection{Paper Organization}

\Cref{sec:overview} gives a technical overview. \Cref{sec:prelims} reviews the
cryptographic primitives, and \Cref{sec:oracle-machines} defines robust oracle-aided
computation and its verification. \Cref{sec:robust-oracle-construction}
constructs privately
verifiable $\SNARG$s for robust computations, and \Cref{sec:impossibility} proves
the lower bound for deterministic public verification.
\Cref{sec:signed-oracles} develops publicly verifiable $\SNARG$s for signed oracles,
including the fully succinct deterministic construction.
\Cref{sec:IVC} defines oracle-aided $\IVC$ and gives constructions in
both settings.
\section{Technical Overview}
\label{sec:overview}

The main difficulty in constructing $\SNARG$s and $\IVC$
for oracle-aided computations is certifying the correctness
of the oracle answers. Although numerous techniques are known
for standard computations, both deterministic and
non-deterministic, the oracle-aided setting presents a
fundamental barrier. As observed in prior work~\cite{cryptoeprint:2024/728}, a cheating
prover can honestly execute a computation with respect to
an oracle that differs from the true oracle on a single query.
This single discrepancy may change the outcome, yet the
verifier can detect it only by querying the oracle at that
point. Thus, without further restrictions, the verifier may
need to check every oracle answer, undermining our goal of making
verification substantially more efficient than the
computation itself.

We overcome this barrier in two settings. First, we consider
\emph{signed oracles}, whose answers come with signatures
authenticating the query-answer pairs. We incorporate
signature verification into the computation being proved
and use local extraction to establish soundness. This yields
publicly verifiable $\SNARG$s and $\IVC$ for arbitrary
oracle-aided computations, with a verifier that makes no
oracle queries. We describe this approach in
\Cref{sec:overview:signed}.

Second, we consider arbitrary oracles but require soundness
only for \emph{robust computations}: for any input $x$ and
any incorrect claimed output $y$, every purported
execution\footnote{A purported execution starts at the
prescribed initial configuration and follows all ordinary
transition rules, but may use arbitrary answers at oracle
steps.} on $x$ that outputs $y$ must contain a substantial
fraction of incorrect oracle answers.
With this relaxed
soundness guarantee, we obtain $\SNARG$s and $\IVC$
with relatively few verifier oracle queries. Our
$\SNARG$s and $\IVC$ in this setting are
\emph{privately verifiable}. We describe this approach in
\Cref{sec:overview:robust}. In \Cref{sec:overview:impossibility},
we rule out deterministic {\em publicly verifiable} $\SNARG$s and $\IVC$ with 
a sublinear number of oracle queries and public parameters of length $o(n)$.

We begin with the signed-oracle setting, whose analysis is simpler and serves as an introduction to our approach. Our main technical innovations
arise in the robust setting.
Our constructions follow the hash-and-$\BARG$ paradigm with local
extraction, and we prove soundness under the $\LWE$ assumption.
The key is to show that local guarantees suffice
to establish the correctness of the oracle answers. For signed oracles, we
adapt the local-extraction analysis to recover a signature
forgery from a false proof. 
For arbitrary oracles, we show that the verifier’s hidden oracle checks reject an incorrect claim with high probability whenever every purported execution supporting that claim contains a substantial fraction of incorrect oracle answers.
The space-dependent constructions assume sub-exponential
hardness of $\LWE$; for deterministic signed-oracle
computations with $T$ polynomially bounded in $\lambda$,
polynomial hardness suffices.

Throughout this overview, $T$ and $S$ denote the time and
space bounds, $n$ is the input length, and $\lambda$ is
the security parameter. The oracle is fixed independently
of setup, and our cryptographic assumptions are required
to hold even for adversaries with access to it.

\subsection{Signed Oracles}\label{sec:overview:signed}

A signed oracle returns an answer $a=\mathcal O(u)$ together
with a signature on $(u,a)$. Its verification key
$\mathsf{vk}$ is public and known to both the prover and
the verifier, while its signing key is known only to the
oracle. We first explain how to use these signatures to
construct a $\SNARG$, and then how to maintain the proof
incrementally.

\subsubsection{From signed answers to a $\SNARG$}

We begin with non-deterministic computations running in time
$T$ and space $S$. Our first construction has proof and public
parameter lengths $\operatorname{poly}(\lambda,S,\log T)$
and relies on sub-exponential hardness of $\LWE$. This matches the best known dependence on the time and space
bounds for oracle-free $\SNARG$s for $\mathsf{NTISP}(T,S)$
from $\LWE$. We later
show how to remove the dependence on the space bound for
deterministic computations.

We follow the hash-and-$\BARG$ approach with local extraction,
used in prior $\LWE$-based $\SNARG$ constructions for
$\mathsf{NTISP}$~\cite{TCC:KalVaiZha21,FOCS:ChoJaiJin21}.   
The prover hashes the computation and uses a batch
argument to certify its local transitions. We augment
these transitions with signature verification and rely on
the local-extraction guarantees.
Here it is essential to recover the signature itself:
a valid signature on an incorrect oracle answer may
exist, so ordinary soundness for the augmented computation
does not suffice. Local extraction allows us to turn
a false proof into an explicit signature forgery.

\paragraph{Soundness.}
Our analysis follows the local-extraction approach of prior
work. We track whether a configuration is reachable at the
prescribed time using correct oracle answers. The initial
configuration is reachable, whereas an accepting configuration
on a false input is not. A transition from a reachable
configuration to an unreachable one must therefore either
violate the machine's transition rules or use an incorrect
oracle answer. The former violates the batch argument's
local-extraction guarantee. In the latter case, the extracted witness
contains a valid signature on an incorrect query-answer
pair, yielding a signature forgery.

To formalize this intuition, we use two
independent batch-argument components to extract neighboring
transitions. Collision resistance ensures that their shared
configuration agrees, while index hiding allows us to move
the extraction indices through the computation. Tracking
reachability across these comparisons connects the prescribed
initial configuration to the claimed endpoint. Consequently,
a successful cheating prover allows us to break one of the
underlying primitives or forge a signature.

\paragraph{Removing the space dependence.}
For deterministic computations, we can remove the dependence
on the space bound using techniques from prior work.
We compress configurations by replacing the work tape with
its Merkle-tree hash. Authentication paths allow us to check
the symbols read by each transition and the resulting change
to the tape hash using a short witness.
The local-extraction guarantee of the batch argument allows
the soundness reduction to recover a signature on an
incorrect oracle answer. Moreover, the soundness analysis
can compare extracted configurations with the unique
execution under the true oracle, avoiding the need to
explore the bounded-space configuration graph.
For $\IVC$, we additionally use an updatable hash family to
authenticate the growing sequence of transition records
without rehashing the preceding records.
For short oracle queries and answers, this yields proof and
public parameter lengths $\operatorname{poly}(\lambda,\log T)$;
for $T$ polynomially bounded in $\lambda$, polynomial hardness
of $\LWE$ suffices.


\subsubsection{Making the proof incremental}

Our $\IVC$ construction follows the same approach, using
the updatable batch arguments of Aldema Tshuva and
Oshman~\cite{AldemaTshuvaO26}. These allow the prover to extend its
proof after each computation step without revisiting
the preceding execution. As above, we apply this machinery
to a computation that also verifies signatures on its
oracle query--answer pairs. The soundness analysis adapts
the same local-extraction argument, with signature
unforgeability ruling out incorrect oracle answers.

This yields publicly verifiable $\IVC$ with proof and
public parameter lengths
$\operatorname{poly}(\lambda,S,\log T)$ and a verifier
that makes no oracle queries. For deterministic
computations with short oracle queries and answers,
the same configuration compression removes the
dependence on $S$.

\subsection{Robust Computations}
\label{sec:overview:robust}

We next consider arbitrary oracles, whose answers need not
carry signatures. Here robustness provides another way to
check correctness: on a false input, every accepting
purported execution contains at least an $\varepsilon$ fraction of
incorrect oracle answers. If we could inspect such an
execution, checking a few randomly chosen oracle answers
would therefore suffice. Our goal is to obtain the same
guarantee from a succinct proof.

\subsubsection{A privately verifiable $\SNARG$}\label{sec:overview:robust-snarg}

We again use hash-and-$\BARG$ with local extraction, but now
give the verifier private extraction trapdoors. These allow
it to recover selected query-answer pairs and check them
against the true oracle. We append a table listing all oracle calls to the
computation transcript, and use the batch argument to certify
that each table entry agrees with the corresponding call in the
transcript. This allows the verifier to sample uniformly from
the oracle calls.

\paragraph{Soundness.}
For signed oracles, extracting one incorrect answer
together with its signature suffices to obtain a
contradiction. For arbitrary oracles, we need a
quantitative guarantee: the verifier's hidden checks
must detect incorrect answers with sufficient probability.

If the proof specified a complete purported execution,
this would follow immediately from robustness. However,
local extraction only gives witnesses for selected
positions, and we cannot assume that these witnesses
come from a single, globally consistent execution.
Moreover, the prover chooses its input and proof after
seeing the public parameters.

We address this using a potential $\Phi_t^x(C)$: the minimum
number of incorrect oracle answers needed to reach configuration
$C$ at time $t$ on input $x$, over all purported executions.
We cap this quantity at $Q$, the number of oracle calls in
every purported execution. The potential is used only in
the analysis; the verifier does not compute it.

A correct transition cannot increase the potential, and an
incorrect oracle answer can increase it by at most one.
The potential starts at zero, while robustness guarantees
that it is at least $\varepsilon Q$ at any accepting
configuration on a false robust input. Along a single
execution, these properties would imply at least
$\varepsilon Q$ incorrect answers. Our argument obtains
an analogous bound in expectation from local extractions
that need not form a single execution.

More precisely, let $p_s$ be the probability that the prover
outputs a false robust instance, all public checks accept,
and the first $s$ private checks accept. We show that, for
each check except the last,
\[
    p_s \leq (1-\varepsilon)p_{s-1}
        + \operatorname{negl}(\lambda).
\]
Although the hidden indices are sampled independently,
the prover chooses its input and proof after seeing the
public parameters, which are generated using these indices.
Thus, this bound does not follow from independence alone.

To account for the preceding checks, let $E$ be the event
defining $p_{s-1}$. We multiply each potential by
$\mathbf{1}_E$ before taking expectations, so outcomes
outside $E$ contribute zero.\footnote{This avoids conditioning on
an event that may have very small probability.}
We compare neighboring local extractions using the current
component and the last component, whose trapdoor is unused
by the preceding checks. Index hiding lets us change one
component's extraction index while evaluating the weighted
potential with the other component's trapdoor. When both
components open the same transcript position, hash binding
ensures that they recover the same configuration, except
with negligible probability. We can therefore switch which
component evaluates the potential and continue changing
indices.

These comparisons allow the expected weighted potential
differences to telescope. The initial-to-final potential
gap then gives a lower bound on the contribution of incorrect
oracle answers. Relating these answers to the oracle-call
table and averaging over its entries yields
\[
    \Pr[E \text{ and the $s$th private check rejects}]
    \geq \varepsilon p_{s-1}-\operatorname{negl}(\lambda),
\]
which implies the desired recurrence. Iterating it over
all but the last check gives negligible soundness error
with $O(\lambda/\varepsilon)$ oracle queries, without
extracting a complete execution.

\subsubsection{Impossibility of succinct deterministic public verification}
\label{sec:overview:impossibility}

We show that robustness alone does not suffice for
deterministic public verification with a sublinear number of oracle queries
and short public parameters. Specifically, there is no
publicly verifiable $\SNARG$ for general robust oracle-aided
computations with oracle-free setup, adaptive completeness
and soundness, public parameters of length $o(n)$, and a
deterministic verifier making $o(Q)$ oracle queries.
Here $n$ is the input length and $Q$ is the number of
oracle calls made by the computation.

Our counterexample is a simple majority computation.
Consider a Boolean oracle whose queries are pairs $(x,i)$,
where $x\in\{0,1\}^n$ and $i\in[Q]$. We view the oracle
as a table: each input $x$ indexes a row consisting of
the $Q$ bits $\calO(x,1),\ldots,\calO(x,Q)$. On input
$x$, the computation reads this row and outputs its
majority:
\[
    C^\calO(x)
    =
    \operatorname{Maj}\bigl(
        \calO(x,1),\ldots,\calO(x,Q)
    \bigr).
\]

For intuition, fix the public parameters and an accepting
proof for some input under the all-one oracle. A
deterministic verifier making only $\ell=o(Q)$ oracle
queries examines very few entries of the input's row.
Preserve those entries and change the remaining entries
of that row to zero. Every answer seen by the verifier
remains unchanged, so it still accepts. However, the row
now contains at most $\ell$ ones, making the claimed
majority incorrect. Moreover, any accepting purported
execution must use incorrect answers on a constant
fraction of its oracle calls. Thus, the false claim
satisfies the robustness requirement.

The difficulty is that the oracle must be fixed before
the public parameters are generated. We use the shortness
of the public parameters to encode every possible
parameter string $\crs$ as a distinct input $x_{\crs}$.
For each such string, we simulate the honest prover
and verifier under the all-one oracle and identify
the entries of row $x_{\crs}$ queried by the verifier.
This allows us to specify the corresponding row for
every possible $\crs$ in advance. After setup, the
cheating prover chooses the input $x_{\crs}$ and
supplies the corresponding proof.

We cannot simply modify all these rows: the verifier
may also query rows other than the one indexed by its
input, and changing those rows could change its
verification decision. Instead, we modify only a sparse
set of rows, selected using a pairwise-independent hash
function. With non-negligible probability, the row
$x_{\crs}$ is modified while every other row visited
by the verifier is left unchanged. On this event, the
verifier sees exactly the same answers as under the
all-one oracle and accepts a false robust claim.
An averaging argument fixes the row selection and
the prover's randomness before setup, yielding a
single oracle for which the attack succeeds with
non-negligible probability over the public parameters.

The argument places no restriction on the proof length.
Thus, the lower bound holds even for non-succinct
non-interactive arguments with these parameter and
query bounds. The same impossibility applies to
$\IVC$, since its proof at termination would give
a $\SNARG$ for the completed computation.
The formal result appears in \Cref{sec:impossibility}.

\subsubsection{Making the proof incremental}\label{sec:overview:robust-IVC}

To obtain $\IVC$, we must maintain this ability to sample
oracle calls as the computation grows. We separate ordinary
computation from oracle calls: an ordinary $\IVC$ certifies
the computation between consecutive calls, while an outer
updatable batch argument aggregates one record per oracle
call. This lets the prover update its proof after every
computation step while allowing the verifier to sample
oracle calls directly.

A further issue is that the number of oracle calls in the
claimed execution is not known at setup. Sampling from the
full query bound could miss a much shorter execution.
We therefore prepare hidden samples for query-count ranges
whose sizes are powers of two. Verification uses a range
within a factor of two of the claimed number of calls,
preserving the sampling guarantee with only logarithmic
overhead.

The same analysis extends to $\IVC$ by evaluating the
potential at oracle-call boundaries. Soundness of the
ordinary $\IVC$ ensures that the intervening computation
cannot increase the potential, while the consistency
guarantees of the updatable batch argument connect
neighboring records. Thus, the quantitative argument
continues to apply as the computation grows.

Under sub-exponential hardness of $\LWE$,
both constructions have proof and public parameter lengths
$\operatorname{poly}(\lambda,S,\log T,1/\varepsilon)$.
Verification uses $O(\lambda/\varepsilon)$ oracle queries,
and $\IVC$ updates require no access to the private
verification key.

%

\section{Preliminaries}
\label{sec:prelims}

\paragraph{Notation.} We let $\secp$ denote the security parameter throughout the paper. We use $\ppt$ to denote probabilistic polynomial-time, and denote the set of all positive integers up to $n$ as $[n] := \{1, \ldots, n\}$.  For any $x\in\{0,1\}^n$ and any subset $J\subset [n]$ we denote by $x_J=(x_j)_{j\in J}$.  For any finite set $S$, $x \gets S$ denotes a uniformly random element $x$ from the set $S$. Similarly, for any distribution $\cD$, $x \gets \cD$ denotes an element $x$ drawn from the distribution $\cD$.

\subsection{Hash Families with Local Opening and Updatability}\label{sec:HT}

\renewcommand{\vk}{\mathsf{vk}}

In this section we recall the definition of a hash family with local opening \cite{C:Merkle87}. These preliminaries are due to \cite{C:BBKLP23}. 

\paragraph{Syntax.}
A hash family ($\HT$) with succinct local opening consists of the following algorithms:
\begin{description}
\item[$\Gen(1^\secp)\to \hk$.] This is a $\PPT$ algorithm that takes as input the security parameter $\secp$ in unary and outputs a hash key $\hk$. 
  \item[$\Hash(\hk,x)\to \rt$.] This is a deterministic poly-time algorithm that takes as input a hash key $\hk$ and an input $x\in\{0,1\}^{N}$ for $N\leq 2^\secp$, and outputs a hash value $\rt$.
  \item[$\Open(\hk,x,j)\to \rho$.]  This is a deterministic poly-time algorithm that takes as input a hash key $\hk$, an input $x\in\{0,1\}^N$ for $N\leq 2^\secpar$, and an index $j\in[N]$, and outputs an opening $\rho$.
  \item[$\Verify(\hk,\rt,j,b,\rho)\to 0/1$.]   This is a deterministic poly-time algorithm that takes as input a hash key $\hk$, a hash value $\rt$, an index $j\in[N]$, a bit $b\in\{0,1\}$ and an opening $\rho$.
  It outputs 1 (accept) or 0 (reject).
  \end{description}

\begin{definition}
\label{def:HT}(Properties of $\HT$) A $\HT$ family $(\Gen,\Hash,\Open,\Verify)$ is required to satisfy the following properties.
 \begin{description}

   \item[~~Opening completeness.] For any $\secpar\in \bbN$, any $N\leq 2^\secpar$, any $x\in\bit^N$, and any index $j \in [N]$, 
    \[ \probcond{
    \Verify(\hk, \rt, j, x_j, \rho) = 1}{
    \hk \gets \Gen(\secparam), \\
    \rt = \Hash(\hk, x), \\
    \rho = \Open(\hk, x, j) \\
    } = 1. \]

    \item[~~Succinctness.] In the completeness experiment above, we have that $|\hk| + |\rt| + |\rho| = \poly(\secp)$. 
      
      \item[~~Collision resistance w.r.t.\ opening.] For any poly-size adversary $\cA$  there exists a negligible function $\negl(\cdot)$ such that for every $\secp \in \bbN$, 
    \[ \probcond{\Verify(\hk,\rt,j,0,\rho_0) =1\\
    \wedge~ \Verify(\hk,\rt,j,1,\rho_1) =1}{
       \hk\gets \Gen(\secparam), \\
        (\rt,j,\rho_0,\rho_1)\gets \cA(\hk)} = \negl(\secpar).\]    
  \end{description}
\end{definition}
\begin{remark}
We say that a hash family with local opening is $T$-secure, for $T=T(\secpar)$, if the collision resistance w.r.t. opening property holds against any $\poly(T)$-size adversary (as opposed to $\poly(\secpar)$-size) and the probability that the adversary finds a collision is $\negl(T)$ (as opposed to $\negl(\secpar)$).
We refer to this property as $T$-collision-resistance w.r.t.\ opening.
\end{remark}

\begin{remark} 
One can naturally extend the definition of a hash family with local opening to allow the $\Open$ algorithm to take as input $(\hk,x,J)$ where $J\subseteq[N]$ consists of a set of indices, as opposed to a single index.  $\Open(\hk,x,J)$ will simply run $\Open(\hk,x,j)$ for every $j\in J$.  $\Verify$ can be extended in a similar way to take as input $(\hk,\rt,J,b_J, \rho_J)$, and accept if and only if $\Verify(\hk,\rt,j, b_j,\rho_j)=1$ for every $j\in J$.

\end{remark}

\begin{theorem}[\cite{C:Merkle87}]
Assuming the existence of a collision resistant hash family there exists a hash family with local opening (according to \Cref{def:HT}). 

\end{theorem}

\paragraph{Updatable hash families.}
For our $\IVC$ constructions (\Cref{sec:IVC}), we use an updatable variant of a hash family
with local openings, as in~\cite{AldemaTshuvaO26}.
This variant is a generalization of \Cref{def:HT}, which allows to authenticate sequences of strings, rather than
only sequences of bits. Specifically, we view a bit
string $x\in\{0,1\}^N$ as a sequence of $N$
strings, each of length one. Under this
identification, the following definition generalizes \Cref{def:HT}.

\begin{definition}[Updatable hash family with local openings]
\label{def:updatable-hash}
Let
$\mathbf z=(z_1,\ldots,z_k)$ be a sequence
of strings of arbitrary lengths.
The algorithms $\HT.\mathsf{Hash}$,
$\HT.\mathsf{Open}$, and $\HT.\mathsf{Verify}$
are extended as follows:
\begin{itemize}
    \item
    $\HT.\mathsf{Hash}(\hk,\mathbf z)\to\val$
    outputs a hash value authenticating
    the sequence $\mathbf z$.

    \item
    $\HT.\mathsf{Open}(\hk,\mathbf z,j)\to\rho$
    outputs an opening of the $j$th string $z_j$.

    \item
    $\HT.\mathsf{Verify}(\hk,\val,j,z,\rho)
    \to\{0,1\}$
    checks that $z$ is the string at position $j$
    in the sequence authenticated by $\val$.
\end{itemize}

We say that $\HT$ is updatable if it additionally
provides the following two algorithms:
\begin{itemize}
    \item
    $\HT.\mathsf{Append}(\hk,\val,z)\to\val'$.
    The algorithm appends the string $z$ to the
    sequence authenticated by $\val$, without
    accessing the preceding strings.

    \item
    $\HT.\mathsf{Size}(\val)\to k$.
    The algorithm returns the number of strings
    in the sequence authenticated by $\val$.
    We set $\HT.\mathsf{Size}(\bot)=0$.
\end{itemize}
We require the following properties.

\begin{itemize}
    \item {\bf Opening completeness.}
    For every $\lambda\in\mathbb N$, every
    $\hk\in\mathrm{Supp}(\HT.\mathsf{Gen}(1^\lambda))$,
    every sequence $\mathbf z=(z_1,\ldots,z_k)$
    of admissible length, and every $j\in[k]$,
    letting
    \[
    \val=\HT.\Hash(\hk,\mathbf z),
    \qquad
    \rho=\HT.\Open(\hk,\mathbf z,j),
    \]
    we have
    \[
    \HT.\Verify
    (\hk,\val,j,z_j,\rho)=1.
    \]

\item {\bf Collision resistance with respect to opening.}
For every polynomial-size adversary $\mathcal A$,
there exists a negligible function $\mu$
such that for every $\lambda\in\mathbb N$,
\[
\Pr\left[
\begin{array}{l}
z\neq\widetilde z,\\
\HT.\mathsf{Verify}(\hk,\val,j,z,\rho)=1,\\
\HT.\mathsf{Verify}
(\hk,\val,j,\widetilde z,\widetilde\rho)=1
\end{array}
:
\begin{array}{l}
\hk\gets\HT.\mathsf{Gen}(1^\lambda),\\
(\val,j,z,\rho,\widetilde z,\widetilde\rho)
\gets\mathcal A(\hk)
\end{array}
\right]\leq\mu(\lambda).
\]

    \item {\bf Append completeness.}
    For every sequence
    $\mathbf z=(z_1,\ldots,z_k)$
    and every additional string $z_{k+1}$,
    satisfying
 \[
k+1\leq 2^\lambda
\qquad \mbox{and} \qquad
\sum_{i=1}^{k+1}(|z_i|+1)\leq 2^{2\lambda},
\]
    we have
    \[
    \HT.\mathsf{Append}
    (\hk,\HT.\Hash(\hk,\mathbf z),
    z_{k+1})=
    \HT.\Hash
    (\hk,(z_1,\ldots,z_{k+1})).
    \]
    Moreover,
    \[
    \HT.\mathsf{Size}
    (\HT.\Hash(\hk,\mathbf z))=k.
    \]
    We set
    \[\HT.\Hash(\hk,\emptyset)=\bot~~\mbox{and}~~\HT.\mathsf{Size}(\bot)=0.
    \]

    \item {\bf Succinctness and efficiency.}
    The lengths of $\hk$, $\val$, and $\rho$
    are bounded by $\mathrm{poly}(\lambda)$.
    The algorithms $\HT.\mathsf{Append}$
    and $\HT.\Verify$ run in time
    $\mathrm{poly}(\lambda,|z|)$,
    independently of the lengths of the
    preceding strings.
    The algorithm $\HT.\mathsf{Size}$
    runs in time $\mathrm{poly}(\lambda)$.
\end{itemize}

\end{definition}

The Merkle-forest construction of~\cite{AldemaTshuvaO26}
provides such an updatable hash family from any collision-resistant
hash functions.  

\subsection{Succinct Non-Interactive Arguments}\label{sec:def:SNARG}

We define succinct non-interactive argument systems ($\SNARG$) \cite{FOCS:Micali94,STOC:GenWic11,STOC:BCCT13} for the computation of a non-deterministic polynomial-time Turing machine $M$.

A $\SNARG$ system for $M$ consists of polynomial-time algorithms $(\SNARG.\Gen,\SNARG.\cP,\SNARG.\cV)$ with the following syntax:

\begin{description}
\item[$\SNARG.\Gen(1^\lambda,n)
      \longrightarrow \crs$.]
The randomized setup algorithm outputs a common reference string $\crs$.
\item[$\SNARG.\cP(\crs,x,w)
      \longrightarrow\pi$.]
The prover is given the $\crs$, an input $x\in\{0,1\}^n$,
and non-deterministic choices $w$ specifying an accepting execution of
$M$ on $x$. It
outputs a proof $\pi$.
\item[$\SNARG.\cV(\crs,x,\pi)
      \longrightarrow 0/1$.]
The verifier is given the $\crs$,  the input~$x$, and the proof~$\pi$. It 
outputs a bit indicating whether it accepts.

\end{description}

\begin{definition}
  A triple of algorithms $(\SNARG.\Gen,\SNARG.\cP,\SNARG.\cV)$ is a $\SNARG$ system for a Turing machine $M$ if the following hold:
  \begin{description}
    \item [~~Completeness.] For every $\secp,n\in \mathbb{N}$ and every $x\in \{0,1\}^n$ and $w \in \{0,1\}^m$ such that $M(x,w) = 1$, 
    \[  
    \probcond{\SNARG.\cV(\crs, x,\pi) = 1}{ \crs \gets \SNARG.\Gen(1^\secp,n),\\ \pi \gets \SNARG.\cP(\crs,x,w)} = 1.~
    \]
    \item [~~Adaptive Soundness.] For every poly-size adversary $\mathsf{Adv}$,  there is a negligible function $\mu$ such that 
    \[  \probcond{x\in\{0,1\}^n\setminus L_M~\wedge~\SNARG.\cV(\crs, x,\pi^*) = 1}{ \crs \gets \SNARG.\Gen(1^\secp,n),\\ (x,\pi^*)\gets\mathsf{Adv}(\crs).} \leq \mu(\secp).~
    \]
    \item [~~Efficiency.] 
    The prover runs in time polynomial in the security parameter and the
length of the execution. We explicitly specify the lengths of the
$\crs$ and proof~$\pi$, as well as the
verifier's running time.  We say that a $\SNARG$ is \emph{fully succinct} if the length of the $\crs$ and the proof~$\pi$ is $\poly(\secp,\log T)$, and the runtime of the verifier is $n\cdot \poly(\secpar, \log T)$.
  \end{description}
\end{definition}


\subsection{Somewhere Extractable Batch Arguments ({\seBARG}s)}
\label{sec:prelim:BARG}

\newcommand{\batchtmsat}{\mathsf{BatchIndexTMSAT}}

A batch argument system ($\mathsf{BARG}$) for an $\NP$ language
$L$ allows proving that $k$ NP statements are true
with communication cost polylogarithmic in $k$. $\mathsf{BARG}$ and their applications
to succinct non-interactive arguments have been
studied in
\cite{FOCS:ChoJaiJin21,TCC:KalVaiZha21,KalaiLVW23}.
In this work, we use the somewhere-extractable
variant of $\mathsf{BARG}$s for ``batch index Turing machine SAT'' ($\batchtmsat$), defined below.

\begin{definition}\label{def:batchtmsat}\cite{C:BBKLP23}
The language $\batchtmsat$ consists of instances of the form $x = (M, z, k, T)$,
    where:
    \begin{itemize}
       \item $M$ is the description of a deterministic
Turing machine that takes as input $(z,i,w)$,
where $w$ is a witness.
        \item $z$ is an input string.
        \item $k$ is a batch size.
        \item $T$ is a running time.
    \end{itemize}   
An instance $x=(M,z,k,T)$ is in
$\batchtmsat$ if for every $i\in[k]$
there exists a witness $w_i$ such that
$M(z,i,w_i)$ accepts within $T$ steps.
\end{definition}
For an instance $x=(M,z,k,T)$, we write
$R(x,i,w)=1$ if and only if $i\in[k]$
and $M(z,i,w)$ accepts within $T$ steps.

\paragraph{Syntax.} A (publicly verifiable and non-interactive) somewhere extractable batch argument system $\seBARG$ for $\batchtmsat$
consists of the following polynomial time algorithms:
\begin{description}
  \item[$\Gen(\secparam, n, 1^m, k, i^*) \to (\crs, \td).$] This is a probabilistic polynomial-time algorithm that takes as input a security parameter~$\secparam$, input length $n$, witness length $1^m$, a parameter $k$, and an index $i^* \in [k]$. It outputs a common reference string \crs along with a trapdoor \td. 
  \item[$\cP(\crs, M, z, 1^T, w_1, \ldots, w_k) \to \pi.$] This deterministic polynomial-time algorithm takes as input a $\crs$, Turing machine $M$, input $z$, runtime $1^T$, and $k$ witnesses $w_1, \ldots, w_k$. It outputs a proof $\pi$.
  \item[$\cV(\crs, x,\pi) \to 0/1.$] This deterministic polynomial-time algorithm takes as input a $\crs$, instance $x=(M, z, k, T)$, and a proof $\pi$. It outputs a bit ($1$ to accept, $0$ to reject). 

  \item[$\Extract(\td,\pi)\rightarrow w^*.$]  This deterministic polynomial-time algorithm takes as input a trapdoor $\td$ and a proof~$\pi$.  It outputs a single witness~$w^*$.
\end{description}

\begin{definition}[\seBARG]
\label{def:barg}
  A somewhere-extractable batch argument system  $\seBARG = (\Gen, \cP, \cV,\Extract)$ for $\batchtmsat$ is required to satisfy the following properties:
  \begin{description}
    \item[~~Completeness.] For any $\secpar \in \bbN,$ any $k(\secpar), n(\secpar), m(\secp), T(\secp) \leq 2^\secpar$, any instance $x=(M, z, k, T)\in \batchtmsat$ with $|M| + |z| = n$, any corresponding witnesses $w_1, \ldots, w_k \in \bit^m$ and any index $i^* \in [k]$,
    \[
      \probcond{\cV(\crs, x, \pi) = 1}{
          (\crs, \td) \gets \Gen(\secparam, n, 1^m, k,i^*), \\
          \pi \gets \cP(\crs, M, z, 1^T, w_1, \hdots, w_k)
      } = 1.
    \]
        \item[~~Efficiency.] 
    In the completeness experiment above,
\[
|\crs|+|\td|+|\pi|
\leq m\cdot\poly(\secpar).
\]
The running time of the verifier is at most
\[
\poly(|\crs|+|\pi|)
+n\cdot \poly(\secpar).
\]
    \item {\bf Extraction completeness.}
For every $\secp\in\mathbb N$,
every $k,n,m,T\leq 2^{\secp}$,
every instance $x=(M,z,k,T)\in
\mathsf{BatchIndexTMSAT}$ with
$|M|+|z|=n$, every corresponding
witnesses $w_1,\ldots,w_k\in\{0,1\}^m$,
and every index $i^*\in[k]$,
\[
\Pr\left[
(x,i^*,w^*)\in \mathsf{R}:
\begin{array}{l}
(\crs,\td)\gets
\mathsf{Gen}(1^{\secp},n,1^m,k,i^*),\\
\pi\gets
\mathcal{P}(\crs,M,z,1^T,w_1,\ldots,w_k),\\
w^*\gets\mathsf{Extract}(\td,\pi)
\end{array}
\right]
\geq 1-\negl(\secp).
\]

   \item[~~Index hiding.] For any poly-size adversary $\cA$ and any polynomials $k(\secp),n(\secp), m(\secp)$, there exists a negligible function $\negl(\cdot)$ such that for every $\secpar \in \bbN$ and every pair of indices $i_0,i_1\in[k]$,   
    \[
      \probcond{\cA(\crs) = b}{
          b \gets \bit, \\
          (\crs, \td) \gets \Gen(\secparam, n, 1^m, k, i_b)
      } \leq \frac{1}{2} + \negl(\secpar).
    \]
   
    \item[~~Somewhere argument of knowledge.] For any poly-size adversary \cA and any  polynomials $k(\secpar), n(\secp), m(\secp), T(\secp)$ there exists a negligible function $\negl(\cdot)$ such that for any index $i^*\in[k]$ and for every $\secpar \in \bbN$,
    \[
      \probcond{
         \cV(\crs, x, \pi) = 1 \\
          \wedge\ (x, i^*, w^*)\not\in \mathsf{R}\\
           \wedge\ |M|+|z|=n
        }{
                   (\crs, \td) \gets \Gen(\secparam, n, 1^m, k, i^*) \\
          (M, z, \pi) =\cA(\crs) \\
          w^* \gets \Extract\left(\td, \pi\right)
      } \leq \negl(\secpar).
    \]
    where above $x=(M,z,k,T)$. 

  \end{description}
\end{definition}

\begin{remark} We say that a $\seBARG$ scheme is $T^*$-secure, for $T^*=T^*(\secpar)$, if the index hiding property and the somewhere argument of knowledge property hold w.r.t. a $\poly(T^*)$-size adversary (as opposed to a $\poly(\secp)$-size), and the advantage probability is $\negl(T^*)$ (as opposed to $\negl(\secpar)$). In addition, the index hiding property and the somewhere argument of knowledge property are required to hold even if the parameters $k(\secp),n(\secp),m(\secp),T(\secp)$ are polynomial in $T^*(\secp)$.   We refer to these properties as $T^*$-index-hiding and $T^*$-somewhere-argument-of-knowledge, respectively.
\end{remark}

\noindent Throughout this paper, when we refer to a $\BARG$ or $\seBARG$, we will implicitly mean a $\seBARG$ for $\batchtmsat$.

\begin{remark}\label{remark:barg-I}  
Given an $\seBARG$, one can naturally extend the definition of the key generation algorithm $\Gen$ to take as input a \emph{vector} $I=(i_1,\ldots,i_\ell)$ where each $i_j\in[k]$, as opposed to a single index.   $\Gen(1^\secpar, n, 1^m, k,I)$ will simply run $\Gen(1^\secpar, n, 1^m, k,i_j)$ for every $j\in[\ell]$. The prover algorithm  $\cP$, given a $\crs$ that encodes the $|I|$ indices, will simply generate $|I|$ proofs (one for each $\crs$), and the verifier will check these $|I|$ proofs independently. In this situation, the $\seBARG$ trapdoor $\td = (\td_1, \hdots, \td_{|I|})$ is written in $|I|$ components, and the scheme inherits a strong form of index hiding: given $\seBARG.\crs$, the $j$th index $i_j$ remains hidden even given all $|I|-1$ other trapdoors
$(\td_{j'})_{j'\in[\ell]\setminus\{j\}}$.
\end{remark}

\begin{theorem}[\cite{FOCS:ChoJaiJin21,C:WatWu22,EC:HJKS22,KalaiLVW23}]\label{thm:cjjbarg}
There exists an $\seBARG$ for $\batchtmsat$ assuming $\LWE$ or $\DLIN$ or (subexponential $\DDH$ and $\QR$).
\end{theorem}

\section{Oracle-Aided Turing Machines}
\label{sec:oracle-machines}

An oracle-aided Turing machine $M$ is a Turing machine equipped with
an oracle query tape and an oracle answer tape. Upon entering a designated
query state, the contents $u$ of the query tape are sent to an oracle
$\mathcal O$, and the contents of the answer tape are replaced by
$\mathcal O(u)$. Queries may be adaptive. Namely, the query made at a
particular step may depend on the answers to previous queries.

We allow $M$ to be non-deterministic. We write $M^{\mathcal O}(x,w)$ for
the outcome of its execution on input $x$, where $w$ specifies its
non-deterministic choices, and define
\[
 L_M^{\mathcal O}
 =\{x:\exists w\text{ such that }M^{\mathcal O}(x,w)=1\}.
\]
The deterministic setting is the special case in which there are no
non-deterministic choices. 

\paragraph{Complexity measures.}
We consider machines with time bound $T=T(n)$,\footnote{We assume throughout that $T\geq\max\{n,S\}$ and $T>\secp$,
by increasing the time bound if necessary.} space bound $S=S(n)$,
and oracle-query bound $Q=Q(n)\leq T(n)$ on inputs of length $n$.
The space bound includes the work tapes and the oracle query and answer
tapes, but not the read-only input tape. In particular, the lengths of
queries and answers are at most $S$. An oracle call counts as one step;
writing a query and subsequently processing its answer are charged in
the usual way. 

A configuration specifies the machine's state, tape contents, and head
positions. It does not include the input tape. The non-deterministic choices are made as the computation proceeds;
we do not assume that a long, random-access witness is available outside
the space bound.
\begin{remark}[Configuration counters]
\label{rem:configuration-counters}
We assume that every configuration $C$ includes
a step counter $\operatorname{time}(C)$ and an oracle-call
counter $\operatorname{queries}(C)$.
Both counters are zero in $C_{\mathrm{init}}(x)$.
At every transition, the step counter increases by one.
The oracle-call counter increases by one at an oracle
transition and remains unchanged at an ordinary transition.
These counter updates are part of the transition rules
and are required also in purported executions, regardless
of whether the oracle answers are correct.
The two counters together require only $O(\log T)$ bits,
so a configuration can be represented using $O(S+\log T)$ bit.
\end{remark}
\begin{definition}[Purported execution]\label{def:puruported-exec}
    A \emph{purported execution} of $M^{\mathcal O}$ on input $x$ is a sequence of configurations
\[
 \tau=(C_0,\ldots,C_T)
\]
that starts in the prescribed initial configuration and respects all
ordinary transition rules. At an oracle step, however, the execution
may use any answer of the prescribed length. In particular, we do not
require that two occurrences of the same query receive the same answer.
The transcript records the answer used at each oracle step, through its
configurations. We say that $\tau$ is accepting if $C_T$ is accepting.
\end{definition}
Let $q(\tau)$ be the number of oracle calls in $\tau$, and let
$(u_1,a_1),\ldots,(u_{q(\tau)},a_{q(\tau)})$ be its query-answer pairs,
in their order of occurrence. Define
\[
 \err_{\mathcal O}(\tau)
 =\bigl|\{j\in[q(\tau)]:a_j\neq\mathcal O(u_j)\}\bigr|.
\]
Notice that queries are evaluated along the purported execution itself,
not along an honest execution with a different sequence of answers.

\begin{definition}[Robust oracle-aided computation]
\label{def:oracle-robustness}
Let $\varepsilon\in(0,1]$. We say that $M^{\mathcal O}$ is
\emph{$\varepsilon$-robust on input $x$} if either $x\in L_M^{\mathcal O}$,
or every accepting purported execution $\tau$ of $M$ on $x$ satisfies
\[
 \operatorname{err}_{\mathcal O}(\tau)\geq\varepsilon q(\tau).
\]
\end{definition}

This is a one-sided notion: changing a rejecting instance into an
accepting execution requires an $\varepsilon$-fraction of incorrect oracle
answers. The quantification is over every accepting purported execution,
including every sequence of non-deterministic choices. The fraction is
measured among oracle calls, and not among all $T$ computation steps.
An accepting purported execution with no oracle calls is already a real
accepting execution, and hence cannot exist for a no-instance.

Our soundness guarantees have an input-wise formulation: a verifier
accepts a false statement only with negligible probability when the
chosen input satisfies the robustness condition. They do not require
the verifier to determine whether that condition holds.

\subsection{Definition of Privately-Verifiable $\SNARG$s for Robust Oracle-Aided Computations}
\label{sec:private-oracle-snargs}

We extend the definition of a $\SNARG$ to oracle-aided computations with
private verification. Fix an oracle-aided machine $M$ and parameters
$(n,T,S,Q,\varepsilon)$ as above. A $\SNARG$ for oracle-aided computations consists of algorithms
\(\SNARG.\Gen\), \(\SNARG.\cP\), and
\(\SNARG.\cV\) with the following syntax.

\begin{description}
\item[$\SNARG.\Gen(1^\lambda,n,\varepsilon)
      \longrightarrow(\crs,\td)$.]
The randomized setup algorithm outputs a common reference string $\crs$
and a private verification key $\td$. $\SNARG.\Gen$ does not query the oracle.

\item[$\SNARG.\cP^{\mathcal O}(\crs,x,w)
      \longrightarrow\pi$.]
The prover is given the $\crs$, an input $x\in\{0,1\}^n$,
and non-deterministic choices $w$ specifying an accepting execution of
$M^{\mathcal O}$ on $x$. It has oracle access to $\mathcal O$ and
outputs a proof $\pi$.

\item[$\SNARG.\cV^{\mathcal O}(\crs,\td,x,\pi)
      \longrightarrow 0/1$.]
The verifier is given the $\crs$, the private verification
key~$\td$, the input~$x$, and the proof~$\pi$. It has oracle access to $\mathcal O$ and
outputs a bit indicating whether it accepts.
\end{description}

\begin{definition}[Privately verifiable oracle-aided $\SNARG$]
\label{def:private-oracle-snarg}
A system as above is a privately verifiable $\SNARG$ for $\varepsilon$-robust
oracle-aided computation~$M$ if the following properties hold for every
oracle $\mathcal O$ fixed independently of setup.

\paragraph{Completeness.}
For every $x\in\{0,1\}^n$ and every $w$ such that $M^{\mathcal O}(x,w)=1$,
\[
 \Pr\!\left[
 \begin{array}{c}
 \SNARG.\cV^{\mathcal O}(\crs,\td,x,\pi)=1
 \end{array}
 \;:\;
 \begin{array}{l}
 (\crs,\td)\leftarrow
\SNARG.\Gen(1^\secp,n,\varepsilon),\\
 \pi\leftarrow\SNARG.\cP^{\mathcal O}(\crs,x,w)
 \end{array}\right]
 \geq1-\negl(\secp).
\]
We also require adaptive completeness: the
probability of rejecting an accepting execution remains negligible when
$(x,w)$ is chosen by a polynomial-size oracle algorithm after seeing
$\mathsf{crs}$.

\paragraph{Adaptive soundness.}
For every polynomial-size oracle adversary $\mathsf{Adv}$, there exists
a negligible function $\mu$ such that
\[
 \Pr\!\left[
 \begin{array}{l}
 x\in \{0,1\}^n\setminus L_M^{\mathcal O},\\
 M^{\mathcal O}\text{ is $\varepsilon$-robust on }x,\\
 \SNARG.\cV^{\mathcal O}(\crs,\td,x,\pi^*)=1
 \end{array}
 \;:\;
 \begin{array}{l}
 (\crs,\td)\leftarrow
       \SNARG.\Gen(1^\secp,n,\varepsilon),\\
 (x,\pi^*)\leftarrow\mathsf{Adv}^{\mathcal O}(\crs)
 \end{array}\right]
 \leq\mu(\secp).
\]

\paragraph{Efficiency.}
The prover runs in time $\poly(\secp,T, 1/\varepsilon)$.
For the space-dependent $\SNARG$s constructed in this paper,
the lengths of $\crs$, the private verification key $\td$,
and the proof $\pi$ are bounded by
$\poly(\lambda,S,\log T,1/\varepsilon)$.
The verifier runs in time
$n\cdot\poly(\lambda,S,\log T,1/\varepsilon)$
and makes $O(\lambda/\varepsilon)$ oracle calls.
We say that a SNARG is fully succinct if it achieves
these efficiency guarantees without the dependence on $S$.
\end{definition}

\begin{remark}
    Note that if we restrict $\td=\bot$ then \Cref{def:private-oracle-snarg} becomes a publicly verifiable $\SNARG$ for an oracle-aided computation. 
\end{remark}

\section{Privately-Verifiable $\SNARG$s for Bounded-Space Robust Oracle-Aided Computations}
\label{sec:robust-oracle-construction}

In this section, we construct a privately-verifiable $\SNARG$ for robust
oracle-aided computations. The idea is to hash the entire computation
transcript, including the oracle answers, and use a $\seBARG$ to prove
consistency of its local transitions. The verifier uses its trapdoor
to extract a few query-answer pairs and checks them against the oracle.


\begin{theorem}
\label{thm:robust-oracle-snarg}
Let $M$ be a non-deterministic oracle-aided machine with time bound $T$ and space bound $S$.  Assume that $M$ always makes exactly $Q\leq T$ oracle queries.
There exists a privately verifiable
$\SNARG$ for $\varepsilon$-robust computations of $M$, with adaptive soundness assuming the $\LWE$ problem is sub-exponentially hard even given oracle access.
Its public parameters, private verification key, and proof have length
\[
 \operatorname{poly}(\lambda,S,\log T,1/\varepsilon).
\]
The verifier runs in time
$n\cdot \poly(\lambda,S,\log T,1/\varepsilon)$ and makes
$O\!\left(\lambda/\varepsilon\right)$
oracle calls. 
\end{theorem}

In what follows we fix an oracle \(\mathcal O\) independently of the setup. Let \(M\) be an oracle-aided machine with time bound \(T\) and space bound \(S\), such that every purported execution on an \(n\)-bit input makes exactly \(Q=Q(n)\geq1\) oracle calls. We assume that $\LWE$ is sub-exponentially hard against non-uniform adversaries with access to \(\mathcal O\).

Our construction uses the hash family from \Cref{sec:HT} and the $\seBARG$ from \Cref{sec:prelim:BARG}  with succinct extraction, instantiated with
$T^*$-security, where
\[
 T^*=2^{O(S)}\poly(T,\lambda,1/\varepsilon).
\]
Thus, the underlying security parameter of the hash family    and the  $\seBARG$, which we denote by $\kappa$, can be chosen as a
sufficiently large polynomial in
$\lambda,S,\log(T), \log(1/\varepsilon)$, by relying on the hardness of $\LWE$ against sub-exponential size adversaries with access to \(\mathcal O\).

\subsection{The Construction}
\label{sec:robust-oracle-protocol}
Let $M$ be a non-deterministic Turing machine running in time~$T$ and space~$S$. In what follows we construct a privately-verifiable $\SNARG$ for $M^{\mathcal{O}}$, for an arbitrary oracle $\mathcal{O}$.  Our construction makes use of the following ingredients:
\begin{itemize}
    \item A hash function with local opening:
    \[
    \HT=(\HT.\Gen,\HT.\Hash,\HT.\Open,\HT.\Verify).
    \]
    \item A somewhere extractable batch argument $\seBARG$:
    \[
    \seBARG=(\seBARG.\Gen,\seBARG.\cP,\seBARG.\cV, \seBARG.\Extract).
    \]
\end{itemize}

We next describe our $\SNARG$ construction for the computation $M^{\mathcal{O}}$.
\begin{itemize}
    \item $\SNARG.\Gen$: This algorithm takes as input $(1^\secp,n,\varepsilon)$, and does the following:
    \begin{enumerate}
        \item Generate $\hk\gets \HT.\Gen(1^\kappa)$, where $\kappa=\poly(\secp, S, \log T, \log \lceil{1/\varepsilon}\rceil)$.
        \item Let $r=\lfloor\lambda/\varepsilon\rfloor$, and sample independently at random  $j_1,\ldots,j_r\gets [Q]$, and let $J=(j_1,\ldots,j_r)$.

\item  \label{item:snarg-gen-sebarg}
Generate
\[
 (\crs_\seBARG,\td_\seBARG)
 \leftarrow\seBARG.\Gen
 (1^\kappa,{n_{\mathrm{loc}}},1^m,k,(T+j)_{j\in J}),
\]
where $k=T+Q$ and where $n_{\mathrm{loc}}$ and $m$ are defined in Step~\ref{item:R_loc} of $\SNARG.\cP$ below.\footnote{Jumping ahead, we note that $n_{\mathrm{loc}}=n+|\hk|+|\rt|+\log (T+Q)$ and $m=\poly(S,\secp,\log(T),\log( 1/\varepsilon))$.}
\item Let
$\crs=(\hk,\crs_\seBARG)$ and $\td=(J,\td_\seBARG)$.
\item Output $(\crs,\td)$.
    \end{enumerate}

\item $\SNARG.\cP$: This algorithm takes as input $(\mathsf{crs},x,w)$, and does the following:
\begin{enumerate}
    \item Run the accepting execution of
$M^{\mathcal O}(x,w)$.

Let $C_0,\ldots,C_T$ be the configurations of this execution,
and let $t_j$ be the step at which its $j$th oracle call occurs. Define
\[
 D_j=(t_j,u_j,a_j),\qquad j\in[Q],
\]
where $u_j$ is the query and $a_j$ is the answer used at that step.
\item Let  $\tau=(C_0,\ldots,C_T,D_1,\ldots,D_Q)$, and compute
\[
\rt=\HT.\Hash(\hk,\tau).
\]
\begin{remark}
    We use fixed-length encodings of configurations
and oracle-table entries, so that their
positions in the encoded transcript are
publicly determined. 
\end{remark}
\item Define $z=(\hk,x,\rt)$.
\item Define the oracle-free $\NP$ relation
$R_\loc$ where $(z,i,w_i)\in R_\loc$ if and only if the following holds:
\begin{itemize}
\item If $i\in[T]$ and the $i$th step is not an oracle step, then 
$w_i=(C_{i-1},C_i,\rho)$ such that the following holds:
\begin{enumerate}
    \item $\Verify(\hk,\rt,I,(C_{i-1},C_i),\rho)=1$, where $I$ are the coordinates in $\tau$ corresponding to $(C_{i-1},C_i)$.
    \item $(C_{i-1},C_i)$ is a valid machine
transition and the configurations $C_{i-1}$ and $C_i$ have the same oracle counter; i.e., $\mathsf{query}(C_{i-1})=\mathsf{query}(C_i)$.
\item If $i=1$, $C_0$ is the initial
configuration and $\mathsf{query}(C_0)=0$, and if $i=T$ then $C_T$ is an accepting configuration. 
\end{enumerate}
\item If $i\in[T]$ and the $i$th step is an oracle step, say the $j$th oracle step,  then $w_i=(C_{i-1},C_i,D_j,\rho)$ such that the following holds:
\begin{enumerate}
    \item $\Verify(\hk,\rt,I,(C_{i-1},C_i,D_j), \rho)=1$, where $I$ are the coordinates in $\tau$ corresponding to $(C_{i-1},C_i,D_j)$.
    \item $(C_{i-1},C_i)$ is a valid machine transition, and
\[
    j=\mathsf{query}(C_i)
     =\mathsf{query}(C_{i-1})+1\in[Q].
\]
    
\item $D_j=(i, u_j, a_j)$, where $(u_j,a_j)$ are the  query and answer in the 
configurations $C_{i-1}$ and $C_i$, respectively. 
\item If $i=1$, $C_0$ is the initial
configuration and $\mathsf{query}(C_0)=0$, and if $i=T$ then $C_T$ is an accepting configuration. 
\end{enumerate}

\item If $i=T+j$ for $j\in[Q]$ then $w_i=(C_{t_j-1}, C_{t_j},D_j,\rho)$ such that the following holds:
\begin{enumerate}
    \item $\Verify(\hk,\rt,I,(C_{t_j-1}, C_{t_j}, D_j), \rho)=1$, where $I$ are the coordinates in $\tau$ corresponding to $(C_{t_j-1}, C_{t_j}, D_j)$.
    \item $D_j=(t_j,u_j,a_j)$ where $t_j\in [T]$.
    \item $(C_{t_j-1}, C_{t_j})$ is a valid machine transition that increases the oracle call from $j-1$ to $j$, i.e., $\mathsf{query}(C_{t_j-1})=j-1$ and $\mathsf{query}(C_{t_j})=j$, and the configurations are associated with query $u_j$ and answer $a_j$.
\end{enumerate}

\end{itemize}

\begin{remark}
Note that the
relation $R_\loc$ does \emph{not} check $a_j=\mathcal O(u_j)$.
Thus it is an ordinary $\NP$ relation, not an oracle-aided one.
\end{remark}

\item \label{item:R_loc} Construct the witnesses $w_1,\ldots,w_{T+Q}$ for the $\NP$ relation
$R_{\mathrm{loc}}$.
Compute
\[
 \pi_\seBARG=\mathsf{seBARG.P}
 (\crs_\seBARG,M_\loc,z,1^{T_\loc},w_1,\ldots,w_{T+Q}),
\]
where $M_{\rm loc}$ is the deterministic
Turing machine that, on input $(z,i,w_i)$,
accepts if and only if
$(z,i,w_i)\in R_{\loc}$,
and $T_{\loc}$ is a bound on its runtime.

Let \(n_{\mathrm{loc}}\) and \(m\) be fixed upper bounds on \(|M_{\mathrm{loc}}|+|z|\) and the length of a local witness, respectively. We pad the common input and witnesses to these lengths.

\item Output $\pi=(\mathsf{rt},\pi_\seBARG)$.
\end{enumerate}

\item $\SNARG.\cV$: This algorithm takes as input a tuple $(\crs,\td,x,\pi)$ and does the following:
\begin{enumerate}
\item Parse $\crs=(\hk,\crs_\seBARG)$, $\td=(J,\td_\seBARG)$ and $\pi=(\rt,\pi_\seBARG)$.
\item Let $z=(\hk,x,\rt)$.
    \item Check that $\seBARG.\cV(\crs_\seBARG,M_{\mathrm{loc}},z,T+Q,T_\loc,\pi_\seBARG)=1$. If this check fails, then reject. 
    \item Otherwise, compute
\[
 (w_{T+j})_{j\in J}=
 \seBARG.\Extract(\td_\seBARG,\pi_\seBARG),
\]
and check that for every $j\in J$ it holds that  $R_\loc(z,T+j,w_{T+j})=1$, and let $(u_j,a_j)$ be the
query-answer pair corresponding to $w_{T+j}$.  Accept if and only if all these checks pass and
\[
 a_j=\mathcal O(u_j)\quad\text{for every }j\in J.
\]
\end{enumerate}
\end{itemize}

\subsection{Analysis}
\label{sec:robust-oracle-analysis}



\paragraph{Completeness.}
Fix an accepting execution of $M^{\mathcal O}(x,w)$, and let $\tau$
be the transcript and oracle table constructed by the honest prover.
By opening completeness, the witnesses constructed from $\tau$
satisfy $R_{\mathrm{loc}}$. By completeness of the $\seBARG$,
all the public verification checks accept.

It remains to consider the extracted witnesses. By the somewhere
argument of knowledge property, all $r$ extracted witnesses satisfy
$R_{\mathrm{loc}}$, except with negligible probability.
Moreover, their opened configurations and table entries agree with
$\tau$, except with negligible probability, as otherwise, an extracted
opening and the corresponding honest opening give a collision with
respect to opening. Every extracted pair $(u_j,a_j)$ therefore
satisfies $a_j=\mathcal O(u_j)$, and the verifier accepts. The same
argument applies when $(x,w)$ is chosen after seeing $\crs$.

\paragraph{Adaptive soundness.}
Fix an oracle $\mathcal O$ satisfying the oracle-security requirement,
an oracle-aided machine $M$, and a robustness parameter $\varepsilon$
whose representation has length $\ell_\varepsilon\leq\poly(T)$.
Fix a polynomial-size oracle adversary $\mathsf{Adv}$.
The adversary receives the $\crs$ and outputs an instance $x$ and a proof
$\pi^*=(\rt,\pi_{\seBARG})$.

For every $s\in\{0,\ldots,r\}$, let $E_s$ be the event that
$x\notin L_M^{\mathcal O}$, $M^{\mathcal O}$ is $\varepsilon$-robust
on $x$, all the public $\seBARG$ verification checks accept, and
the private checks in components $1,\ldots,s$ accept. A private check
includes both checking the extracted local witness and checking its
oracle answer. Let
\[
 p_s=\Pr[E_s]
\]
in the real experiment. Thus, $p_r$ is the soundness error.

We prove that for every $s\in[r-1]$,
\begin{equation}
 p_s\leq(1-\varepsilon)p_{s-1}+\negl(T^*).
 \label{eq:robust-one-more-check}
\end{equation}
Note that
iterating~\Cref{eq:robust-one-more-check}, and using the fact that $p_0\leq 1$ and $r=\lfloor{\lambda/\varepsilon}\rfloor$, implies that 
\[
p_r\leq p_{r-1}\leq (1-\varepsilon)^{r-1}\cdot p_0 +(r-1)\cdot \negl(T^*)= 
(1-\varepsilon)^{\left({\lfloor{\secp/\varepsilon}\rfloor}-1\right)}\cdot p_0+\left(\lfloor{\lambda/\varepsilon}\rfloor-1\right)\cdot\negl(T^*)=\negl(\secp),
\]
as desired. To complete the adaptive soundness proof, it thus suffices to prove~\Cref{eq:robust-one-more-check}.


\paragraph{Proving \Cref{eq:robust-one-more-check} using a potential function.}
For each input $x$, 
consider the configuration graph of all purported executions (\Cref{def:puruported-exec}) of $M$
on $x$. We include an oracle-call counter in each configuration.
An ordinary edge corresponds to a valid ordinary transition of
$M$, including any permitted non-deterministic choice. At an
oracle step with query $u$, we include an edge for every answer
$a$ of the prescribed length, regardless of whether
$a=\mathcal{O}(u)$.

Assign each edge $e$ the cost
\[
    c(e)=
    \begin{cases}
        0, & \text{if $e$ is an ordinary transition},\\
        \mathbf{1}[a\ne\mathcal{O}(u)],
          & \text{if $e$ uses answer $a$ to query $u$}.
    \end{cases}
\]
Thus, the cost of a path starting at the initial configuration
equals the number of incorrect oracle answers in the corresponding
purported execution.

For every $t\in\{0,\ldots,T\}$ and configuration $C$, let
$d^x_t(C)$ be the minimum cost of a path of length $t$ from the
initial configuration to $C$, with the minimum of an empty set
defined to be $+\infty$. These values can be computed by dynamic
programming as follows:
\[
    d^x_0(C)=
    \begin{cases}
        0, & C=C_{\mathrm{init}}(x),\\
        +\infty, & \text{otherwise},
    \end{cases}
\]
and, for each $t<T$,
\[
    d^x_{t+1}(C')
      = \min_{e:C\to C'} \bigl(d^x_t(C)+c(e)\bigr).
\]
Define the bounded potential
\[
 \Phi_t^x(C)=\min\{d_t^x(C),Q\},
\]
where $\min\{+\infty,Q\}=Q$.

For every valid purported transition $C\longrightarrow C'$ at
step $t$, of cost $c$, we have
\begin{equation}
 \Phi_t^x(C')\leq\Phi_{t-1}^x(C)+c.
 \label{eq:robust-local-potential}
\end{equation}
Also, $\Phi_0^x(C_{\mathrm{init}}(x))=0$. If $x$ is an $\varepsilon$-robust
no-instance, then every accepting configuration $C$ satisfies
\begin{equation}
 \Phi_T^x(C)\geq\varepsilon Q.
 \label{eq:robust-terminal-potential}
\end{equation}
Indeed, every accepting purported execution has $Q$ oracle calls
and at least $\varepsilon Q$ incorrect answers. If $C$ is unreachable, then
$d_T^x(C)=+\infty$, and hence
$\Phi_T^x(C)=Q\geq\varepsilon Q$.

\begin{claim}\label{clm:testing-robustness}
The entire table of potentials, as well as the predicate that $x$
is an $\varepsilon$-robust no-instance, can be computed with oracle access to
$\mathcal O$ in time $\poly(2^S,n,T,\ell_\varepsilon)\leq \poly(T^*)$.
\end{claim}

The proof of \Cref{clm:testing-robustness} follows by using a standard dynamic
programming algorithm. \\

To prove \Cref{eq:robust-one-more-check}, recall that $p_s$ is the
probability that $E_{s-1}$ occurs and the $s$th
private check accepts. Thus, it suffices to prove that
\[
\Pr[E_{s-1}~\land~ \text{the $s$th private check rejects}]
\geq \varepsilon p_{s-1}-\negl(T^*).
\]
We prove this using a sequence of hybrids, where
we exploit index hiding to change the extraction
indices of two $\seBARG$ components. We first show,
using the potential function, that the expected
total cost of the extracted transitions is at
least $\varepsilon Qp_{s-1}$, up to a negligible error.
We then relate this cost to incorrect oracle
answers in the table, implying that the $s$th
private check rejects with probability at least
$\varepsilon p_{s-1}$, up to a negligible error.

\paragraph{Proof of \Cref{eq:robust-one-more-check}.}
Fix $s\in[r-1]$, and let $E=E_{s-1}$.
We use the $s$th and $r$th $\seBARG$ components,
whose trapdoors are not used in defining $E$.
For $a,b\in[T+Q]$, let $H_{a,b}$ denote the
experiment in which these components have
extraction indices $a$ and $b$, respectively.
Denote by $\star$ a uniformly
sampled index in
$\{T+1,\ldots,T+Q\}$.

For every $t\in[T]$, consider $H_{\star,t}$,
and let $(C_{t-1},C_t)$ be the configurations
extracted from component~$r$.
Let $c_t\in\{0,1\}$ be the cost of this transition (as defined above).  If the witness extracted from component $r$
does not satisfy $R_{\rm loc}$, we set the
potential values and transition cost associated
with this extraction to zero.
Thus, all the random variables below are
well-defined even when extraction fails. Set
\[
\begin{aligned}
F_t^- &=
\mathbb{E}_{H_{\star,t}}
[1_E\cdot\Phi_{t-1}^x(C_{t-1})],\\
F_t^+ &=
\mathbb{E}_{H_{\star,t}}
[1_E\cdot\Phi_t^x(C_t)],\\
W_t &=
\mathbb{E}_{H_{\star,t}}[1_E\cdot c_t].
\end{aligned}
\]
By \Cref{eq:robust-local-potential},
\[
F_t^+\leq F_t^-+W_t+\negl(T^*).
\]
Moreover, the initial and terminal checks,
together with \Cref{eq:robust-terminal-potential}, imply
\[
F_1^-=0,\qquad
F_T^+\geq\varepsilon Qp_{s-1}-\negl(T^*).
\]

For every $j\in[Q]$, consider
$H_{T+j,\star}$, and let
$D_j=(t_j,u_j,a_j)$ be the oracle-table
entry extracted from component $s$.
Set
\[
b_j=1[a_j\neq O(u_j)],
\qquad
G_j=\mathbb{E}_{H_{T+j,\star}}[1_E\cdot b_j],
\]
where $b_j=0$ if extraction is invalid.

We next argue that the following two
equations hold:
\[
\begin{aligned}
F_t^+ &= F_{t+1}^- \pm \negl(T^*),
&& t\in[T-1],\\
\sum_{t=1}^T W_t
&=\sum_{j=1}^Q G_j\pm\negl(T^*).
\end{aligned}
\]
This follows from $T^*$-index hiding, the
somewhere argument-of-knowledge property,
and collision resistance with respect to opening,
as follows. By \Cref{clm:testing-robustness}, the event $E$
and the potential values used in these equations
can be computed in time $\poly(T^*)$ with oracle
access to $O$. Thus, $T^*$-index hiding allows
us to change an extraction index whenever
the quantity being evaluated does not depend
on the corresponding trapdoor. Moreover,
whenever $E$ occurs, the extracted witnesses
are valid, and any two valid openings of the
same position agree, except with probability
$\negl(T^*)$.

To prove the first equality, we consider the
sequence of hybrids
\[
H_{\star,t},\quad H_{t+1,t},\quad
H_{t+1,t+1},\quad H_{\star,t+1}.
\]
We first change the extraction index of
component $s$ from $\star$ to $t+1$,
while evaluating the potential using
component $r$. In $H_{t+1,t}$, component
$r$ opens $C_t$ as the configuration after
step $t$, whereas component $s$ opens
$C_t$ as the configuration before step $t+1$.
By collision resistance, these configurations
agree except with negligible probability.
We may therefore evaluate the potential
using component $s$, change the extraction
index of component $r$ to $t+1$, and then
switch back to component $r$, since both
components now open the configuration
before step $t+1$. Finally, we change the
index of component $s$ back to $\star$.
Each change incurs an additive error
of at most $\negl(T^*)$, proving
the first equality.

For the second equality, let $A_{t,j}$
indicate that the transition extracted at
time $t$ is valid, is the $j$th oracle
transition, and uses an incorrect oracle
answer. We set $A_{t,j}=0$ if extraction
is invalid. Consider the hybrid
$H_{T+j,t}$, where component $s$ opens
the $j$th oracle-table entry and component
$r$ opens the transition at time $t$.
By the local checks in $R_{\rm loc}$,
if either witness identifies this transition
as the $j$th oracle call, then the two
witnesses open the same configurations
and oracle-table entry. By collision
resistance, these openings agree except
with negligible probability. Consequently,
\[
A_{t,j}=1[t_j=t]\cdot b_j
\]
except with probability $\negl(T^*)$
on the event $E$. By index hiding,
we can change the index of component $s$
back to $\star$ when evaluating $A_{t,j}$,
and change the index of component $r$
back to $\star$ when evaluating
$1[t_j=t]\cdot b_j$. Thus,
\[
\mathbb{E}_{H_{\star,t}}[1_E A_{t,j}]=
\mathbb{E}_{H_{T+j,\star}}
[1_E\cdot 1[t_j=t]\cdot b_j]
\pm\negl(T^*).
\]
Summing over $t$ and $j$ gives the second
equality, since each incorrect oracle
transition contributes to exactly one
$A_{t,j}$, and each oracle-table entry
has exactly one time index $t_j$.
The total error remains negligible,
since $TQ\leq (T^*)^2$.

Combining these inequalities, we obtain
\[
\begin{aligned}
\sum_{j=1}^Q G_j
&\geq \sum_{t=1}^T W_t-\negl(T^*)\\
&\geq \sum_{t=1}^T(F_t^+-F_t^-)
       -\negl(T^*)\\
&\geq F_T^+-F_1^--\negl(T^*)\\
&\geq \varepsilon Qp_{s-1}-\negl(T^*).
\end{aligned}
\]

Finally, in the real experiment the $s$th
private check samples a uniformly random
oracle-table index. Therefore,
\[
\Pr[E~\land~\text{the $s$th check rejects}]\geq \frac1Q\sum_{j=1}^Q G_j\geq \varepsilon p_{s-1}-\negl(T^*).
\]
It follows that
\[
p_s\leq(1-\varepsilon)p_{s-1}+\negl(T^*),
\]
which proves \Cref{eq:robust-one-more-check}.




\paragraph{Efficiency.}
A local witness contains a constant number of configurations and
table entries, together with their local openings. Thus its length
is $m=\poly(\kappa,S,\log T)$. The common input has length
$n+\poly(\kappa,S,\log T)$. Applying the succinctness and succinct
extraction bounds of the underlying primitives to the $r$
components, with the stated choice
$\kappa=\poly(\lambda,S,\log(nT),\log(1/\varepsilon))$, gives
\[
 |\crs|+|\td|+|\pi|
 \leq \poly(\lambda,S,\log(nT),1/\varepsilon).
\]
The verifier checks the $r$ public proofs, extracts $r$ local
witnesses, checks their local relations, and makes at most
$r=O(\lambda/\varepsilon)$ oracle calls. Its running time is
\[
 n\cdot\poly(\lambda,S,\log(nT),1/\varepsilon).
\]
Under the convention $T\geq n$, these are the stated bounds
with $\log T$ in place of $\log(nT)$. The prover runs in time
polynomial in the execution length, the security parameter,
and $1/\varepsilon$.

The argument above concerns the fixed-query-count construction.
An extension to a varying number of oracle calls requires a
corresponding construction and is not obtained by changing the
denominator in the robustness condition to the upper bound $Q$.

\section{Impossibility of Publicly-Verifiable $\SNARG$s for Robust Oracle-Aided Computations}\label{sec:impossibility}

In this section, we prove a lower bound on the oracle-query complexity
of a $\SNARG$ with deterministic public verification
whose public parameters are shorter than the input length.  

\begin{theorem}\label{thm:impossibility}
    There does not exist a $\SNARG$ with adaptive completeness and adaptive soundness for robust oracle-aided computations with deterministic public verification, with $|\crs|=o(n)$, where $n$ is the input length, and where the verifier makes $o(q)$ queries, where $q$ is the number of oracle gates.
\end{theorem}

 More
specifically, we show that a verifier that makes sufficiently fewer
oracle queries than the computation cannot have negligible soundness
error. Our argument does not use any bound on the proof length, and thus holds even for {\em non-succinct} non-interactive arguments, as long as the verifier makes $o(q)$ oracle calls, where $q$ is the number of oracle gates in the computation. It also holds for  deterministic computations that make
non-adaptive oracle queries. 

\begin{proof}

Let $n=n(\lambda)$ and $N=N(\lambda)$ be polynomially bounded, and
consider the oracle-aided computation
\[
 C^\calO(x)=\mathsf{Maj}\bigl(\calO(x,1),\ldots,\calO(x,N)\bigr),
    \qquad x\in\{0,1\}^n.\footnote{Assume that $N$ is odd.}
\]
We refer
to $(\calO(x,1),\ldots,\calO(x,N))$ as the row indexed by $x$. 

Suppose that there exists a  $\SNARG$ $\Pi=(\Gen,\mathcal{P},\mathcal{V})$ for the computation $C^\calO$  where $\mathcal{V}$ makes $\ell\leq N/8$ queries to $\calO$. 
To get a contradiction, the idea is to start with the oracle $\calO_1$ that answers $1$ on every query, so that $C^{\calO_1}(x)=1$ for every $x\in\{0,1\}^n$, and then change some rows $(\calO(x,1),\ldots,\calO(x,N))$ so that the majority is~$0$ without the verifier noticing, thus causing it to accept a false statement.  Intuitively, this is done by changing the oracle on every $(x,i)$ that the verifier does not query.  The tricky part is doing this in a way that is independent of the $\crs$, since the oracle $\calO$  should be fixed before $\crs$ was generated.

To this end, we define $x$ to depend on $\crs$ in a way that $x$ determines $\crs$.  For example, for every $\crs$, let 
\[x_\crs=\crs\,\|\,1 \|\, 0^{n-|\crs|-1}.\footnote{Here we use our assumption that $n> |\crs|$.}
\]
Thus, $\crs\mapsto x_\crs$ is an efficiently invertible encoding into the
input space. 
Fix randomness~$r$ for the prover, and let
\[
    \pi_{\crs,r}=\mathcal{P}^{\calO_1}(\crs,x_\crs;r).
\]
Let $Q_{\crs,r}$ be the set of queries made by  $\mathcal{V}$ in the execution
\[
    \mathcal{V}^{\calO_1}(\crs,x_\crs,\pi_{\crs,r}),
\]
and let
\[
    I_{\crs,r}=\{i\in[N]:(x_\crs,i)\in Q_{\crs,r}\}.
\]
Since $C^{\calO_1}(x)=1$ for every $x$, 
\begin{equation}
\label{eq:short-pp-completeness}
    \Pr_{\crs\leftarrow\Gen(1^\lambda),\,r}
    \bigl[\mathcal{V}^{\calO_1}(\crs,x_\crs,\pi_{\crs,r})=1\bigr]
    \geq 1-\negl(\secp),
\end{equation}
where $r$ is uniform.\footnote{We assume adaptive completeness, since 
$x_\crs$ depends on $\crs$.}

\paragraph{Defining the oracle.}
We cannot modify every row $x_{\crs}$, since
the verifier may query the oracle on rows
other than the one corresponding to its input.
Instead, we modify only a small fraction
of rows, selected using a pairwise-independent
hash function.
Specifically, let $k=\log_2(2\ell)$,\footnote{Assume without loss of generality that $\ell$ is a power of~$2$.} and sample a pairwise-independent
function
\[
    h:\{0,1\}^n\longrightarrow\{0,1\}^k.\footnote{For example, we can take $h(x)=Ax+b$ over
$\mathbb{F}_2$, for uniformly random $A\gets\{0,1\}^{k\times n}$ and $b\gets\{0,1\}^k$. }
\]
We say that a row $x$ is marked if $h(x)=0^k$.
Let $p=2^{-k}=\frac{1}{2\ell}$.
For every $\crs$ and every $i\in[N]$, define
\begin{equation}
\label{eq:short-pp-oracle}
    \calO_{h,r}(x_\crs,i)=
    \begin{cases}
        1, & \text{if }h(x_\crs)\neq 0^k\text{ or }i\in I_{\crs,r},\\
        0, & \text{otherwise}.
    \end{cases}
\end{equation}
All queries not covered by this definition are answered with $1$.
Thus, unmarked rows are all ones. A marked row of the form $x_\crs$
contains ones only at the positions in $I_{\crs,r}$.

We emphasize that this defines an oracle before the $\crs$ is sampled. 

\paragraph{Robustness.}
Note that the computation $C^{\calO_{h,r}}$ is $1/4$-robust
on every input.  This is the case since every row of $\calO_{h,r}$ is either all ones or contains at most
$\ell\leq N/8$ ones. Thus, changing at most $N/4$ oracle answers leaves a row
of the first type with at least $3N/4$ ones, and a row of the second
type with at most $3N/8$ ones, and hence does not change the outcome of the computation. 

\paragraph{The cheating prover.}
Given $\crs$, the adversary $\mathcal{P}^*_r$ outputs
\[
    (x,\pi)=(x_\crs,\pi_{\crs,r}).
\]
That is, it simulates the honest prover on input $x_\crs$, using random
tape $r$ and answering every simulated oracle query with $1$. It does
not query the actual oracle, and does not need to know $h$.

\paragraph{Analysis.}
Fix $\crs,r$ for which
$\mathcal{V}^{\calO_1}(\crs,x_\crs,\pi_{\crs,r})=1$. Let $B_{\crs,r}$ be the set
of other rows visited in this execution.  Namely,
\[
    B_{\crs,r}=\bigl\{u\in\{0,1\}^n\setminus\{x_\crs\}:
        \exists i\in[N]\text{ such that }(u,i)\in Q_{\crs,r}\bigr\}.
\]
We have $|B_{\crs,r}|\leq \ell$. Consider the event
\[
    E_{\crs,r}=\bigl\{h(x_\crs)=0^k\text{ and }
        h(u)\neq 0^k\text{ for every }u\in B_{\crs,r}\bigr\}.
\]
It is easy to see that on this event, the verifier accepts a false statement.
Thus, it remains to argue that event $E_{\crs,r}$ holds with non-negligible probability. To this end, note that the set $B_{\crs,r}$
is defined using $\calO_1$, and is therefore independent of $h$. By
pairwise independence and a union bound,
\begin{align*}
    \Pr_h[E_{\crs,r}]
    &\geq \Pr_h[h(x_\crs)=0^k]
      -\sum_{u\in B_{\crs,r}}
        \Pr_h[h(x_\crs)=0^k~\land~ h(u)=0^k]\\
     &=p-|B_{\crs,r}|\cdot p^2\\
     &\geq p-\ell p^2\\
     &=p-\frac{\ell}{4\ell^2}
    = \frac{1}{2\ell}-\frac{1}{4\ell}=\frac{1}{4\ell}.
     \end{align*}
Together with Equation~\eqref{eq:short-pp-completeness}, this gives
\[
\Pr_{h,r,\,\crs\leftarrow\Gen(1^\lambda)}
\left[
    C^{\calO_{h,r}}(x_\crs)=0\ \land\
    \mathcal{V}^{\calO_{h,r}}(\crs,x_\crs,\pi_{\crs,r})=1
\right]
\geq \frac{1-\negl(\secp)}{4\ell}.
\]
Here $h,r,\crs$ are sampled independently.
By averaging, there exist fixed choices $h^*,r^*$ for which the same
lower bound holds over $\crs$ alone. Setting
$\calO=\calO_{h^*,r^*}$ and
$\mathcal{P}^*=\mathcal{P}^*_{r^*}$ gives us the desired contradiction.

\end{proof}

\section{Publicly-Verifiable $\SNARG$s for Oracle-Aided Computations with Signed  Oracles}\label{sec:signed-oracles}

\begin{definition}[Signed oracle]
\label{def:signed-oracle}
Let $\mathcal O:\{0,1\}^*\to\{0,1\}^*$ be an oracle, and let
\[
    \mathsf{Sig}
    =(\mathsf{Sig.Gen},\mathsf{Sig.Sign},\mathsf{Sig.Verify})
\]
be a signature scheme.
A signed oracle associated with $\mathcal O$ is obtained by generating
\[
    (\vk,\sk)\gets
    \mathsf{Sig.Gen}(1^{\kappa_{\mathrm{sig}}}),
\]
where $\kappa_{\mathrm{sig}}$ is the signature security parameter,
and defining
\[
    \mathcal O^{\mathsf{sig}}(u)=(a,\sigma),
    \qquad
    a=\mathcal O(u),
    \qquad
    \sigma\gets\mathsf{Sig.Sign}(\sk,(u,a)).
\]
The verification key $\vk$ is public, whereas $\sk$ is held only
by the oracle and is unknown to the prover or the verifier.
We require the signature scheme to have perfect completeness
and to be existentially unforgeable against adaptive
chosen-message attacks, even given access to $\mathcal O$.

\end{definition}

Notice that the signature authenticates both the query and the answer.
Also, the signature need not be unique: different calls on the same
query may return different signatures, but they return the same answer.

\begin{remark}[Signature security for larger time bounds]
\label{rem:signed-oracle-security}
Throughout our signed-oracle results with security parameter
$\lambda$ and time bound $T$, we require the oracle's signature
scheme to be existentially unforgeable against adaptive
chosen-message attacks by non-uniform adversaries of size
$\operatorname{poly}(\lambda,T)$, even given access to $\mathcal O$,
with forgery probability negligible in $T$.\footnote{We assume that $T\geq \secp$.}
We require this guarantee to hold with security
parameter
\[
    \kappa_{\mathrm{sig}}
    =\operatorname{poly}(\lambda,\log T).
\]
\end{remark}

\begin{theorem}
\label{thm:signed-oracle-snarg}
Let $M$ be a non-deterministic oracle-aided machine with time bound $T$,
space bound $S$, and oracle-query bound $Q\leq T$. Let
$\mathcal O_{\mathsf{sig}}$ be a signed oracle as in~\Cref{def:signed-oracle}, with the security parameter $ \kappa_{\mathrm{sig}}
    =\operatorname{poly}(\lambda,\log T)$.
Assume that $\LWE$ is sub-exponentially
hard even given access to $\mathcal O$.

There exists a (publicly verifiable) $\SNARG$ for $M^{{\mathcal O}}$, with
adaptive soundness, whose public parameters and proof have length
\[
    \poly(\secp,S,\log T).
\]
The verifier runs in time
$n\cdot\poly(\secp,S,\log T)$ and makes no oracle calls.
Setup also makes no oracle calls. The prover makes at most $Q$ calls to
$\mathcal O_{\mathsf{sig}}$ and runs in time polynomial in the security
parameter and the length of the signed execution.\footnote{We emphasize that we make no robustness assumption.}
\end{theorem}

As in the polynomial-time setting of \Cref{sec:def:SNARG}, we first take
$n,T,S$ to be polynomially bounded in $\lambda$, and take $T\geq n$.
We discuss the security requirements for larger time bounds at the end
of the section. In \Cref{sec:short-signed-oracles}, we give a
fully succinct variant for deterministic computations.

\subsection{The Construction}\label{sec:construction:signed}

As in the construction in  \Cref{sec:robust-oracle-construction}, our construction uses the following ingredients:
\begin{itemize}
    \item A hash function with local opening (\Cref{sec:HT}):
    \[
    \HT=(\HT.\Gen,\HT.\Hash,\HT.\Open,\HT.\Verify).
    \]
    \item A somewhere extractable batch argument (\Cref{sec:prelim:BARG}):
    \[
    \seBARG=(\seBARG.\Gen,\seBARG.\cP,\seBARG.\cV, \seBARG.\Extract).
    \]
\end{itemize}
We assume that the hash family $\HT$ and the $\seBARG$ are $T^*$-secure,
where
\[
    T^*=\poly(2^S,n,T,\lambda).
\]
As in \Cref{sec:robust-oracle-construction}, sub-exponential hardness allows their underlying
security parameter to be chosen as
\[
    \kappa=\poly(\secp,S,\log T).
\]
Similar to the (privately verifiable) $\SNARG$ in \Cref{sec:robust-oracle-construction}, our $\SNARG$ construction also has the hash-and-$\seBARG$ structure.  However, unlike the construction in \Cref{sec:robust-oracle-construction}, the $\SNARG$ in this section is publicly verifiable and the verifier does not extract any witnesses
or check any answers against the oracle. Instead, the local relation
checks a signature at every oracle transition. Extraction is used only
in the soundness proof.

It is important that we use the somewhere extraction property of the underlying $\seBARG$, rather than only ordinary soundness, since a verifying signature on an incorrect answer may
exist. We must show that a successful adversary allows us to
\emph{produce} such a signature, and not merely prove its existence.

We assume without loss of generality that every execution halts in exactly $T$ steps, by padding if needed. In contrast to the construction in \Cref{sec:robust-oracle-construction}, here there is no need to fix the number of oracle calls
or to include a separate oracle table.

\paragraph{The local relation $R_\sig$.}
In what follows, let $\vk$ be the verification key corresponding to the oracle $\mathcal{O}_\sig$, let $\hk$ be a hash key, and $\rt$ be a hash value (supposedly hashing the tableau of an execution of $M^{\mathcal{O}_\sig}$).
An instance of  the oracle-free relation $R_{\sig}$ is a string of the form $(z,i)$ where $i\in[T]$ and 
\[
    z=(\vk,\hk,x,\rt).
\]
A local
witness corresponding to $(z,i)$ is a string of the form
\[
    w_i=(C_{i-1},C_i,\rho_i,\sigma_i).
\]
The relation checks the following:
\begin{enumerate}
    \item The opening $\rho_i$ authenticates $(C_{i-1},C_i)$ at their
    prescribed positions in a transcript with hash value $\mathsf{rt}$.

    \item If this is an ordinary step, $(C_{i-1},C_i)$ is a valid
    ordinary transition of $M$ on $x$. In this case $\sigma_i=\bot$.

    \item If this is an oracle step, $(C_{i-1},C_i)$ is a valid
    purported oracle transition. Let $u_i$ and $a_i$ be the query and
    answer specified by these configurations. Check that
    \[
        \Sig.\Verify
        (\vk,(u_i,a_i),\sigma_i)=1.
    \]

    \item If $i=1$, then $C_0$ is the prescribed initial configuration.
    If $i=T$, then $C_T$ is accepting. All configurations and
    transitions respect the prescribed bounds.
\end{enumerate}

Let $M_\sig$ be the deterministic machine for this
relation. Our $\SNARG$ construction uses the $\seBARG$ from \Cref{sec:prelim:BARG} for the universal language $\batchtmsat$ 
(\Cref{def:batchtmsat}), with respect to instances of the form $(M_\sig,z,k,T_\loc)$, where $T_{\loc}$ is a bound on the running time of $M_{\sig}$, and $z$ contains the
instance-dependent data.
Let $n_{\loc}$ bound $|M_{\sig}|+|z|$, and
let $m$ bound the length of a local witness.
We pad inputs and witnesses to their prescribed lengths.

We next describe our $\SNARG$ construction for the computation $M^{\mathcal{O}_\sig}$.

\begin{itemize}
    \item $\SNARG.\Gen$: On input
    $(1^\secp,n,\vk)$, do the following:
    \begin{enumerate}
        \item Generate $\hk\leftarrow\HT.\Gen(1^\kappa)$.
        \item For each $b\in\{1,2\}$, independently generate
        \[
            (\crs_b,\td_b)
            \leftarrow
            \seBARG.\Gen
            (1^\kappa,{n_{\mathsf{loc}}},1^m,T,1).
        \]
        \item Output
        $\crs=(\vk,\hk,
        \crs_1,\crs_2)$.
        Discard both trapdoors.
    \end{enumerate}
\begin{remark}
    Jumping ahead, the reason we use two independent $\seBARG$ common reference strings is that the second component allows us, in the analysis, to compare
extractions at neighboring transitions.
\end{remark}
    \item $\SNARG.\cP$: On input $(\crs,x,w)$, with
    access to $\mathcal O_{\sig}$, do the following:
    \begin{enumerate}
        \item Run the accepting execution of $M^{\mathcal O}(x,w)$,
        obtaining the signatures for each query-answer pair. Let
        $\tau=(C_0,\ldots,C_T)$ be its transcript.
        \item Compute
        $\rt=\HT.\Hash(\hk,\tau)$, and let $z=(\vk,\hk,x,\rt)$.
        \item For every $i\in[T]$ compute
        the local witness $w_i$ corresponding to $(z,i)$ in       $R_{\sig}$.
        \item For each $b\in\{1,2\}$, compute
        \[
            \pi_b=
            \seBARG.\cP
            (\crs_b,M_{\sig},z,
             1^{T_{\loc}},w_1,\ldots,w_T).
        \]
        \item Output $\pi=(\rt,\pi_1,\pi_2)$.
    \end{enumerate}

    \item $\SNARG.\cV$: On input $(\mathsf{crs},x,\pi)$, do the following:
    \begin{enumerate}
        \item Parse $\crs = (\vk, \hk, \crs_1, \crs_2)$.
       \item  Parse $\pi=(\mathsf{rt},\pi_1,\pi_2)$.
        \item Let $z = (\vk, \hk, x, \rt)$.
        
        \item Accept if and only if,
    for each $b\in\{1,2\}$,
    \[
        \mathsf{seBARG.\cV}
        (\mathsf{crs}_b,(M_{\mathsf{sig}},z,
         T,T_{\mathsf{loc}}),\pi_b)=1.
    \]
    \end{enumerate}
\end{itemize}

\subsection{Analysis}\label{sec:signed:analysis}

In this section we prove that the $\SNARG$ construction from \Cref{sec:construction:signed} satisfies the guarantees of \Cref{thm:signed-oracle-snarg}.

\paragraph{Completeness.}
Every oracle answer in an honest execution is accompanied by a
verifying signature. Thus all local witnesses satisfy
$R_{\sig}$, and completeness follows from opening completeness,
signature correctness, and $\seBARG$ completeness. The same reasoning
holds when $(x,w)$ is chosen after seeing the public parameters.

\paragraph{Adaptive soundness.}  We prove adaptive soundness via a reachability argument.  
For every $x$, $t\in\{0,\ldots,T\}$, and configuration $C$, define
\[
    \Phi_t^x(C)=
    \begin{cases}
        0, & \text{if $C$ is reachable in $t$ steps of
                    $M^{\mathcal O}$ on $x$},\\
        1, & \text{otherwise}.
    \end{cases}
\]
Reachability here allows all non-deterministic choices, but only
correct oracle answers. The table of these values can be computed
with oracle access in time
$\poly(2^S,n,T)\leq\poly(T^*)$,
by dynamic programming on the configuration graph. The same
computation determines whether $x\in L_M^{\mathcal O}$.

For any valid ordinary transition, or any valid oracle transition
using a correct answer, we have
\begin{equation}
    \Phi_t^x(C_t)\leq\Phi_{t-1}^x(C_{t-1}).
    \label{eq:signed-potential-step}
\end{equation}
Indeed, a genuine execution reaching $C_{t-1}$ can be extended by this
transition. Also,
\[
    \Phi_0^x(C_{\mathsf{init}}(x))=0,
\]
and, if $x\notin L_M^{\mathcal O}$, then
$\Phi_T^x(C)=1$ for every accepting configuration $C$.\footnote{Note that no robustness assumption is used in any of these statements.}

\medskip
Fix a polynomial-size oracle adversary $\Adv$. For
$a,b\in[T]$, let $H_{a,b}$ be the experiment in which the two $\seBARG$
components have extraction indices $a$ and $b$. Everything else,
including the distribution of the oracle's signing key, is unchanged.
Let $E$ be the event that the adversary outputs a false statement and
both public verification checks accept. Thus the soundness error in
the real experiment is $\Pr_{H_{1,1}}[E]$.

In $H_{t,t}$, extract the witness from the first component and denote
its configurations by $(\widehat C_{t-1},\widehat C_t)$. Set
\[
    F_t^-=
    \mathbb E_{H_{t,t}}
    [\mathbf 1_E\cdot \Phi_{t-1}^x(\widehat C_{t-1})],
    \qquad
    F_t^+=
    \mathbb E_{H_{t,t}}
    [\mathbf 1_E\cdot \Phi_t^x(\widehat C_t)].
\]

Throughout the following argument, whenever an extracted witness fails
its corresponding local relation $R_{\sig}$, we set all
potential contributions computed from that witness to zero.
This convention is applied separately to each extracted witness.

\begin{claim}\label{claim:signed-ext}
\[F_t^+ - F_t^-\leq \negl(\secp).
\]

   \end{claim}

\paragraph{Proof of \Cref{claim:signed-ext}.}

Fix $t\in[T]$. In $H_{t,t}$, let $\mathsf{Bad}_t$ be the event
that the witness extracted from the first component is invalid.
We first show that
\[
    \Pr_{H_{t,t}}[E\land\mathsf{Bad}_t]
    \leq \negl(\lambda).
\]
By the somewhere argument of knowledge property of the $\seBARG$
(\Cref{def:barg}), and the fact that $E$ implies that the first public
verification check accepts, the probability that $E$ occurs and
the extracted witness is invalid is at most
$\negl(T^*)$, which is
$\negl(\secp)$.

To finish the claim, let $\mathsf{Wrong}_t$ be the event that
$\neg\mathsf{Bad}_t$ holds and the extracted witness contains an
incorrect oracle answer. It remains to prove that
\[
    \Pr_{H_{t,t}}[\mathsf{Wrong}_t]
    \leq \negl(\secp).
\]
This follows from the security of the underlying signature scheme,
as follows. Suppose for the sake of contradiction that this event
occurs with non-negligible probability $\delta(\secp)$.
We construct a forger that breaks the existential unforgeability
of the underlying signature scheme. Specifically, the forger,
given $\vk$, does the following:
\begin{enumerate}
    \item Emulate
    $\SNARG.\Gen(1^\secp,n,\vk)$ to generate
    a hash key $\hk$ and
    $(\crs_b,\td_b)_{b\in\{1,2\}}$, except that
    both $\seBARG$ components are generated with extraction index
    $t$ rather than $1$. Retain both trapdoors, and let
    $\crs=(\vk,\hk,
    \crs_1,\crs_2)$.

    \item Run $\Adv(\crs)$ to generate
    $(x,\pi)$, where $\pi=(\rt,\pi_1,\pi_2)$, while
    simulating each query $u$ to $\mathcal{O}_{\sig}$
    by querying $\mathcal{O}(u)$ and requesting a signature on
    $(u,\mathcal{O}(u))$ from its signing oracle. Return the
    answer and signature to $\Adv$.
    Abort if the adversary's output is malformed.

    \item Extract
    $w^*=\seBARG.\Extract(\td_1,\pi_1)$.

    \item Let $z=(\vk,\hk,x,\rt)$.
    If $R_{\sig}(z,t,w^*)=1$ and this witness contains
    an incorrect oracle query/answer pair $(u,a)$ along with
    a verifying signature $\sigma$, then output $((u,a),\sigma)$.
    Otherwise, output $\bot$.
\end{enumerate}
The forger simulates $H_{t,t}$ exactly. Note that the message
$(u,a)$ was never submitted to the signing oracle, since every
submitted message contained a correct oracle answer, whereas
$a\neq\mathcal{O}(u)$. Moreover, the signature verifies by the
validity of the extracted witness. Thus, whenever
$\mathsf{Wrong}_t$ occurs, the forger outputs a valid forgery.

The forger runs in polynomial time in $\secp$ under the stated
polynomial bounds on $n,T,S$. In particular, it does not compute
$E$ or the reachability potential. Its success probability is
therefore $\delta(\secp)$, contradicting existential
unforgeability even given access to $\mathcal{O}$. This proves
the claimed bound on $\Pr_{H_{t,t}}[\mathsf{Wrong}_t]$.

On a valid extracted witness outside $\mathsf{Wrong}_t$, the
transition is either a valid ordinary transition or a valid
oracle transition using the correct answer. By \Cref{eq:signed-potential-step},
the potential cannot increase. On an invalid extracted witness,
both potential contributions are zero by our convention.
Since the potential lies in $[0,1]$, we conclude that
\[
    F_t^+-F_t^-
    \leq \Pr_{H_{t,t}}[E\land\mathsf{Wrong}_t]
    \leq \Pr_{H_{t,t}}[\mathsf{Wrong}_t]
    \leq \negl(\secp),
\]
as desired.

\qed

\begin{claim}\label{claim:signed-neighbors} 
For every $t<T$,
\[
|F_t^+-F_{t+1}^-|\leq\negl(\secp).
\]
      
\end{claim}

\paragraph{Proof of \Cref{claim:signed-neighbors}.}
Fix $1\leq t<T$. Start with $F_t^+$ and modify it by taking the
expectation over $H_{t,t+1}$, as opposed to over $H_{t,t}$, while
continuing to compute the potential using the witness from the
first component. Note that both $E$ and this potential contribution
can be computed without the trapdoor of the second component.
Moreover, they can be computed in time
$\poly(T^*)$ with oracle access to $\mathcal{O}$.
The signed oracle can be simulated using $\mathcal{O}$ and an
independently generated signing key. Since the weighted potential
lies in $[0,1]$, the $T^*$-index hiding property implies that this
changes the expectation by at most $\negl(T^*)$.

In $H_{t,t+1}$, the witness from the first component opens the
configuration after step $t$, and the witness from the second
component opens the configuration before step $t+1$. Both
therefore open the $t$th configuration. Let $\mathsf{Bad}$ be
the event that either extracted witness is invalid, or that both
are valid but open different configurations at this position.
Since $E$ implies that both public verification checks accept,
the $T^*$-somewhere argument of knowledge property and
$T^*$-collision resistance with respect to opening imply
\[
    \Pr_{H_{t,t+1}}[E\land\mathsf{Bad}]
    \leq \negl(T^*).
\]
Indeed, two valid witnesses opening different configurations at
the same position give a collision with respect to opening.
On $E\land\neg\mathsf{Bad}$, the two witnesses give the same
value of $\Phi_t^x$. We may therefore compute the
$\mathbf{1}_E$-weighted potential using the witness from the
second component instead, changing its expectation by at most
\[
    \Pr_{H_{t,t+1}}[E\land\mathsf{Bad}]
    \leq \negl(T^*).
\]

Next, change the extraction index in the first component from
$t$ to $t+1$, while continuing to compute the potential using
the witness from the second component. Since the weighted
potential does not use the trapdoor of the first component,
index hiding again preserves its expectation up to an additive
$\operatorname{negl}(T^*)$ term.

In $H_{t+1,t+1}$, both extracted witnesses open the configuration
immediately before step $t+1$. By the same extraction-validity
and collision-resistance argument, we may use the witness from
the first component again, changing the weighted expectation
by at most $\negl(T^*)$.
The resulting expectation is $F_{t+1}^-$. Summing the negligible
additive differences, and using the fact that $T^*\geq\secp$,
gives
\[
    \left|F_t^+-F_{t+1}^-\right|
    \leq \negl(\secp),
\]
as desired.

\qed

These index-hiding reductions compute $E$ and the reachability
potential, and hence may take time $\poly(T^*)$.
This is why we instantiate the hash family and $\seBARG$ with the stated
security parameter. Their oracle queries can be simulated using
$\mathcal O$ and an independently generated signing key.

The initial and terminal checks give
\[
    F_1^-\leq\negl(\secp),
    \qquad
    F_T^+\geq\Pr_{H_{T,T}}[E]-\negl(\secp).
\]
Telescoping the equations in ~\Cref{{claim:signed-ext}}
and~\Cref{claim:signed-neighbors} therefore gives
\[
    \Pr_{H_{T,T}}[E]
    \leq O(T)\cdot \negl(\secp)
    =\negl(\secp).
\]
Finally, index hiding lets us change both extraction indices from $T$
back to $1$, preserving $\Pr[E]$ up to a negligible additive term.
This proves adaptive soundness.

\paragraph{Efficiency.}
A local witness contains two configurations, their openings, and at
most one signature. Thus
\[
    m=\operatorname{poly}(\kappa,S,\log T).
\]
The common input has length
$n+\operatorname{poly}(\kappa,S,\log T)$. The efficiency bounds from
\Cref{sec:prelim:BARG}, together with
$\kappa=\operatorname{poly}(\lambda,S,\log T)$, give the claimed
parameter lengths and verification time. There are only two $\seBARG$
components. Neither setup nor verification queries the oracle, and
the honest prover obtains one signature per oracle call in its
execution.

\subsection{A Fully Succinct Variant for Deterministic Computations}
\label{sec:short-signed-oracles}

For deterministic computations we can avoid the dependence on $S$.  We start by assuming that the oracle queries and answers are short. Specifically, let $M$ be a deterministic Turing machine running in time bounded by $T$, and let $\ell\leq T$ be a public bound
on the bit lengths of the queries and answers occurring in executions
of $M^{\mathcal O}$. We assume that
\[
    \ell=\poly(\secp,\log T).
\]
We later show how to avoid this assumption  (see \Cref{remark:long-queries}).

We use the signed
oracle from \Cref{def:signed-oracle}, which signs each query/answer pair.

\begin{theorem}
\label{thm:short-signed-oracles}
Let $M$ be a deterministic oracle-aided machine with time bound $T$, and oracle-query bound $Q\leq T$, whose
queries and answers have bit length at most
$\ell=\poly(\secp,\log T)$.
 Let
$\mathcal O_{\mathsf{sig}}$ be a signed oracle as in~\Cref{def:signed-oracle}, with the security parameter $ \kappa_{\mathrm{sig}}
    =\operatorname{poly}(\lambda,\log T)$.
There exists a (publicly
verifiable) $\SNARG$ for $M^{{\mathcal O}}$,
whose public parameters and proof have length $\poly(\secp,\log T)$, and the verifier runs in time
$n\cdot\poly(\secp,\log T)$ and makes no oracle
calls. Setup also makes no oracle calls. The prover makes at most
$Q$ calls to $\mathcal O_{\sig}$ and runs in time $\poly(\secp,T)$.\footnote{We emphasize that no robustness assumption is
needed.}  For polynomially bounded $T$ the $\SNARG$ is adaptively sound assuming that $\LWE$ is polynomially hard even given oracle access to $\mathcal{O}$. For larger $T$, the $\SNARG$ is adaptively sound assuming that $\LWE$ is $\poly(T)$-hard even given oracle access to $\mathcal{O}$.  
\end{theorem}

We next give the fully succinct $\SNARG$ construction.  
For simplicity, and without loss of generality, we assume that the query and answer tapes each
consist of $\ell$ cells throughout the execution; any additional information is kept on the work tape.  In particular, before each oracle query, the query tape contains only the query, which is of length~$\ell$, and after each oracle call the answer tape contains only the answer, which is also of length~$\ell$. 

\subsubsection{Our Construction}
\label{sec:construction:fully-succ}
As in \Cref{sec:construction:signed} and \Cref{sec:robust-oracle-construction}, our construction makes use of the following ingredients:
\begin{itemize}

    \item A hash function with local opening (\Cref{sec:HT}):
    \[
    \HT=(\HT.\Gen,\HT.\Hash,\HT.\Open,\HT.\Verify).
    \]
    \item A somewhere extractable batch argument (\Cref{sec:prelim:BARG}):
    \[
    \seBARG=(\seBARG.\Gen,\seBARG.\cP,\seBARG.\cV, \seBARG.\Extract).
    \]

\end{itemize}
However, our construction requires the hash function $\HT$ to have the following additional property:
\begin{definition}[Verification of local updates]\label{def:local-updates}
For $x\in\{0,1\}^N$, $j\in[N]$, and $b'\in\{0,1\}$, let
$x[j\leftarrow b']$ denote the string obtained from $x$ by replacing
its $j$th bit by $b'$. We say that $\HT$ {\em supports succinct
verification of local updates} if it additionally provides a
deterministic algorithm
\[
  \HT.\mathsf{VerifyUpdate}
    (\hk,\rt,\rt',j,b,b',\rho)\longrightarrow\{0,1\}.
\]
Here $\rt$ and $\rt'$ are the old and new hash values, $b$ and $b'$
are the old and new bits at position $j$, and $\rho$ is a local
opening. The string length $N$ is public and implicit in the
verification algorithm. We require the following properties.

\begin{description}
  \item[Update completeness.]
  For every $\secp\in\mathbb N$, $N\leq 2^{\secp}$,
  $x\in\{0,1\}^N$, $j\in[N]$, and $b'\in\{0,1\}$, let
  $\hk\gets\HT.\Gen(1^{\secp})$ and set
  \[
    \begin{aligned}
      \rt&=\HT.\mathsf{Hash}(\hk,x),\\
      \rt'&=\HT.\mathsf{Hash}(\hk,x[j\leftarrow b']),\\
      \rho&=\HT.\mathsf{Open}(\hk,x,j).
    \end{aligned}
  \]
  Then, with probability one,
  \[
    \HT.\mathsf{VerifyUpdate}
      (\hk,\rt,\rt',j,x_j,b',\rho)=1.
  \]

  \item[Succinctness.]
  The update check uses the ordinary local opening $\rho$ and runs
  in time $\poly(\secp,\log N)$, without receiving either of the
  underlying strings. The opening length satisfies the succinctness
  requirement of $\HT$.

  \item[Update soundness.]
  For every polynomial-size adversary $\mathcal A$, there exists a
  negligible function $\operatorname{negl}$ such that, in the
  experiment
  \[
    \begin{aligned}
      \hk&\gets\HT.\Gen(1^{\secp}),\\
      (x,j,b,b',\rt',\rho)&\gets\mathcal A(\hk),\\
      \rt&=\HT.\mathsf{Hash}(\hk,x),
    \end{aligned}
  \]
  we have
  \[
    \Pr\!\left[
      \begin{array}{c}
        \HT.\mathsf{VerifyUpdate}
          (\hk,\rt,\rt',j,b,b',\rho)=1\\[2pt]
        {}\land\bigl(
          b\neq x_j\ \lor\
          \rt'\neq\HT.\mathsf{Hash}
            (\hk,x[j\leftarrow b'])
        \bigr)
      \end{array}
    \right]
    \leq\operatorname{negl}(\secp).
  \]
 \end{description}
\end{definition}
The Merkle-tree construction satisfies these requirements assuming the existence of a collision resistant hash function. The update checker verifies the old root using
$\rho$, and checks the new root using the same authentication path
with the leaf changed from $b$ to $b'$. An accepted incorrect update
from the hash of $x$ yields a collision by comparison with the honest
path for $x$.

\begin{remark}
    In the oracle-aided setting, update soundness is required to hold even for
adversaries with access to $\mathcal O$.
\end{remark}

\begin{remark}\label{remark:HT:eff}
For our application, we additionally require that for every string $x\in\{0,1\}^{\leq N}$,
$\HT.\Hash(\hk,x\Vert 0^{N-|x|})$ can be computed from $(\hk,x,N)$
in time $\poly(\secp,|x|,\log N)$, where blank tapes are encoded as
all-zero strings. The Merkle-tree construction also satisfies this
requirement.
\end{remark}

As in \Cref{sec:construction:signed}, we first take $n,T$ to be polynomially
bounded in $\secp$, with $T\geq n$, and set $\kappa=\secp$.
In contrast to the preceding constructions, polynomial
security of our primitives suffices here. All security
properties, including update soundness, are required to
hold even given access to $\mathcal O$.



\begin{definition}[Compressed configurations]\label{def:compress}
A compressed configuration
$\widetilde C$ consists of the machine state, head positions,
oracle-call counter, the full contents of the query and answer tapes,
and a hash of the work tape, using the hash function with local opening~$\HT$. By our simplifying assumption,
the two explicit query and answer tapes each have $\ell$ cells. The read-only input $x$
remains part of the common input.
\end{definition}

\paragraph{The local relation $R_{\mathrm{short}}$.}
Let $\mathsf{vk}$ be the oracle's verification key, let $\hk$ be a key for $\mathsf{HT}$, and let $\rt$ be
a purported hash of a sequence of compressed configurations.
An instance of the oracle-free relation $R_{\mathrm{short}}$ is of
the form $(z,i)$, where $i\in[T]$ and
\[
  z=(\mathsf{vk},\hk,x,\rt).
\]
A corresponding witness is of the form
\[
  w_i=(\widetilde C_{i-1},\widetilde C_i,\rho_i,\eta_i),
\]
where $\rho_i$ contains the transcript openings and $\eta_i$ contains
the information needed to check the transition.
The relation $R_{\mathrm{short}}$ checks the following:
\begin{enumerate}
  \item The opening $\rho_i$ authenticates
  $(\widetilde C_{i-1},\widetilde C_i)$ at their prescribed positions
  with respect to the hash value $\rt$, under the hash key $\hk$.

\item If this is an ordinary step, let $\rt_{i-1},\rt_i$ be the work-tape
hash values in $\widetilde C_{i-1},\widetilde C_i$, respectively,
and let $j$ be the work-tape head position in
$\widetilde C_{i-1}$. In this case, $\eta_i=(b,\rho)$ contains
the bit read from the work tape and its local opening.
The relation $R_{\mathrm{short}}$ checks that
\[
    \HT.\Verify(\hk,\rt_{i-1},j,b,\rho)=1.
\]
Using $b$, the input $x$, and the explicit fields of
$\widetilde C_{i-1}$, it determines the prescribed ordinary
transition of $M$. It checks that all explicit fields of
$\widetilde C_i$ follow this transition, including the machine
state, head positions, and contents of the query and answer
tapes. In particular, cells of these explicit tapes that are
not written remain unchanged, and the oracle-call counter
does not change.

If the transition writes a bit $b'$ to work-tape position $j$,
the relation additionally checks that
\[
    \HT.\mathsf{VerifyUpdate}(\hk,\rt_{i-1},\rt_i,j,b,b',\rho)=1,
\]
where $b'$ is the bit prescribed by the transition.
If the transition does not write to the work tape, it instead
checks that $\rt_{i-1}=\rt_i$.
 
  \item If this is an oracle step, $\eta_i$ contains a
signature $\sigma_i$. The relation $R_{\mathrm{short}}$
checks that the work-tape hash values in
$\widetilde C_{i-1}$ and $\widetilde C_i$ are identical,
and that all explicit fields follow the prescribed
oracle transition, using the answer in $\widetilde C_i$.
This includes the machine state, head positions, query
and answer tapes, and oracle-call counter.

In addition, letting $u_i$ be the query in
$\widetilde C_{i-1}$ and $a_i$ be the answer in
$\widetilde C_i$, the relation checks that
\[
    \Sig.\Verify(\vk,(u_i,a_i),\sigma_i)=1.
\]

  \item If $i=1$, then $\widetilde C_0$ is the prescribed initial
  compressed configuration. If $i=T$, then $\widetilde C_T$ is
  accepting.\footnote{By \Cref{remark:HT:eff}, the initial compressed configuration
can be computed in time $\poly(\secp,\log T)$. Whether a
given final compressed configuration is accepting can
be checked within the same time bound.}

\end{enumerate}
Note that the relation $R_{\mathrm{short}}$ does not query $\mathcal O$.

Let $M_{\rm short}$ be the deterministic
Turing machine that, on input $(z,i,w_i)$,
checks whether $w_i$ is a valid local
witness for $(z,i)$ according to
$R_{\rm short}$. Let
$n_{\mathrm{loc}}$ bound $|M_{\mathrm{short}}|+|z|$, let $m$ bound
the length of a local witness, and let $T_{\mathrm{loc}}$ bound the
running time of $M_{\mathrm{short}}$.

We next describe our $\SNARG$ construction for the computation
$M^{\mathcal O}$, using the signed oracle $\mathcal O_{\sig}$.
\begin{itemize}
  \item $\SNARG.\Gen$: On input $(1^{\secp},n,\vk)$,
  do the following:
  \begin{enumerate}
    \item Generate
    \[
     \hk\gets\mathsf{HT}.\Gen(1^\kappa).
    \]

    \item For each $b\in\{1,2\}$, independently generate
    \[
      (\crs_b,\td_b)
      \gets\mathsf{seBARG}.\Gen
        (1^\kappa,{n_{\mathrm{loc}}},1^m,T,1).
    \]

    \item Output
    $\crs=(\vk,\hk,
                  \crs_1,\crs_2)$.
    Discard both trapdoors.
  \end{enumerate}

  \item $\SNARG.\cP$: On input $(\crs,x)$, with access
  to $\mathcal O_{\sig}$, do the following:
  \begin{enumerate}
    \item Run the accepting execution of $M^{{\mathcal O}}(x)$,
    obtaining the signatures for its oracle answers. Let
    $\widetilde\tau=(\widetilde C_0,\ldots,\widetilde C_T)$ be
    its compressed transcript.

    \item Compute $\rt=\HT.\Hash(\hk,\widetilde\tau)$,
    and let $z=(\vk,\hk,x,\rt)$.

    \item For every $i\in[T]$, compute the local witness $w_i$
    corresponding to $(z,i)$ in $R_{\mathrm{short}}$.

    \item For each $b\in\{1,2\}$, compute
    \[
      \pi_b=\seBARG.\cP
        (\crs_b,M_{\mathrm{short}},z,
         1^{T_{\mathrm{loc}}},w_1,\ldots,w_T).
    \]

    \item Output $\pi=(\rt,\pi_1,\pi_2)$.
  \end{enumerate}

  \item $\SNARG.\cV$: On input $(\crs,x,\pi)$, parse
  $\crs=(\vk,\hk,
                \crs_1,\crs_2)$ and
  $\pi=(\rt,\pi_1,\pi_2)$, and let
  $z=(\vk,\hk,x,\rt)$.
  Accept if and only if,
  for each $b\in\{1,2\}$,
  \[\seBARG.\cV(\crs_b,
(M_{\rm short},z,T,T_{\loc}),\pi_b)=1.
\]
\end{itemize}
As in \Cref{sec:construction:signed}, the second $\seBARG$ component allows us to compare
extractions at neighboring transitions in the analysis. Neither setup
nor verification queries the oracle.

\subsubsection{Analysis}

In this section we prove that the $\SNARG$ construction from \Cref{sec:construction:fully-succ} satisfies \Cref{thm:short-signed-oracles}.

\paragraph{Completeness.}
We first note that every honest local witness $w_i$ is a valid witness for $(z,i)$ with respect to the relation $R_{\mathrm{short}}$.  This follows from opening
completeness, updated opening completeness, correctness of the hash function with local openings, and signature
correctness. Completeness therefore follows from $\seBARG$ completeness.
The same reasoning holds when $x$ is chosen after seeing the public
parameters.

\paragraph{Adaptive soundness.}
For every input $x$, let
\[
  \widetilde C_0^x,\ldots,\widetilde C_T^x
\]
be the compressed configurations of the real execution of
$M^{\mathcal O}$ on input $x$, using the sampled hash key $\hk$. Since $M$ is
deterministic, these configurations and their underlying tapes can
be computed in time $\poly(\secp,T)$ with access to $\mathcal O$.
For every $t\in\{0,1,\ldots,T\}$, define
\[
  \Psi_t^x(\widetilde C)
  =\mathbf{1}[\widetilde C\neq\widetilde C_t^x].
\]
This potential is zero at the honest initial configuration. If
$x\notin L_M^{\mathcal O}$, then $\Psi_T^x(\widetilde C)=1$ for
every accepting compressed configuration $\widetilde C$.

Fix a polynomial-size oracle adversary $\Adv$. Define
$H_{a,b}$ and $E$ as in \Cref{sec:signed:analysis}, using $R_{\mathrm{short}}$. Namely, 
in $H_{a,b}$, the two $\seBARG$ components have extraction indices $a$
and $b$, and $E$ is the event that the adversary outputs a false
statement and both public verification checks accept. Everything else
is distributed as in the real experiment. The soundness error is
$\Pr_{H_{1,1}}[E]$.

In $H_{t,t}$, extract the witness from the first component and denote
its configurations by $(\widehat C_{t-1},\widehat C_t)$. Set
\[
  F_t^-=\mathbb E_{H_{t,t}}
    [\mathbf{1}_E\cdot\Psi_{t-1}^x(\widehat C_{t-1})],
  \qquad
  F_t^+=\mathbb E_{H_{t,t}}
    [\mathbf{1}_E\cdot\Psi_t^x(\widehat C_t)].
\]
As before, all potential contributions from an invalid extracted
witness are set to zero.

\paragraph{Bounding the increase in potential.}
Let $\mathsf{Depart}_t$ be the event in $H_{t,t}$ that the extracted
witness is valid,
$\widehat C_{t-1}=\widetilde C_{t-1}^x$, and
$\widehat C_t\neq\widetilde C_t^x$.
A valid witness can increase the potential only on this event.

\begin{claim}\label{claim:depart-neg}
    There exists a negligible function $\negl$ such that for every $t\in[T]$,
    \[
    \Pr_{H_{t,t}}[\mathsf{Depart}_t]=\negl(\secp).
    \]
\end{claim}

\paragraph{Proof of \Cref{claim:depart-neg}.} 

Fix $t\in[T]$. On $\mathsf{Depart}_t$, the extracted witness
is valid and
$\widehat C_{t-1}=\widetilde C_{t-1}^x$.
Suppose first that this is an ordinary step. Let $W$ be the
contents of the work tape in the real execution at time $t-1$,
let $j$ be its head position, and let $(b,\rho)$ be the
work-tape bit and opening in the extracted witness. Let
$\rt_{t-1},\rt_t$ be the work-tape hash values in
$\widehat C_{t-1},\widehat C_t$, respectively. Since the
preceding compressed configuration is honest,
\[
    \rt_{t-1}=\HT.\Hash(\hk,W).
\]

If $b\neq W_j$, then $\rho$ and the honest opening
$\HT.\Open(\hk,W,j)$ are two valid openings of different bits
at position $j$ under the same hash value. This violates
collision resistance with respect to opening.

Otherwise, $b=W_j$. By determinism and the checks in Step~2
of $R_{\mathrm{short}}$, all explicit fields of
$\widehat C_t$ agree with those of $\widetilde C_t^x$.
If the transition does not write to the work tape, the
relation checks that $\rt_t=\rt_{t-1}$, so the work-tape
hash value is also correct. If the transition writes a bit
$b'$ at position $j$, then $b'$ is the bit written in the
real execution, and the relation checks that
\[
    \HT.\mathsf{VerifyUpdate}(\hk,\rt_{t-1},\rt_t,j,b,b',\rho)=1.
\]
Thus, a departure in this case requires
\[
    \rt_t\neq\HT.\Hash(\hk,W[j\leftarrow b']),
\]
which violates update soundness (\Cref{def:local-updates}).
Consequently, every ordinary-step departure gives either
a collision with respect to opening or a violation of
update soundness.

At an oracle step, the preceding configuration explicitly contains
the honest query $u$. If the following answer is $\mathcal O(u)$,
the transition checks force the honest next compressed configuration.
Thus a departure gives an answer $a$ and a signature $\sigma$ such that
\[
  a\neq\mathcal O(u)
  \quad\text{and}\quad
  \Sig.\Verify(\vk,(u,a),\sigma)=1.
\]
We next formalize the reductions. For either of the two
hash-security reductions, take $\hk$ from the corresponding
security experiment and independently generate the oracle's
signing key pair. Generate both $\seBARG$ components with
extraction index $t$, retaining their trapdoors, and run
$\mathsf{Adv}$ on the resulting public parameters. Simulate
each signed-oracle query $v$ by computing $\mathcal O(v)$
and signing $(v,\mathcal O(v))$ using the generated signing
key. Extract the witness from the first component, check
its validity, and run $M^{\mathcal O}(x)$ to compute the
real configurations and work tapes.

This simulates $H_{t,t}$ exactly. In the first ordinary-step
case above, the opening-binding reduction outputs the two
conflicting openings. In the second case, the update-soundness
reduction outputs
\[
    (W,j,b,b',\rt_t,\rho).
\]
Since $\rt_{t-1}=\HT.\Hash(\hk,W)$, this is an accepted
incorrect update in the experiment of~\Cref{def:local-updates}.
Thus each of these two cases has negligible probability.
Notice that $W$ is obtained by running the real computation;
the reductions do not extract an arbitrary work tape from
its hash.

For the signature reduction, take $\vk$ from the
unforgeability experiment, and generate $\hk$ and both $\seBARG$
components internally, again with extraction index $t$.
As in the proof of \Cref{claim:signed-ext}, simulate each signed-oracle
query $v$ by requesting a signature on
$(v,\mathcal O(v))$. If the extracted witness is valid and
contains an incorrect oracle answer $a$ to a query $u$,
output its signature $\sigma$ on $(u,a)$. This signature
verifies, and $(u,a)$ was never submitted to the signing
oracle, since $a\neq\mathcal O(u)$. Hence the probability
of an oracle-step departure is negligible.

All these reductions run in time $\poly(\secp,T)$ with
access to $\mathcal O$, which is polynomial in $\secp$
under our assumption that $T$ is polynomially bounded.
The three cases exhaust $\mathsf{Depart}_t$, and therefore
\[
    \Pr_{H_{t,t}}[\mathsf{Depart}_t]
    \leq \negl(\secp),
\]
as desired.

\qed

\Cref{claim:depart-neg} implies that 
\[
  F_t^+-F_t^-
  \leq\Pr_{H_{t,t}}[E\wedge\mathsf{Depart}_t]
  \leq\Pr_{H_{t,t}}[\mathsf{Depart}_t]
  \leq\operatorname{negl}(\secp).
\]

\begin{claim}\label{claim:succinct:signed-neighbors}
     For every $t<T$,
\[
  \bigl|F_t^+-F_{t+1}^-\bigr|
  \leq\operatorname{negl}(\secp).
\]
\end{claim}
\paragraph{Proof of \Cref{claim:succinct:signed-neighbors}.}
The proof of \Cref{claim:succinct:signed-neighbors} is similar to the proof of \Cref{claim:signed-neighbors}. In each hybrid, where we switch between extracted witnesses,
the probability that $E$ occurs and either the witness is invalid or
their transcript openings disagree at a common position, is negligible.
This follows from the somewhere argument of knowledge property and
collision resistance with respect to opening of $\HT$.

The index-hiding tests compute $E$ and $\Psi$ by running the real
deterministic computation, in time $\poly(\secp,T)$ with access to
$\mathcal O$. As before, each test uses only the trapdoor of the
component whose index is not being changed, and the signed oracle
is simulated using an independently generated signing key. Thus
polynomial security suffices, and for every $t<T$,
\[
  \bigl|F_t^+-F_{t+1}^-\bigr|
  \leq\operatorname{negl}(\secp),
\]
as desired.

The initial and terminal checks, together with the somewhere argument
of knowledge property, give
\[
  F_1^-=0,
  \qquad
  F_T^+\geq\Pr_{H_{T,T}}[E]-\operatorname{negl}(\secp).
\]
Telescoping yields
\[
  \Pr_{H_{T,T}}[E]
  \leq O(T)\cdot\operatorname{negl}(\secp)
  =\operatorname{negl}(\secp),
\]
since $T$ is polynomially bounded in $\secp$. Finally, index hiding
lets us change both extraction indices back to $1$, preserving
$\Pr[E]$ up to a negligible additive term. This proves adaptive
soundness without extracting an entire transcript or an arbitrary
committed work tape.

\paragraph{Efficiency.}
A compressed configuration has length
$O(\ell)+\poly(\kappa,\log T)$. A local witness contains two such
configurations, their transcript openings, and either a constant
number of work-tape authentication paths or one signature. Thus
\[
  m=\poly(\kappa,\ell,\log T)=\poly(\secp,\log T).
\]
The common input has length $n+\poly(\secp,\log T)$; the local
relation takes time $\poly(n,\secp,\ell,\log T)$. The $\seBARG$
efficiency bounds give public parameters and proof length
$\poly(\secp,\log T)$ and verification time
$n\cdot\poly(\secp,\log T)$. The prover runs in time
$\poly(\secp,T)$ and makes at most $Q$ calls to
$\mathcal O_{\sig}$, obtaining one signature per oracle call.
Setup and verification are oracle-free. None of the reductions
enumerates the configuration graph, so no security parameter depending
on $S$ is needed.

\paragraph{Larger time bounds.}
For super-polynomial $T$, the same argument requires the security
properties of $\HT$, $\seBARG$, and
$\Sig$ against $\poly(\secp,T)$-time algorithms, with errors
negligible in $T$. If these guarantees are available with
underlying security parameter $\kappa=\poly(\secp,\log T)$ then the stated succinctness bounds are preserved.
For the non-deterministic construction in \Cref{sec:construction:signed}, the hash family $\HT$ and the $\seBARG$ must additionally withstand
the $\poly(2^S,n,T,\secp)$-time reachability reductions. Ordinary
polynomial security alone does not justify either extension to
super-polynomial time bounds.

    \begin{remark}[Long queries and answers]
\label{remark:long-queries}
The assumption that the queries and answers are short can be removed
if the oracle signs their hashes, rather than the query/answer pairs
themselves.  Specifically, let $\kappa = poly(\lambda,\log T)$
be the security parameter used for the hash family, chosen sufficiently large that $2^\kappa$
exceeds the maximum encoded length of any oracle query or answer, and provide the required security against $poly(T)$ time adversaries. Generate $\hk\gets\HT.\Gen(1^{\kappa})$ as part of the oracle's setup, independently of $\mathcal O$ and of the signing key, and include
$\hk$ in the oracle's authenticated public parameters.

Define
\[
 \mathcal O_{\sig}(u)=(a,\sigma),\qquad
 a=\mathcal O(u),\qquad
 \sigma\gets\mathsf{Sig.Sign}\bigl(\sk,
       (\HT.\Hash(\hk,u),\HT.\Hash(\hk,a))\bigr).
\]
Here the hashes use the prescribed tape encodings, including their
lengths. The $\SNARG$ setup uses this hash key, rather than generating
a new one.

A compressed configuration now contains hashes of the query and
answer tapes, in addition to a hash of the work tape. At an ordinary
step, the local relation checks the symbols read and the updates to
all accessed tapes using $\HT.\Verify$ and $\HT.\mathsf{VerifyUpdate}$.
All other tape roots remain unchanged. At an oracle step, it checks
a signature on the query root in the preceding configuration and
the answer root in the following configuration. It also checks that
the remaining fields follow the prescribed oracle transition.
In particular, the work-tape root remains unchanged. Thus a local
witness still has length $\poly(\secp,\log T)$, independently of the
query and answer lengths.

To see why the soundness argument continues to hold, consider the
first departure from the real deterministic execution. At an
ordinary step, such a departure violates opening binding or update
soundness. At an oracle step, the preceding query root is the hash
of the real query $u$. If the verified pair of roots was never
signed, the extracted signature is a forgery. Otherwise, let $v$
be an oracle query on which this pair of roots was signed.
If $v\ne u$, then $u$ and $v$ have the same hash, contradicting
collision resistance. If $v=u$, the signed answer root is
$\HT.\Hash(\hk,\mathcal O(u))$, so this step does not depart from
the real compressed execution. The reduction knows the real query
and records all queries used to simulate the signed oracle; it
does not need to extract an arbitrary string from its hash.
\end{remark}

\section{Incrementally Verifiable Computations} \label{sec:IVC}

In this section we define the notion of $\IVC$ for oracle-aided computations and provide constructions.  We provide two definitions:  The first is for signed-oracles and is publicly verifiable. It is essentially the standard notion of $\IVC$, except that the prover may use authenticated oracle
responses. The second considers general, unsigned oracles. In this
case verification is private, and soundness is required only for
robust computations.

\subsection{Definitions}

Fix an oracle-aided Turing machine $M$ (deterministic or non-deterministic) with time bound $T$
and space bound $S$, and an oracle $\calO$.
For an input $x$, let $C_{\mathrm{init}}(x)$ denote the initial
configuration, and let $\mathrm{Reach}^\calO_t(x)$ be the set of
configurations reachable after exactly $t$ steps of $M^\calO$ on $x$.\footnote{If $M$ is deterministic, this set contains a single configuration, whereas for a non-deterministic machine it may contain several configurations. We assume that halted executions are padded when necessary.}

As in \Cref{sec:oracle-machines}, \Cref{rem:configuration-counters}, throughout \Cref{sec:IVC}, we assume that every configuration $C$ includes
a step counter $\operatorname{time}(C)$ and an oracle-call
counter $\operatorname{queries}(C)$.
Both counters are zero in $C_{\mathrm{init}}(x)$.
At every transition, the step counter increases by one.
The oracle-call counter increases by one at an oracle
transition and remains unchanged at an ordinary transition.
These counter updates are part of the transition rules
and are required also in purported executions, regardless
of whether the oracle answers are correct.

\begin{definition}[Publicly verifiable $\IVC$ for signed oracles]
\label{def:public-IVC}
Let $\calO_{\mathsf{sig}}$ be a signed oracle associated with oracle~$\calO$ with verification key $\vk$,
as in Definition~\ref{def:signed-oracle}. 
A publicly verifiable $\IVC$ scheme for $M^\calO$ consists of algorithms
\[
(\mathsf{IVC.Gen},\mathsf{IVC.Update},\mathsf{IVC.V})
\]
with the following syntax.
\begin{itemize}
\item
$\mathsf{IVC.Gen}(1^\lambda,n,T,\vk)\longrightarrow\crs$.
The randomized setup algorithm outputs a common reference string
$\crs$, which includes the authenticated key $\vk$. Here $n$ is
the input length, and $T$ is a bound on the runtime, both represented in binary.

$\mathsf{IVC.Gen}$ does not query the oracle.

\item {\bf
$\mathsf{IVC.Update}^{\calO_{\mathsf{sig}}}
(\crs,x,t,C_t,\pi_t,w_{t+1})
\longrightarrow (C_{t+1},\pi_{t+1})$.}
The prover is given the $\crs$, input $x$,
time $t$, configuration $C_t$, its proof
$\pi_t$, and a local witness $w_{t+1}$.
The witness specifies the non-deterministic
choice, if any, and may also contain
auxiliary information needed to compute
the next configuration, such as
authentication openings.
The algorithm outputs the next
configuration $C_{t+1}$ and an updated
proof $\pi_{t+1}$.
Initially, $C_0=C_{\mathrm{init}}(x)$
and $\pi_0=\bot$.

\item
$\mathsf{IVC.V}(\crs,x,t,C_t,\pi_t)\longrightarrow\{0,1\}$.
The verifier checks the claim that $C_t$ is reachable after $t$
steps of $M^\calO$ on input $x$. It uses no private verification key
and makes no oracle calls.
\end{itemize}
The following properties are required, where $\kappa_{\mathsf{sig}}=\operatorname{poly}(\secp,\log T)$
denotes the security parameter of the underlying signature scheme, as in
\cref{rem:signed-oracle-security}.
\begin{itemize}
\item {\bf Adaptive completeness.}  
Sample
\[
(\mathsf{vk},\mathsf{sk})
    \gets \mathsf{Sig.Gen}(1^{\kappa_{\mathsf{sig}}}),\qquad
\crs\leftarrow\mathsf{IVC.Gen}(1^\lambda,n,T,\vk).
\]
Let $\calO_{\mathsf{sig}}$ be the corresponding signed oracle.
Then for every poly-size circuit that takes as input $\crs$ and has oracle access to  $\calO_{\mathsf{sig}}$, and outputs an instance $x\in\{0,1\}^n$, a time bound~$t\in\{0,1,\ldots,T\}$, and  
a sequence of local witnesses $w=(w_1,\ldots,w_t)$ corresponding to an honest execution $(C_0,\ldots,C_t)$, the following holds:  

Starting with $C_0=C_{\mathrm{init}}(x)$ and $\pi_0=\bot$, for every $i\in[t]$ let
\[
(C_i,\pi_i)\leftarrow \mathsf{IVC.Update}^{\calO_{\mathsf{sig}}}
(\crs,x,i-1,C_{i-1},\pi_{i-1},w_i),
\]
where $w_i$ specifies the corresponding transition. Then,
for every $i\in\{0,\ldots,t\}$,
\[
\Pr\big[\mathsf{IVC.V}(\crs,x,i,C_i,\pi_i)=1\big]
\geq 1-\negl(\lambda).
\]

\item {\bf Adaptive soundness.}
For every polynomial-size signed-oracle adversary $\mathcal A$,
there exists a negligible function $\mu$ such that, sampling
\[
\begin{aligned}
(\mathsf{vk},\mathsf{sk})
    \gets \mathsf{Sig.Gen}(1^{\kappa_{\mathsf{sig}}}),\\
\crs&\leftarrow\mathsf{IVC.Gen}(1^\lambda,n,T,\vk),\\
(x,t,C,\pi)&\leftarrow\mathcal A^{\calO_{\mathsf{sig}}}(\crs),
\end{aligned}
\]
where $\calO_{\mathsf{sig}}$ is defined using $\mathsf{sk}$ as in
Definition~\ref{def:signed-oracle}, we have
\[
\Pr\left[
\begin{array}{c}
|x|=n,\quad 0\leq t\leq T,\\
C\notin\mathrm{Reach}^\calO_t(x),\\
\mathsf{IVC.V}(\crs,x,t,C,\pi)=1
\end{array}
\right]\leq\mu(\lambda).
\]
The underlying oracle $\calO$ is fixed independently of setup.

\item {\bf Efficiency.}
The lengths of $\crs$ and $\pi_t$, and the running time of
setup, are bounded by $\poly(\lambda,S,\log T)$.
Proof update and verification take time
$\poly(n,\lambda,S,\log T)$, independently of the number of
preceding steps, except through $\log T$. The verifier makes
no oracle calls. We say that the scheme is fully succinct if
these bounds have no dependence on $S$.
\end{itemize}
\end{definition}

\paragraph{General Oracles.} We next consider general oracles, without signatures.
For any input~$x\in\{0,1\}^n$, any time $t\in\{0,1,\ldots,T\}$, and any configuration $C$ that is unreachable in time~$t$ on input~$x$, we say that $C$ is
$\varepsilon$-robust at time $t$ on input $x$ (extending \Cref{def:oracle-robustness}) if every purported
execution $\tau$ (\Cref{def:puruported-exec}) from $C_{\mathrm{init}}(x)$ to $C$ in exactly
$t$ steps satisfies
\[
\mathrm{err}_\calO(\tau)\geq\varepsilon q(\tau),
\]
where $q(\tau)$ denotes the number of oracle calls in $\tau$.

\begin{definition}[Privately verifiable $\IVC$ for robust computations]
\label{def:private-IVC}
A privately verifiable $\IVC$ scheme for $\varepsilon$-robust
computations of $M^\calO$ consists of algorithms
\[
(\mathsf{IVC.Gen},\mathsf{IVC.Update},\mathsf{IVC.V})
\]
with the following syntax.
\begin{itemize}
\item
$\mathsf{IVC.Gen}(1^\lambda,n,T,\varepsilon)
\longrightarrow(\crs,\td)$.
The randomized setup algorithm outputs a common reference string
$\crs$ and a private verification key $\td$. 

$\mathsf{IVC.Gen}$ does not query the oracle.

\item
$\mathsf{IVC.Update}^{\calO}
(\crs,x,t,C_t,\pi_t,w_{t+1})
\longrightarrow (C_{t+1},\pi_{t+1})$.
The prover updates its proof as in Definition~\ref{def:public-IVC},
using the current configuration and  the non-deterministic choice $w_{t+1}$, if any.

\item
$\mathsf{IVC.V}^{\calO}(\crs,\td,x,t,C_t,\pi_t)
\longrightarrow\{0,1\}$.
The verifier checks the claim that $C_t$ is reachable after
$t$ steps of $M^\calO$ on input $x$, using its private verification
key $\td$ and oracle access to $\calO$.
\end{itemize}
The following properties are required.
\begin{itemize}
\item {\bf Adaptive completeness.}  
 Sample
\[
(\crs,\td)\leftarrow
\mathsf{IVC.Gen}(1^\lambda,n,T,\varepsilon).
\]
Then for every poly-size circuit that takes as input $\crs$ and has oracle access to $\calO$, and outputs an instance $x\in\{0,1\}^n$, a time bound~$t\in\{0,1,\ldots,T\}$ and a witness $w$ (in case $M$ is non-deterministic), corresponding to an honest execution $(C_0,\ldots,C_t)$, the following holds: 

Starting with $C_0=C_{\mathrm{init}}(x)$ and $\pi_0=\bot$, for every $i\in[t]$ let
\[
(C_i,\pi_i)\leftarrow \mathsf{IVC.Update}^{\calO}
(\crs,x,i-1,C_{i-1},\pi_{i-1},w_i),
\]
where $w_i$ specifies the corresponding transition. Then,
for every $i\in\{0,\ldots,t\}$,
\[
\Pr\big[\mathsf{IVC.V}^{\calO}(\crs,\td,x,i,C_i,\pi_i)=1\big]
\geq 1-\negl(\lambda).
\]

\item {\bf Adaptive soundness.}
For every polynomial-size oracle adversary $\mathcal A$,
there exists a negligible function $\mu$ such that, sampling
\[
\begin{aligned}
(\crs,\td)&\leftarrow
\mathsf{IVC.Gen}(1^\lambda,n,T,\varepsilon),\\
(x,t,C,\pi)&=\mathcal A^{\calO}(\crs),
\end{aligned}
\]
we have
\[
\Pr\left[
\begin{array}{c}
|x|=n,\quad 0\leq t\leq T,\\
C\notin\mathrm{Reach}^\calO_t(x),\\
C\text{ is }\varepsilon\text{-robust at time }t~\mbox{on input}~x,\\
\mathsf{IVC.V}^{\calO}(\crs,\td,x,t,C,\pi)=1
\end{array}
\right]\leq\mu(\lambda).
\]
The oracle $\calO$ is fixed independently of setup.

\item \textbf{Efficiency.}
The lengths of $\mathsf{crs}$, $\mathsf{td}$, and $\pi_t$,
and the running time of setup, are bounded by
$\poly(\secp,S,\log T,1/\varepsilon)$.
Proof update and verification take time
$\poly(n,\secp,S,\log T,1/\varepsilon)$,
independently of the number of preceding steps,
except through $\log T$.
In each invocation, the verifier makes
$O(\secp/\varepsilon)$ queries to $\calO$.

\end{itemize}
\end{definition}

\subsection{Preliminaries: Updatable Batch Arguments}\label{sec:updatableBARG}
Both constructions use as a building block an updatable $\BARG$, defined by  Aldema
Tshuva and Oshman~\cite{AldemaTshuvaO26}, and formally defined below.
In what follows we recall a slightly specialized
version of the definition of an updatable $\BARG$ \cite{AldemaTshuvaO26}, that captures the properties needed in this
work.

Fix an $\NP$ relation $R$, and a polynomial-time consistency predicate
$\mathsf{Cons}$ on pairs of statements.  Intuitively, an updatable
$\BARG$ allows the prover to maintain a short proof for an adaptively
growing sequence of $\NP$ statements
$z_1,\ldots,z_k$,
together with witnesses $w_1,\ldots,w_k$ such that
$R(z_i,w_i)=1$ for every $i$, and
\[
\mathsf{Cons}(z_i,z_{i+1})=1
\]
for every $i<k$.  The proof can be updated efficiently upon receiving
$(z_{k+1},w_{k+1})$, without accessing the preceding witnesses.

\begin{definition}[Updatable batch argument]
\label{def:upbarg} \cite{AldemaTshuvaO26}
Fix an $\NP$ relation $R$ and a polynomial-time consistency predicate
$\mathsf{Cons}$.  We use the convention that
$\mathsf{Cons}(\bot,z)=1$ for every valid first statement~$z$.

An updatable batch argument $\mathsf{UpBARG}$ for
$(R,\mathsf{Cons})$ is associated with an updatable hash family with
local openings (\Cref{def:updatable-hash})
\[
\HT
=
(\HT.\mathsf{Gen},
 \HT.\mathsf{Hash},
 \HT.\mathsf{Open},
 \HT.\mathsf{Verify},
  \HT.\mathsf{Append},
  \HT.\mathsf{Size}),
\]
and consists of algorithms
\[
(\mathsf{Gen},\mathsf{TGen},\mathsf{Update},\mathsf{Ver},\mathsf{Extract})
\]
with the following syntax.

\begin{itemize}

\item $\mathsf{Gen}(1^\lambda)$ outputs a common reference string
$\mathsf{crs}$.

\item $\mathsf{TGen}(1^\lambda,j)$ outputs a common reference string
$\mathsf{crs}$ and a trapdoor $\mathsf{td}$ for an extraction index
$j$.  The scheme supports any $j\le 2^\lambda$.

\item
$\mathsf{Update}
(\mathsf{hk},\mathsf{crs},\val,\pi,z^{-},z,w)
\longrightarrow
(\val',\pi')$
takes a hash key $\mathsf{hk}$, a common reference string
$\mathsf{crs}$, the current hash value $\val$, the current proof $\pi$,
the last statement $z^{-}$, and a new statement-witness pair $(z,w)$.
It outputs the updated hash value $\val'=\HT.\mathsf{Append}(\hk,\val,z)$ and proof~$\pi'$.

In particular, $\mathsf{Update}$ does not receive any of the preceding
statements or witnesses.

\item
$\mathsf{Ver}
(\mathsf{hk},\mathsf{crs},\val,\pi,z^{-})
\longrightarrow
\{0,1\}$ takes, in addition to $(\hk,\mathsf{crs})$, a current hash value, proof, and claimed last statement, and
outputs an acceptance bit.

\item
$\mathsf{Extract}
(\mathsf{hk},\mathsf{td},\val,\pi)
\longrightarrow
(z,\rho,w,z^{\mathsf{prev}},\rho^{\mathsf{prev}})$
extracts the statement $z$ and witness $w$ at the hidden index $j$,
together with an opening $\rho$ authenticating $z$ at position $j$.
If $j>1$, it additionally extracts the preceding statement
$z^{\mathsf{prev}}$ and an opening $\rho^{\mathsf{prev}}$
authenticating it at position $j-1$.  When $j=1$, we interpret
$z^{\mathsf{prev}}=\bot$ and
$\rho^{\mathsf{prev}}=\bot$.

\end{itemize}

The scheme satisfies the following properties.

\begin{itemize}

\item {\bf Incremental completeness.}
For every $\lambda\in\mathbb{N}$, every
$\mathsf{hk}\in
\operatorname{Supp}\bigl(\HT.\Gen(1^\lambda)\bigr)$,
every
$\mathsf{crs}\in
\operatorname{Supp}\bigl(\mathsf{Gen}(1^\lambda)\bigr)$,
every $k<2^\lambda$, every
$(\val,\pi,z^-)$ such that
$\HT.\mathrm{Size}(\val)=k$,
and every statement-witness pair $(z,w)$, such that $R(z,w)=1$ and
$\mathsf{Cons}(z^{-},z)=1$, the following holds.

\begin{enumerate}
    \item 
The verifier accepts the empty state:
\[
\mathsf{Ver}
(\mathsf{hk},\mathsf{crs},\bot,\bot,\bot)
=1.
\]
\item Suppose that
\[
\mathsf{Ver}
(\mathsf{hk},\mathsf{crs},\val,\pi,z^{-})
=1,
\]
and let
\[
(\val',\pi')
=
\mathsf{Update}
(\mathsf{hk},\mathsf{crs},
 \val,\pi,z^{-},z,w).
\]
Then 
\[
\mathsf{Ver}
(\mathsf{hk},\mathsf{crs},\val',\pi',z)
=1.
\]
\end{enumerate}
\item {\bf Efficiency and succinctness.}
After $k<2^\lambda$ successive updates, let
$m$ be the maximum length of any statement or witness added so far.
The common reference string and hash value have length
$\poly(\lambda)$, and the proof has length
$O(m)+\poly(\lambda)$.

The update, verification, and extraction algorithms run in time
polynomial in their explicit inputs.  In particular, their running
times are independent of the number $k$ of preceding statements,
except through the succinct state $(\val,\pi,z^{-})$.

\item {\bf Index hiding.}
For every $j\le 2^\lambda$, the distributions
\[
\mathsf{crs}\leftarrow\mathsf{Gen}(1^\lambda)
\]
and
\[
(\mathsf{crs},\mathsf{td})
\leftarrow
\mathsf{TGen}(1^\lambda,j)
\]
are computationally indistinguishable when the trapdoor is hidden.
In particular, the common reference string computationally hides the
extraction index $j$.

\item {\bf Somewhere argument of knowledge.}
For every polynomial-size adversary $\mathcal A$
and every polynomially bounded $K=K(\lambda)$,
there exists a negligible function $\mu$ such that
for every $\lambda\in\mathbb N$ and every $j\in[K]$,
the following holds.

Sample
\[
\hk \gets \mathsf{HT.Gen}(1^\lambda),
\qquad
(\crs,\td)\gets\mathsf{TGen}(1^\lambda,j),
\]
and let
\[
(\val,\pi,z^-)=\mathcal A(\hk,\crs),
\]
\[
(z,\rho,w,z^{\rm prev},\rho^{\rm prev})
\gets\mathsf{Extract}(\hk,\td,\val,\pi).
\]
Then
\[
\Pr\left[
\begin{aligned}
&\mathsf{Ver}(\hk,\crs,\val,\pi,z^-)=1\\
&\land\ j\leq\mathsf{HT.Size}(\val)\leq K\\
&\land\big(
R(z,w)\neq 1\\
&\qquad\lor\ \mathsf{Cons}(z^{\rm prev},z)\neq 1\\
&\qquad\lor\
\mathsf{HT.Verify}(\hk,\val,j,z,\rho)\neq 1\\
&\qquad\lor\
\big(j>1\ \land\
\mathsf{HT.Verify}(
\hk,\val,j-1,z^{\rm prev},\rho^{\rm prev})\neq 1
\big)
\big)
\end{aligned}
\right]\leq\mu(\lambda).
\]
For $j=1$, we use the convention
$z^{\rm prev}=\rho^{\rm prev}=\bot$.

\item {\bf Last-statement collision resistance.}
For every polynomial-size adversary $\mathcal A$
and every polynomially bounded $K=K(\lambda)$,
there exists a negligible function $\mu$ such that
\[
\Pr\left[
\begin{array}{l}
1\leq \mathsf{HT.Size}(\val)\leq K,\\
\mathsf{Ver}(\hk,\crs,\val,\pi,z^-)=1,\\
\widetilde z\neq z^-,\\
\mathsf{HT.Verify}(
\hk,\val,\mathsf{HT.Size}(\val),\widetilde z,\rho)=1
\end{array}
:
\begin{array}{l}
\hk\gets\mathsf{HT.Gen}(1^\lambda),\\
\crs\gets\mathsf{Gen}(1^\lambda),\\
(\val,\pi,z^-,\widetilde z,\rho)
=\mathcal A(\hk,\crs)
\end{array}
\right]\leq\mu(\lambda).
\]

\end{itemize}
\end{definition}

We emphasize that the somewhere argument-of-knowledge property will be
important in the signed-oracle setting.  Ordinary soundness would not
suffice there: for an incorrect oracle answer $a\neq \calO(u)$, a valid
signature on $(u,a)$ exists under the signing key.  Security follows
from the fact that an accepting proof allows us to \emph{extract} such
a signature and thereby obtain a forgery.

\begin{remark}[$T^\ast$-security]
\label{rem:upbarg-security}
For $T^\ast=T^\ast(\lambda)\geq\lambda$, we say that
$\mathsf{UpBARG}$, instantiated with security parameter
$\kappa$, is $T^\ast$-secure if its index-hiding,
somewhere argument-of-knowledge, and last-statement
collision-resistance properties hold against non-uniform
adversaries of size $\operatorname{poly}(T^\ast)$, with
advantage or failure probability $\operatorname{negl}(T^\ast)$.
These guarantees hold for every batch bound
$K\leq\operatorname{poly}(T^\ast)$ with $K<2^\kappa$,
uniformly over extraction indices $j\in[K]$.
In our oracle-aided applications, they must hold even given
access to $\mathcal O$, fixed independently of setup.
\end{remark}

\begin{theorem}[\cite{AldemaTshuvaO26}]
\label{thm:upbarg}
Updatable hash-and-$\BARG$ schemes ($\mathsf{UpBARG}$) for $\NP$ satisfying
Definition~\ref{def:upbarg} exist assuming either polynomial hardness
of $\LWE$, or subexponential hardness of $\DDH$, or polynomial hardness of $\DDH$
and $k$-$\mathsf{Lin}$.
\end{theorem}

In our oracle-aided applications, we require
the security properties of the $\mathsf{UpBARG}$ to hold
even against adversaries with oracle access
to $\mathcal O$.

\subsection{$\IVC$ Construction for Signed Oracles}\label{sec:IVC-signed}

We construct a publicly verifiable $\IVC$ scheme for oracle-aided
computations with signed oracles. Our construction is based on
the $\mathsf{UpBARG}$ of \Cref{sec:updatableBARG}, following the approach of
Aldema Tshuva and Oshman~\cite{AldemaTshuvaO26} of constructing $\IVC$ for standard (non-oracle) computations.

The main observation is that signatures make oracle
transitions locally verifiable without querying the oracle.
Indeed, if an oracle transition uses query $u$ and answer $a$,
then a signature $\sigma$ satisfying
\[
\mathsf{Sig.Verify}(\vk,(u,a),\sigma)=1
\]
serves as a witness for this transition.

There is, however, an important subtlety. Even when
$a\neq \calO(u)$, a valid signature on $(u,a)$ exists under
the oracle's signing key. Thus, ordinary soundness of
the $\mathsf{UpBARG}$ does not suffice. Instead, we use its
somewhere argument-of-knowledge property, which allows
us to extract a signature from an accepting proof.
We show that an accepting proof for an unreachable
configuration yields a forgery of the signature scheme.

\paragraph{The local relation.}
Fix an oracle-aided Turing machine $M$ with time bound $T$ and space bound $S$, and let
$\calO_{\mathsf{sig}}$ be a signed oracle associated with $\calO$
with verification key $\vk$.

We define an oracle-free $\NP$ relation
$R_\sig$ consisting of instance-witness
pairs $(D_i,\eta_i)$, where
\[
D_i=(i,\vk,\hk,\val_x,C_{i-1},C_i)
\]
is a transition record, $\val_x=\HT.\Hash(\hk,x)$ is a
hash value authenticating the input~$x$,
and $C_{i-1},C_i$ are consecutive configurations of
$M$ (excluding the read-only input tape).\footnote{The verification and hash keys are
included in the record to ensure that $R_\sig$ is an ordinary, fixed $\NP$ relation.} 
\begin{remark}[Hashing the input]
We represent $x$ by its authenticated hash $\val_x$ to keep
the instance of the local relation succinct. Since the
read-only input tape does not count toward the space bound,
the input length may be much larger than $S$.
Local openings allow each input-tape access to be checked
using a short witness.
This is important because the $\mathsf{UpBARG}$ proof-length bound
depends on the lengths of the local instances and witnesses.
The $\IVC$ verifier still receives $x$ and computes
$\val_x=\Hash(\hk,x)$ itself.
\end{remark}
A valid witness $\eta_i$ corresponding to~$D_i$ satisfies the following:

\begin{enumerate}
\item
If $C_{i-1}\to C_i$ is a valid ordinary transition
of $M$, then the witness $\eta_i$ specifies any non-deterministic
choice made in this transition. If the transition
reads a bit of the input tape, the witness also
contains the bit and a local opening showing that
it agrees with the authenticated input $\val_x$.

\item
If $C_{i-1}\to C_i$ is a valid oracle
transition, denoting by $u_i$ and $a_i$ the query and
answer specified by these configurations,
the witness $\eta_i$ contains a signature $\sigma_i$ such that
\[
\mathsf{Sig.Verify}(\vk,(u_i,a_i),\sigma_i)=1.
\]
\end{enumerate}

In both cases, the relation checks that the transition also
respects the machine's prescribed time and space bounds,
including the oracle-call counter. Importantly,
$R_{\mathsf{sig}}$ makes no oracle calls.

We also define a polynomial-time consistency
predicate $\mathsf{Cons}$.
For two consecutive records
\[
D=(i,\vk,\hk,\val,C,C')
\]
and
\[
D'=(i',\vk',\hk',\val',C'',C'''),
\]
we define $\mathsf{Cons}(D,D')=1$ if and only if
\[
i'=i+1,\qquad
(\vk',\hk',\val')=(\vk,\hk,\val),
\qquad C''=C'.
\]
For the first record, $\mathsf{Cons}(\bot,D)=1$
if and only if its time index is $i=1$ and its
starting configuration is the prescribed initial
configuration for inputs of length $n$.
Here the read-only input tape is represented
externally, as above.

Notice that both $R_{\mathsf{sig}}$ and
$\mathsf{Cons}$ are oracle-free.

\paragraph{The $\IVC$ construction.}
Our construction uses the following ingredients:
\begin{itemize}
    \item An updatable hash family with local openings:
    \[
        \mathsf{HT}
        =
        (\mathsf{HT.Gen},
         \mathsf{HT.Hash},
         \mathsf{HT.Open},
         \mathsf{HT.Verify},
         \mathsf{HT.Append},
         \mathsf{HT.Size}).
    \]

    \item An updatable batch argument for
    $(R_{\mathrm{sig}},\mathsf{Cons})$,
    as defined in \Cref{sec:updatableBARG}, associated with the hash
    family $\mathsf{HT}$:
    \[
           \mathsf{UpBARG}
        = (\mathsf{UpBARG.Gen},
            \mathsf{UpBARG.TGen},
            \mathsf{UpBARG.Update},         \mathsf{UpBARG.Ver},
            \mathsf{UpBARG.Extract}).
      \]
\end{itemize}
The signature scheme is the one underlying the given
signed oracle, whose verification key is $\mathsf{vk}$.
We use two independent $\mathsf{UpBARG}$ instances,
which will be useful in the soundness proof.

\begin{itemize}
\item
{\bf $\mathsf{IVC.Gen}(1^\lambda,n,T,\vk)\longrightarrow\crs$:}
\begin{enumerate}
    \item 
Generate a hash key
$\hk\gets\mathsf{HT.Gen}(1^\kappa)$, where the security parameter $\kappa$ is chosen to be $\kappa=\mathrm{poly}(\lambda,S,\log T)$ (as discussed in the analysis below).
\item Generate two independent $\mathsf{UpBARG}$ reference strings
$\crs_1,\crs_2
\gets\mathsf{UpBARG.Gen}(1^\kappa)$.

\item Output
$\crs=(\vk,\hk,\crs_1,\crs_2)$.
\end{enumerate}
Note that $\IVC.\Gen$ makes no oracle calls.
\item {\bf
$\mathsf{IVC.Update}^{\calO_{\mathsf{sig}}}
(\crs,x,t,C_t,\pi_t,w_{t+1})
\longrightarrow (C_{t+1},\pi_{t+1})$:}

\begin{enumerate}
    \item Parse $\crs=(\vk,\hk,\crs_1,\crs_2)$
    and compute
    $\val_x=\HT.\mathsf{Hash}(\hk,x)$.

\item If $t=0$, check that
$C_0=C_{\mathrm{init}}(x)$ and $\pi_0=\bot$,
and set
$\val_0=\Pi_{1,0}=\Pi_{2,0}=D_0=\bot$.

Otherwise, parse
$\pi_t=(\val_t,\Pi_{1,t},\Pi_{2,t},D_t)$
and check that
\[
\HT.\mathsf{Size}(\val_t)=t
~~\mbox{ and }~~
D_t=(t,\vk,\hk,\val_x,C',C_t)
\]
for some configuration $C'$.

    \item \label{step:3} Compute $C_{t+1}$ by applying
    the next transition of $M$ to $C_t$,
    using the non-deterministic choice
    $w_{t+1}$, if any.
    If this is an oracle transition
    with query $u$, obtain
    \[
       (a,\sigma)\gets
       \calO_{\mathsf{sig}}(u),
    \]
    and use $a$ as the oracle answer
    when computing $C_{t+1}$.

    \item Define the new transition record
    \[
    D_{t+1}=
    (t+1,\vk,\hk,\val_x,C_t,C_{t+1}).
    \]

    \item Construct a witness $\eta_{t+1}$ for
    $R_{\mathsf{sig}}(D_{t+1},\eta_{t+1})=1$.

If this is an ordinary transition, the witness
contains the non-deterministic choice
$w_{t+1}$, if any, together with the local
opening of the input bit read by the
transition, when needed.
If this is an oracle transition, the witness
contains the signature $\sigma$ obtained
in \Cref{step:3}.

    \item For each $b\in\{1,2\}$, compute
    \[
    (\val_{b,t+1},\Pi_{b,t+1})
    \gets \mathsf{UpBARG.Update}(
\hk,\crs_b,\val_t,\Pi_{b,t},D_t,D_{t+1},\eta_{t+1}).
    \]

    By the correctness of the updatable hash
    family, the two resulting hash values
    are identical:
    \[
    \val_{1,t+1}=\val_{2,t+1}
    =\HT.\mathsf{Append}
      (\hk,\val_t,D_{t+1}).
    \]
    Denote this common value by $\val_{t+1}$.

\item Output $(C_{t+1},\pi_{t+1})$, where
\[
\pi_{t+1} =
(\val_{t+1},\Pi_{1,t+1},\Pi_{2,t+1},D_{t+1}).
\]

\end{enumerate}

\item
{\bf $\mathsf{IVC.V}(\crs,x,t,C,\pi_t)$.}

\begin{enumerate}
    \item If $t=0$, accept if and only if
$C=C_{\mathrm{init}}(x)$ and $\pi_t=\bot$.
\item 
If $t\geq 1$, parse
$\pi_t=(\val_t,\Pi_{1,t},\Pi_{2,t},D_t)$ and check that
\[
\mathsf{HT.Size}(\val_t)=t~~\wedge~~
D_t=(t,\vk,\hk,\val_x,C',C)
\]
for some configuration $C'$, where
$\val_x=\mathsf{HT.Hash}(\hk,x)$.
\item 
Accept if and only if, for both
$b\in\{1,2\}$,
\[
\mathsf{UpBARG.Ver}
(\hk,\crs_b,\val_t,\Pi_{b,t},D_t)=1.
\]
\end{enumerate}
The verifier makes no oracle calls.
\end{itemize}

\begin{theorem}\label{thm:IVC-signed}
Let $M$ be an oracle-aided Turing machine with time
bound $T$ and space bound $S$, and let
$\mathcal O_{\mathsf{sig}}$ be a signed oracle as in~\Cref{def:signed-oracle}, with the security parameter $ \kappa_{\mathrm{sig}}
    =\operatorname{poly}(\lambda,\log T)$.

Let $T^\ast=\operatorname{poly}(2^S,T,\secp)$.
Assume that $\mathsf{UpBARG}$ and $\mathsf{HT}$, instantiated
with $\kappa=\operatorname{poly}(\secp,S,\log T)$, are
$T^\ast$-secure even given access to $\mathcal O$, and that the
signature scheme satisfies \cref{rem:signed-oracle-security}.
Then there exists a publicly
verifiable $\IVC$ scheme for $M^\calO$ with adaptive
soundness, according to \Cref{def:public-IVC}, with the following properties:
\begin{itemize}
    \item The public parameters and maintained $\IVC$ proof
have length
$\mathrm{poly}(\lambda,S,\log T)$.
\item 
Proof update and verification take time
$\mathrm{poly}(n,\lambda,S,\log T)$.
\item 
Neither setup nor verification makes oracle calls.
\item The authenticated oracle responses needed by the
prover are exactly those used by the underlying
execution.
\end{itemize}
\end{theorem}

\begin{proof}
Completeness follows from the correctness of the
signature scheme, the completeness of the hash
family with local openings and updatability, and incremental
completeness of the $\mathsf{UpBARG}$.

The efficiency bounds follow from the fact that
each record contains a constant number of
configurations, together with public parameters
and a time index. Thus, both a record and its
witness have length
\[
\mathrm{poly}(\kappa,S,\log T)=\poly(\lambda,S,\log T).
\]
The two $\mathsf{UpBARG}$ instances increase the proof
length and update time by only a constant factor.

\paragraph{Adaptive soundness.}
As in \Cref{sec:construction:signed}, let
\[
    T^*=\operatorname{poly}(2^S,T,\lambda)
\]
be large enough to bound the running times of the reductions
below, including the configuration-graph computations.
We assume that $\mathsf{UpBARG}$ and $\mathsf{HT}$ are secure against
adversaries of size $\operatorname{poly}(T^*)$, even given
oracle access to $\calO$, with error
$\operatorname{negl}(T^*)$. We assume that the signature scheme is $T$-secure (see 
\cref{rem:signed-oracle-security}). Fix an oracle $\calO$ independently of setup,
and let $\mathcal A$ be an adversary with access
to $\calO_{\mathsf{sig}}$.
We first analyze adversaries that output a
fixed time $t\in[T]$.  At the end of the proof we show how to deal with adaptively chosen $t\in[T]$.

Let $E$ be the event that $\mathcal A$ outputs
$(x,t,C,\pi_t)$ such that
\[
C\notin\mathrm{Reach}^{\calO}_t(x)
\qquad\text{and}\qquad
\mathsf{IVC.V}(\crs,x,t,C,\pi_t)=1.
\]
We prove that 
\[\Pr[E]=\negl(\secp).
\]
To this end, parse $\pi_t=(\val,\Pi_1,\Pi_2,D_t)$, and note that if event $E$ holds then 
\[
\HT.\mathsf{Size}(\val)=t~~\mbox{ and }~~
D_t=(t,\vk,\hk,\val_x,C',C)
\]
for some $C'$, where
$\val_x=\HT.\mathsf{Hash}(\hk,x)$.

For every $i\in\{0,\ldots,t\}$ and
configuration $C'$, define
\[
\Phi_i^x(C')=
\begin{cases}
0, & C'\in\mathrm{Reach}^{\calO}_i(x),\\
1, & \text{otherwise}.
\end{cases}
\]
For every valid transition
$C_{i-1}\to C_i$ using the correct
oracle answer, we have
\[
\Phi_i^x(C_i)\leq
\Phi_{i-1}^x(C_{i-1}).
\]

For $a,b\in[t]$, let $H_{a,b}$
denote the experiment in which the
two $\mathsf{UpBARG}$ reference strings are
generated using
$\mathsf{TGen}(1^\kappa,a)$ and
$\mathsf{TGen}(1^\kappa,b)$,
respectively.
By index hiding, the probability of
$E$ in each of these experiments
differs negligibly from its probability
in the real experiment.

For every $i\in[t]$, consider $H_{i,i}$,
and use the trapdoor of the first
$\mathsf{UpBARG}$ to extract
\[
(\widehat D_i,\rho_i,\eta_i,
 \widehat D_{i-1},\rho_{i-1}),
\]
where
\[
\widehat D_i=
(i',\vk',\hk',\val_x',B_i,A_i).
\]
We say that the extraction is \emph{valid}
if
\[
R_\sig(\widehat D_i,\eta_i)=1,
\qquad
\mathsf{Cons}(\widehat D_{i-1},
             \widehat D_i)=1,
\]
and the extracted hash openings $\rho_i$ and $\rho_{i-1}$ verify
at positions $i$ and $i-1$, respectively
(the latter only if $i>1$).
We say that the extraction is \emph{aligned} if
\[
(i',\vk',\hk',\val_x')
=(i,\vk,\hk,\val_x).
\]
We write $\mathsf{Valid}_i$ and
$\mathsf{Align}_i$ for the events that
the extraction is valid and aligned,
respectively. If either event fails,
we set both potential contributions
associated with the extraction to zero.

Define
\[
F_i^-=
\mathbb E_{H_{i,i}}
\left[
\mathbf 1_E
\mathbf 1_{\mathsf{Valid}_i}
\mathbf 1_{\mathsf{Align}_i}
\Phi_{i-1}^x(B_i)
\right]
\]
and
\[
F_i^+=
\mathbb E_{H_{i,i}}
\left[
\mathbf 1_E
\mathbf 1_{\mathsf{Valid}_i}
\mathbf 1_{\mathsf{Align}_i}
\Phi_i^x(A_i)
\right].
\]
By the somewhere argument-of-knowledge
property, for every $i\in[t]$,
\[
\Pr_{H_{i,i}}
[E\wedge\neg\mathsf{Valid}_i]
\leq\operatorname{negl}(T).
\]

\begin{claim}\label{claim:ivc-step}
    For every $i\in[t]$,
\[
F_i^+\leq F_i^-+\operatorname{negl}(T).
\]
\end{claim} 
\paragraph{Proof of \Cref{claim:ivc-step}.} Since invalid or unaligned extractions
contribute zero to both potentials,
it suffices to bound the probability
that a valid and aligned extraction
increases the reachability potential.

If this is an ordinary
transition, the local relation verifies
every input bit read by the transition
against $\val_x$.
An incorrect bit would yield two
different valid openings of the same
position under
$\HT.\mathsf{Hash}(\hk,x)$,
contradicting collision resistance
with respect to opening.
Consequently, by collision resistance
with respect to opening,
\[
\Pr_{H_{i,i}}\left[
\mathsf{Valid}_i\wedge\mathsf{Align}_i
\wedge\mathsf{OrdBad}_i
\right]
\leq\operatorname{negl}(T),
\]
where $\mathsf{OrdBad}_i$ is the event
that the extracted transition is ordinary
but is not a valid transition of $M$
on the actual input $x$.

If this is an oracle transition with
query $u$ and answer $a$, validity
implies that the extracted witness
contains a signature $\sigma$
satisfying
\[
\mathsf{Sig.Verify}(\vk,(u,a),\sigma)=1.
\]
If $a\neq\calO(u)$, this signature
constitutes an existential forgery.
Indeed, the reduction simulates
$\calO_{\mathsf{sig}}$ by requesting
signatures only on messages
$(u,\calO(u))$.
Thus $(u,a)$ was never submitted
to its signing oracle.
Hence, by collision resistance with respect
to opening and signature unforgeability,
\[
\Pr_{H_{i,i}}\left[
\mathsf{Valid}_i\wedge\mathsf{Align}_i
\wedge
\left(
\Phi_i^x(A_i)>
\Phi_{i-1}^x(B_i)
\right)
\right]
\leq \operatorname{negl}(T).
\]
Since the reachability potential takes
values in $\{0,1\}$, we conclude that
\[
F_i^+-F_i^-
\leq
\Pr_{H_{i,i}}\left[
E\wedge\mathsf{Valid}_i
\wedge\mathsf{Align}_i
\wedge
\left(
\Phi_i^x(A_i)>
\Phi_{i-1}^x(B_i)
\right)
\right]\\
\leq\operatorname{negl}(T),
\]
proving 
\Cref{claim:ivc-step}.

\qed

We next prove the following:

\begin{claim}\label{claim:ivc-neighbors}
    For every $i<t$,
\[
|F_i^+-F_{i+1}^-|
\leq\operatorname{negl}(T).
\]
\end{claim}

\paragraph{Proof of \Cref{claim:ivc-neighbors}.}
Start with $F_i^+$ in $H_{i,i}$.
By index hiding, we may change the
extraction index of the second $\mathsf{UpBARG}$
from $i$ to $i+1$, without changing
the expectation by more than a
negligible amount, since the quantity
being evaluated uses only the
trapdoor of the first $\mathsf{UpBARG}$.

Consider the resulting hybrid
$H_{i,i+1}$.
The first extractor opens the record
at position $i$, whereas the second
extractor opens the record at position
$i+1$, together with its predecessor
at position $i$. By the somewhere
argument-of-knowledge property
the probability that $E$ holds and one of the extractions
is invalid is negligible.
Moreover, both extracted records at
position $i$ are authenticated with
respect to the same hash value $\val$.
By collision resistance with respect
to opening, these records agree,
except with negligible probability.

Since the second extraction is valid,
the consistency predicate holds
between its extracted record at
position $i+1$ and its predecessor.
Thus these records have consecutive
time indices, identical public
parameters, and matching boundary
configurations.
Consequently, the record extracted
by the first component at position
$i$ is aligned if and only if the
record extracted by the second
component at position $i+1$ is
aligned, except with negligible
probability.
Furthermore, whenever both are
aligned, the configuration at the
end of step $i$ extracted by the
first component equals the
configuration at the beginning
of step $i+1$ extracted by the
second component. 
Therefore, up to a negligible
additive error, we may replace
the potential contribution from
the first extraction at position
$i$ by the potential contribution
from the second extraction at
position $i+1$.

We may now change the extraction
index of the first $\mathsf{UpBARG}$ from
$i$ to $i+1$, while evaluating
the potential using the second
extractor.
By index hiding, this changes
the expectation negligibly.

In $H_{i+1,i+1}$, both extractors
open the record at position $i+1$.
By the somewhere argument-of-knowledge
property and collision resistance with
respect to opening, the probability that
$E$ holds and either extraction is invalid,
or their authenticated records disagree,
is negligible.
Their alignment conditions therefore
agree, except on an event whose
intersection with $E$ has negligible
probability.

We may consequently evaluate the
potential using the first extractor again,
changing the expectation by at most a
negligible amount.
The resulting expectation is $F_{i+1}^-$,
proving
\Cref{claim:ivc-neighbors}.

\qed

Combining 
\Cref{claim:ivc-step} and
\Cref{claim:ivc-neighbors}, we obtain
\[
F_t^+\leq F_1^-+t\cdot\operatorname{negl}(T).
\]
Thus, to complete the soundness proof,
it suffices to show that
\[
F_1^-=0
\qquad\text{and}\qquad
F_t^+\geq
\Pr_{H_{t,t}}[E]-\operatorname{negl}(T).
\]
We establish these two bounds next.

For $i=1$, validity implies
$\mathsf{Cons}(\bot,\widehat D_1)=1$.
Thus, whenever the extraction is
valid and aligned, its starting
configuration is
$C_{\mathrm{init}}(x)$.
Consequently,
\[
F_1^-=0.
\]
For the terminal condition, recall that $D_t$ is the claimed
last record explicitly included in the adversary's proof
$\pi_t=(\val,\Pi_1,\Pi_2,D_t)$, whereas $\widehat{D}_t$ denotes
the record extracted from the first $\mathsf{UpBARG}$ component.
Let $\mathsf{Acc}$ be the event that
$\mathsf{IVC.V}(\mathsf{crs},x,t,C,\pi_t)=1$.
 $\mathsf{Acc}$ implies that the verifier's checks pass. Namely, 
 \[
    \mathsf{HT.Size}(\val)=t
    \qquad\text{and}\qquad
    D_t=(t,\mathsf{vk},\hk,\val_x,C',C)
\]
for some $C'$, where $\val_x=\HT.\Hash(\hk,x)$.

Consider the hybrid $H_{t,\mathrm{gen}}$ in which the first
$\mathsf{UpBARG}$ reference string is generated using
$\mathsf{UpBARG.TGen}(1^\kappa,t)$, and the second is
generated using $\mathsf{UpBARG.Gen}(1^\kappa)$.
All other parts of the experiment remain unchanged.
By the somewhere argument-of-knowledge property, the
probability that $\mathsf{Acc}$ holds and the first
extraction is invalid is negligible.

Suppose that $\mathsf{Acc}$ holds and the first extraction
is valid. Then its extracted opening $\rho_t$ satisfies
\[
    \mathsf{HT.Verify}
    (\hk,\val,t,\widehat{D}_t,\rho_t)=1.
\]
Moreover, acceptance of the second component implies that
\[
    \mathsf{UpBARG.Ver}
    (\hk,\mathsf{crs}_2,\val,\Pi_2,D_t)=1.
\]
Since $\mathsf{HT.Size}(\val)=t$, if
$\widehat{D}_t\neq D_t$, these two conditions give a violation
of last-statement collision resistance for the second
$\mathsf{UpBARG}$ component. Formally, the reduction generates the
first component together with its extraction trapdoor,
runs the adversary, and outputs
$(\val,\Pi_2,D_t,\widehat{D}_t,\rho_t)$.
Consequently,
\[
    \Pr_{H_{t,\mathrm{gen}}}
    \left[
        \mathsf{Acc}\wedge
        \bigl(
            \neg\mathrm{Valid}_t
            \,\vee\,
            \widehat{D}_t\neq D_t
        \bigr)
    \right]
    \leq \operatorname{negl}(T).
\]

The event in this probability can be evaluated using only
the public data and the first component's extraction
trapdoor. Thus, by index hiding, replacing the second
reference string by one generated using
$\mathsf{UpBARG.TGen}(1^\kappa,t)$ changes its probability
by at most a negligible amount. We obtain
\[
    \Pr_{H_{t,t}}
    \left[
        \mathsf{Acc}\wedge
        \bigl(
            \neg\mathrm{Valid}_t
            \,\vee\,
            \widehat{D}_t\neq D_t
        \bigr)
    \right]
    \leq \operatorname{negl}(T).
\]

Finally, $E\subseteq\mathsf{Acc}$.
On the event
$E\wedge\mathrm{Valid}_t\wedge(\widehat{D}_t=D_t)$,
the verifier's checks imply that the extraction is aligned
and that its ending configuration satisfies $A_t=C$.
Since $E$ also implies
$C\notin\operatorname{Reach}_t^\calO(x)$, we have
$\Phi_t^x(A_t)=1$.
Therefore, by the definition of $F_t^+$,
\[
    F_t^+
    \geq
    \Pr_{H_{t,t}}
    \left[
        E\wedge\mathrm{Valid}_t
        \wedge(\widehat{D}_t=D_t)
    \right]\geq
    \Pr_{H_{t,t}}[E]
    -\operatorname{negl}(T),
\]
as desired.

Finally, we allow the adversary to choose $t$ adaptively.
Guessing $t\in[T]$ loses a factor of at most $T$, so the
soundness error remains
$T\cdot\operatorname{negl}(T)=\operatorname{negl}(T)$,
which is negligible in $\secp$ since $T>\secp$.
The case $t=0$ follows from the initial-state check.
\end{proof}

\paragraph{Fully succinct deterministic computations.}
For deterministic computations, we can remove the
dependence on $S$ by using the compressed
configurations of \Cref{sec:short-signed-oracles}. We assume
that the oracle queries and answers have length
at most $\ell=\mathrm{poly}(\lambda,\log T)$.

For a full configuration $C$, let
$\widetilde C=\mathsf{Comp}_{\hk}(C)$ denote its compressed
representation, as in \Cref{def:compress}, in which
the work tape is represented by its hash.
In this variant, the $\IVC$ certifies compressed
configurations
$\widetilde C_t=\mathsf{Comp}_{\hk}(C_t)$, rather than full configurations $C_t$.\footnote{
    We use the same hash key $\mathsf{hk}$ for the
Merkle-forest construction authenticating the sequence
of transition records and the Merkle-tree construction
authenticating the input and work tape.
Both constructions use the same underlying
collision-resistant hash function, with distinct
encodings for the two uses.
The latter supports verification of local updates and
efficient hashing of blank tapes, as in \Cref{sec:construction:fully-succ}.
We use $\mathsf{HT}$ for both interfaces, with the
intended interface determined by context.}
We modify the construction above as follows.
The transition records have the form
\[
D_i=(i,\vk,\hk,\val_x,
\widetilde C_{i-1},\widetilde C_i),
\qquad
\val_x=\HT.\mathsf{Hash}(\hk,x).
\]
We use the same $\mathsf{UpBARG}$ construction, with a
local relation $R_{\mathrm{sig}}^{\mathrm{comp}}$
for these records and the same consistency
predicate on consecutive records.

The local relation checks transitions between
compressed configurations, as in \Cref{sec:short-signed-oracles}.
For an ordinary transition, the witness contains
the work-tape bit being read and its local
opening. The relation verifies this opening
with respect to the work-tape hash in
$\widetilde C_{i-1}$.
If the transition writes a bit to the work tape,
the relation checks the new work-tape hash using
$\HT.\mathsf{VerifyUpdate}$; otherwise, it
checks that the work-tape hash remains unchanged.
It also checks that all explicit fields follow
the prescribed transition.
Whenever the machine reads an input bit,
the witness additionally contains a local opening
authenticating this bit with respect to $\val_x$.

For an oracle transition, the relation checks
that the explicit fields follow the prescribed
oracle transition, that the work-tape hash remains
unchanged, and that the oracle answer is
accompanied by a verifying signature.
The initial consistency check requires the
prescribed initial compressed configuration,
whose blank work-tape hash can be computed
efficiently by \Cref{remark:HT:eff}.
Unlike $R_{\mathrm{short}}$ of \Cref{sec:short-signed-oracles},
this relation does not authenticate an entire
transcript or require the final configuration
to be accepting.

The update algorithm has the same syntax as
in \Cref{def:public-IVC}, with $\widetilde C_t$ in
place of $C_t$.
At an ordinary step, the local witness $w_{t+1}$ contains
the work-tape bit being read and its authentication path,
together with an authenticated input bit when needed.
If the prescribed transition writes to the work tape,
$w_{t+1}$ additionally contains the proposed updated
work-tape root.
The updater verifies the local openings and checks the
proposed root using $\mathsf{HT.VerifyUpdate}$, with the
written bit determined by the prescribed transition.
If no work-tape cell is written, it retains the current root.
It then forms $\widetilde C_{t+1}$ from this root and
the prescribed updates to the explicit fields, without
accessing the entire work tape.
At an oracle step, the updater obtains the answer and
its signature from $\calO_{\mathsf{sig}}$, as in the
construction above.
It then updates the two $\mathsf{UpBARG}$ proofs and
outputs $(\widetilde C_{t+1},\pi_{t+1})$.

The honest prover can generate the required local openings
and updated roots by maintaining a dynamic Merkle tree
alongside the computation.
Since the work tape is initially blank,
the tree can be initialized implicitly,
without processing all $S$ cells.
Each read or write accesses or updates only
$O(\log T)$ authentication nodes, and hence
requires $\mathrm{poly}(\lambda,\log T)$ time.
This auxiliary data structure is not part of
the maintained $\IVC$ proof.

For polynomially bounded $T$, we set the
underlying security parameter to $\kappa=\lambda$.
For larger time bounds, we use
$\kappa=\mathrm{poly}(\lambda,\log T)$,
assuming the corresponding strengthened
security guarantees for the $\mathsf{UpBARG}$, hash family,
and signature scheme, as in \Cref{sec:short-signed-oracles}.
Consequently, the public parameters,
compressed configurations, and $\IVC$ proofs
have length $\mathrm{poly}(\lambda,\log T)$,
and proof update and verification take time
$n\cdot\mathrm{poly}(\lambda,\log T)$.
Neither setup nor verification makes oracle calls.

Formally, define the compressed reachability set
\[
\widetilde{\mathrm{Reach}}^{\calO}_t(x;\hk)
=
\left\{
\mathsf{Comp}_{\hk}(C):
C\in\mathrm{Reach}^{\calO}_t(x)
\right\}.
\]
Soundness of this variant is defined with respect
to membership in this set.
Since $M$ is deterministic, this set contains
the compressed configuration of the unique honest
execution. As in \Cref{sec:short-signed-oracles}, soundness follows
the unique honest execution, rather than
enumerating the configuration graph.
The proof uses the same neighboring-record
consistency argument as above, together with
the local-update soundness argument of
\Cref{sec:short-signed-oracles}.

\subsection{Privately Verifiable $\IVC$ for Robust Computations}
\label{sec:IVC-robust}

We next turn to the setting of \Cref{sec:robust-oracle-construction}.  Here the preceding argument
does not apply.  An unsigned oracle answer cannot be checked inside an
ordinary $\NP$ relation, and the verifier must query $\calO$ itself.

The construction is based on grouping the computation into blocks, one
block for each oracle call.  More precisely, let $A_{j-1}$ denote the
configuration immediately after the $(j-1)$st oracle call, let $B_j$
denote the configuration immediately before the $j$th oracle call, and
let $A_j$ denote the configuration immediately after it.  The $j$th
block therefore has the form
\[
A_{j-1}
\leadsto
B_j
\xrightarrow{(q_j,a_j)}
A_j,
\]
where the first part consists only of ordinary transitions, and the
last transition is the $j$th oracle call, with query $q_j$ and purported
answer $a_j$.

Importantly, the computation from configuration $A_{j-1}$ to configuration $B_j$ uses only ordinary transitions. Hence we can use a standard $\IVC$ proof to certify that $B_j$
is reachable from $A_{j-1}$.  Thus, each block proof contains the two boundary configurations of the
ordinary interval, a short proof
for the ordinary interval, and the following oracle transition.

This grouping is useful because robustness is measured with respect
to the number of oracle calls, rather than the total number of
computation steps.  Hence, if a purported execution reaches a false
$\varepsilon$-robust endpoint, then at least an $\varepsilon$-fraction of
these blocks contain an incorrect oracle answer.  The private verifier
can therefore detect cheating by extracting and checking the query-answer pairs in a small
number of hidden, randomly chosen blocks.

\paragraph{Ordinary intervals.}

Fix an oracle-aided Turing machine $M$ with time bound $T$
and space bound $S$ on inputs of length $n$, and fix an input
$x\in\{0,1\}^n$.
We define the ordinary-execution relation $R_{\mathrm{ord}}^x$
as follows, suppressing its dependence on $M$, $T$, and $S$
in the notation:
\[
R_{\mathrm{ord}}^x((A,B,\ell),\tau)=1
\]
if and only if $0 \leq \ell \leq T$ and
$\tau=(C_0,\ldots,C_\ell)$ is a sequence of configurations such that
$C_0=A$, $C_\ell=B$, and every transition
$C_{i-1}\to C_i$ is a valid ordinary transition of $M$ on input $x$.
In particular, none of these transitions is an oracle transition.
Define the corresponding reachability language
\[
L_{\mathrm{ord}}^x
=
\left\{
(A,B,\ell):
\exists \tau \text{ such that }
R_{\mathrm{ord}}^x((A,B,\ell),\tau)=1
\right\}.
\]
Thus, $(A,B,\ell)\in L_{\mathrm{ord}}^x$ if and only if $B$ can
be reached from $A$ in exactly $\ell$ ordinary transitions.
We allow $\ell=0$, in which case membership requires $A=B$.

We use the publicly verifiable $\IVC$ scheme from \Cref{sec:IVC-signed} restricted to ordinary transitions, denoted by 
$\mathsf{OrdIVC}$.\footnote{In
\Cref{sec:IVC-signed}, the computation starts at
$C_{\mathrm{init}}(x)$, whereas here it starts at an
arbitrary configuration $A$. The same construction applies
by carrying $A$ along in the transition records and checking
that the computation starts at $A$.}  As in  \Cref{sec:IVC-signed}, the input
$x$ is represented by its authenticated hash \[\val_x=\Hash(\hk,x),\]
and we authenticate input-tape accesses using local openings.
Thus, writing
$\theta=(\mathsf{crs}_{\mathrm{ord}},\mathsf{hk},\val_x)$, where $\mathsf{crs}_{\mathrm{ord}}$ denotes the common reference string of $\mathsf{OrdIVC}$,  which we assume
includes the prescribed input length~$n$, time bound~$T$,
and space bound~$S$,
the verifier checks a proof $\pi$ for the claim that $(A,B,\ell)\in L_{\mathrm{ord}}^x$ by checking that
\[
\mathsf{OrdIVC.Verify}(\theta,(A,B,\ell),\pi)=1.
\]
The proof~$\pi$ has length 
$\mathrm{poly}(\lambda,S,\log T)$ and can be updated after
each ordinary transition without redoing the computation (or remembering the execution trace).

\paragraph{Oracle-call records.}
Consider a purported execution on input $x$, and set
$A_0=C_{\mathrm{init}}(x)$.
Let $B_j$ be the configuration immediately before the $j$th
oracle call, and let $A_j$ be the configuration immediately
after it. Define
\begin{equation}\label{eqn:records}
D_j=(A_{j-1},B_j,A_j)
\qquad\text{and}\qquad
\ell_j=
\mathsf{time}(B_j)-\mathsf{time}(A_{j-1}).\footnote{Recall \Cref{rem:configuration-counters} which assumes that each configuration includes a step (or time) counter as well as an oracle-call counter.}
\end{equation}
Thus, $\ell_j$ is determined by $D_j$.

The statement supplied to the $\mathsf{UpBARG}$ is $(\theta,D_j)$.
Its witness is an ordinary-$\IVC$ proof $\pi_j$ for the interval
$(A_{j-1},B_j,\ell_j)$.
More precisely, we define an oracle-free $\NP$ relation
$R_{\mathrm{block}}$ as follows.
For $D=(A^-,B,A)$, the relation
\[
R_{\mathrm{block}}((\theta,D),\pi)=1
\]
if and only if all of the following hold:
\begin{enumerate}
    \item The configurations in $D$ are well formed and respect the
    prescribed space and time bounds. 
    \item Setting
   $\ell=\mathsf{time}(B)-\mathsf{time}(A^-)$,
    we have $0\leq\ell\leq T$.
    \item The oracle-call counters
    in $A^-$ and $B$ agree; namely $\mathsf{queries}(A^-)=\mathsf{queries}(B)$.

    \item The ordinary-$\IVC$ proof verifies:
    \[
    \mathsf{OrdIVC.Verify}(\theta,(A^-,B,\ell),\pi)=1.
    \]

    \item $B$ is an oracle-query configuration, and $A$ is
    obtained from $B$ by applying the prescribed oracle
    transition using the answer recorded in $A$.
    In particular,
    \[
    \mathsf{time}(A)=\mathsf{time}(B)+1~~\mbox{ and }~~\mathsf{queries}(A)=\mathsf{queries}(B)+1.
    \]
\end{enumerate}

Let $(u,a)$ be the query-answer pair specified by $(B,A)$.
The fifth condition checks the machine's transition using
the supplied answer~$a$; it does not check that $a=\calO(u)$.
Consequently, $R_{\mathrm{block}}$ makes no oracle calls.

Define the corresponding $\NP$ language
\[
L_{\mathrm{block}}
=
\left\{
(\theta,D):
\exists \pi \text{ such that }
R_{\mathrm{block}}((\theta,D),\pi)=1
\right\}.
\]

\paragraph{Consistency.}
For records $D=(A^-,B,A)$ and
$D'=({A'}^-,B',A')$, define
\[
\mathsf{Cons}((\theta,D),(\theta',D'))=1
\]
if and only if
\[
\theta=\theta'
\qquad\text{and}\qquad
A={A'}^-.
\]
For the first record,
$\mathsf{Cons}(\bot,(\theta,D))=1$ if and only if $A^-$ is
the prescribed initial configuration for inputs of length $n$,
with both the step counter and the oracle-call counter equal
to zero. As in \Cref{sec:IVC-signed}, the read-only input is represented
externally by $\val_x$.

\subsubsection{The privately verifiable $\IVC$ scheme}

Our construction uses the following ingredients:
\begin{itemize}
    \item An updatable hash family with local openings
    $\mathsf{HT}$, as in \Cref{sec:IVC-signed}.

    \item The publicly verifiable $\IVC$ scheme
    $\mathsf{OrdIVC}$ for ordinary intervals described above,
    with algorithms
    \[
        \mathsf{OrdIVC}
        =
        (\mathsf{OrdIVC.Gen},
         \mathsf{OrdIVC.Update},
         \mathsf{OrdIVC.Verify}).
    \]
    This scheme authenticates input-tape accesses using
    local openings to $\val_x$.

    \item An updatable batch argument $\mathsf{UpBARG}$
    for $(R_{\mathrm{block}},\mathsf{Cons})$,
    as defined in \Cref{sec:updatableBARG}, associated with the hash
    family $\mathsf{HT}$.
\end{itemize}

\paragraph{Setup.}
Fix a robustness parameter $\varepsilon\in(0,1]$, and let
$Q\ge 1$ be an upper bound on the number of oracle calls.
The algorithm
$\mathsf{IVC.Gen}(1^\lambda,n,T,\varepsilon)$
proceeds as follows:
\begin{enumerate}
    \item Set
    $r=\left\lceil\frac{2\lambda}{\varepsilon}\right\rceil$ and $     H=\lceil\log_2 Q\rceil$.

   \item  Set $\kappa
    =
    \poly
    (\lambda,S,\log T,\log(1/\varepsilon))$.

    \item Generate
   $\hk\leftarrow\HT.\Gen(1^\kappa)$.
   \item Generate public parameters
    $\crs_{\mathrm{ord}}\gets \mathsf{OrdIVC}.\Gen(1^\kappa,n,T)$.    
    \item For every $h\in\{0,\ldots,H\}$:
    \begin{enumerate}
        \item Set $K_h=2^h$.

        \item Independently sample
        $j_{h,1},\ldots,j_{h,r}\leftarrow[K_h]$.
        \item For every $i\in[r]$, generate
        $(\crs_{h,i},\td_{h,i})
            \leftarrow
            \mathsf{UpBARG.TGen}(1^\kappa,j_{h,i})$.        
            \item Set
        $\mathrm{crs}_h
            =
            (\crs_{h,i})_{i\in[r]},
            \qquad
            \td_h
            =
            \bigl((j_{h,i},\mathrm{td}_{h,i})\bigr)_{i\in[r]}$.

    \end{enumerate}

    \item Output
    $\crs
        =
        \bigl(
            \hk,
            \crs_{\mathrm{ord}},
            (\crs_h)_{h=0}^{H}
        \bigr)$ and 
        $\td
        =
        (\td_h)_{h=0}^{H}$.
  
\end{enumerate}
Setup makes no oracle calls.

\begin{remark}
We use several scales because the number $q$ of oracle calls made by
the current prefix is not known at setup time and may itself be chosen
adaptively after seeing the public parameters.
\end{remark}

\paragraph{The maintained $\IVC$ proof.}
For an input $x$, let
\[
    \val_x=\mathsf{HT.Hash}(\hk,x),
    \qquad
    \theta=
    (\crs_{\mathrm{ord}},\hk,\val_x).
\]
For each record $D_j$, denote the corresponding $\mathsf{UpBARG}$ statement by
\[
    Z_j=(\theta,D_j).
\]
After a prefix ending at configuration $C_t$ and containing
$q$ oracle calls, the prover maintains $(C_t,\pi_t)$ where
\[
    \pi_t=
    \bigl(
        \val,
        (\Pi_{h,i})_{\substack{0\le h\le H\\ i\in[r]}},
        D_{\mathrm{last}},
        \eta
    \bigr)
\]
satisfies the following properties:
\begin{itemize}
    \item $\val$ is the hash of the sequence
    $Z_1,\ldots,Z_q$, and $\Pi_{h,i}$ is a $\mathsf{UpBARG}$
    proof that $Z_1,\ldots,Z_q\in L_{\mathsf{block}}$ and that they satisfy the  consistency condition $\mathsf{Cons}$, under $\crs_{h,i}$.

    \item $D_{\mathrm{last}}=D_q$ when $q>0$, and
    $D_{\mathrm{last}}=\bot$ when $q=0$.

    \item $\eta$ is an ordinary-$\IVC$ proof for
    \[
        \bigl(A_q,C_t,
        \operatorname{time}(C_t)-\operatorname{time}(A_q)\bigr),
    \]
    with respect to~$\theta$, where
    $A_0=C_{\mathrm{init}}(x)$.
\end{itemize}

Initially, $C_0=C_{\mathrm{init}}(x)$ and $\pi_0=\bot$.
We interpret this initial proof as the state in which
$\val=\bot$, every $\Pi_{h,i}=\bot$,
$D_{\mathrm{last}}=\bot$, and $\eta$ is the ordinary-$\IVC$
proof for the zero-length interval
$(A_0,A_0,0)$.

\paragraph{Proof updates.}
The algorithm
$\mathsf{IVC.Update}^{\calO}
(\mathrm{crs},x,t,C_t,\pi_t,w_{t+1})$
updates the maintained proof as follows:
\begin{itemize}
    \item \textbf{Ordinary transition.}
    Compute $C_{t+1}$ using the prescribed transition
    and the local non-deterministic choice, if any.
    Update $\eta$ using the ordinary-$\IVC$ update algorithm.
    The hash $\val$, the $\mathsf{UpBARG}$ proofs, and
    $D_{\mathrm{last}}$ remain unchanged.

    \item \textbf{Oracle transition.}
    Let $B_{q+1}=C_t$. Obtain the answer to the query
    in $B_{q+1}$ from $\calO$, and let $A_{q+1}=C_{t+1}$
    be the resulting configuration. Define
    \[
        D_{q+1}=(A_q,B_{q+1},A_{q+1}),
        \qquad
        Z_{q+1}=(\theta,D_{q+1}).
    \]
    The current proof $\eta$ is a witness for
    $R_{\mathrm{block}}(Z_{q+1},\eta)=1$.

    Set $Z_{\mathrm{last}}=\bot$ if $q=0$, and
    $Z_{\mathrm{last}}=(\theta,D_q)$ otherwise.
    For every $h\in\{0,\ldots,H\}$ and $i\in[r]$, compute
    \[
            (\val'_{h,i},\Pi'_{h,i})
        \leftarrow
        \mathsf{UpBARG.Update}\bigl(
            \hk,\crs_{h,i},
            \val,\Pi_{h,i},
            Z_{\mathrm{last}},Z_{q+1},\eta
        \bigr).
       \]
    All resulting hash values equal
    \[
        \val'=\mathsf{HT.Append}(\hk,\val,Z_{q+1}).
    \]
    Replace $\val$ by $\val'$, replace each $\Pi_{h,i}$ by
    $\Pi'_{h,i}$, and set $D_{\mathrm{last}}=D_{q+1}$.
    Initialize $\eta$ to be the ordinary-$\IVC$ proof for the
    zero-length interval $(A_{q+1},A_{q+1},0)$.
\end{itemize}
The update uses no private verification key and does not
require the preceding execution trace.

\paragraph{Verification.}
The algorithm
$\mathsf{IVC.V}^{\calO}(\crs,\td,x,t,C,\pi)$
proceeds as follows. It rejects if any parsing operation
or required check fails.
\begin{enumerate}
    \item Check that $|x|=n$ that $0\le t\le T$.
    \item Check that $C$ is a well-formed configuration
respecting the prescribed space bound and satisfying
$\operatorname{time}(C)=t$.
Set $q=\operatorname{queries}(C)$ and check that
$0\le q\le\min\{Q,t\}$.
    
  \item Parse $\crs
        =
        \bigl(
            \hk,
            \crs_{\mathrm{ord}},
            (\crs_h)_{h=0}^{H}
        \bigr)$.

\item     Compute
    \[
        \val_x=\mathsf{HT.Hash}(\hk,x),
        \qquad
        \theta=
        (\crs_{\mathrm{ord}},\hk,\val_x).
    \]

    \item If $t=0$, accept if and only if $C=C_{\mathrm{init}}(x)$ and $\pi=\bot$.  Otherwise, continue.

    \item Parse
    $\pi=
        \bigl(
            \val,
            (\Pi_{h,i})_{\substack{0\le h\le H\\ i\in[r]}},
            D_{\mathrm{last}},
            \eta
        \bigr)$    and check that
    $\mathsf{HT.Size}(\val)=q$.

    \item If $q=0$, check that
    $\val=\bot$ and that $
        D_{\mathrm{last}}=\bot$ and $\Pi_{h,i}=\bot$ for every $h\in\{0,\ldots,H\}$ and $i\in[r]$, and accept if and only if all these checks pass and 
    \[
        \mathsf{OrdIVC.Verify}
        \bigl(
            \theta,
            (C_{\mathrm{init}}(x),C,t),
            \eta
        \bigr)=1.
    \]

    \item If $q>0$, parse
   $D_{\mathrm{last}}=(A_{q-1},B_q,A_q)$, 
    and check that these configurations are well formed, satisfy the space bound restriction,  and satisfy the following conditions: 
    \[
        \operatorname{queries}(A_{q-1})
        =
        \operatorname{queries}(B_q)
        =
        q-1,
        \qquad
        \operatorname{queries}(A_q)=q
    \]
    and
    \[
        0\le
        \operatorname{time}(A_{q-1})
        \le
        \operatorname{time}(B_q),
        \qquad
        \operatorname{time}(A_q)
        =
        \operatorname{time}(B_q)+1
        \le t.
    \]
    \item Set
    $Z_{\mathrm{last}}=(\theta,D_{\mathrm{last}})$.
    \item For every $h\in\{0,\ldots,H\}$ and $i\in[r]$,
    check that
    \[
        \mathsf{UpBARG.Ver}
        \bigl(
            \hk,\crs_{h,i},
            \val,\Pi_{h,i},Z_{\mathrm{last}}
        \bigr)=1.
    \]

    \item Set
    $\ell_{\mathrm{tail}}
        =
        t-\operatorname{time}(A_q)$,
    and check that
    \[
        \mathsf{OrdIVC.Verify}
        \bigl(
            \theta,
            (A_q,C,\ell_{\mathrm{tail}}),
            \eta
        \bigr)=1.
    \]

    \item Set
   $h^*=\lceil\log_2 q\rceil$ and 
        $K=2^{h^*}$.
Thus $q\le K<2q$.
\item Parse 
        $\td
        =
        (\td_h)_{h=0}^{H}$, and parse $\td_{h^*}
            =
            \bigl(j_{h^*,i},\mathrm{td}_{h^*,i}\bigr)_{i\in[r]}$.  
   \item  For each $i\in[r]$, perform the following private check:
    \begin{enumerate}
        \item Read
        $(j,\td_{h^*,i})$
        from the $i$th entry of $\td_{h^*}$.
        If $j>q$, skip this component.

        \item Otherwise, compute
        \[
        (\widehat Z,\rho,\widehat\pi,
              \widehat Z^{\mathrm{prev}},
              \rho^{\mathrm{prev}})
            \leftarrow
            \mathsf{UpBARG.Extract}
            \bigl(
                \hk,\td_{h^*,i},
                \val,\Pi_{h^*,i}
            \bigr).
            \]

        \item Check that
        $\widehat Z=(\theta,\widehat D)$ for some $\widehat D=(U,B,A)$          with
        \[
            \operatorname{queries}(U)
            =
            \operatorname{queries}(B)
            =
            j-1,
            \qquad
            \operatorname{queries}(A)=j.
        \]
        Check that
        \[
            \mathsf{HT.Verify}
            (\mathrm{hk},\val,j,\widehat Z,\rho)=1~~\mbox{ and }~~
            R_{\mathrm{block}}
            (\widehat Z,\widehat\pi)=1.
        \]

        \item If $j=1$, check that
       $\widehat Z^{\mathrm{prev}}
            =
            \rho^{\mathrm{prev}}
            =
            \bot$.
           
            If $j>1$, check that
        $\widehat Z^{\mathrm{prev}}$ begins with
        $\theta$ and that
        \[
            \mathsf{HT.Verify}
            \bigl(
                \hk,\val,j-1,
                \widehat Z^{\mathrm{prev}},
                \rho^{\mathrm{prev}}
            \bigr)=1.
        \]
        In either case, check that
        $\mathsf{Cons}
            (\widehat Z^{\mathrm{prev}},\widehat Z)=1$.

        \item If $j=q$, also check that
        $\widehat Z=Z_{\mathrm{last}}$.

        \item Let $(u,a)$ be the query-answer pair
        specified by the oracle transition from $B$ to $A$.
        Query $\calO(u)$ and check that
        $a=\calO(u)$.
        \end{enumerate}

    \item Accept if all checks pass.
\end{enumerate}

The verifier makes at most
$r=O(\lambda/\varepsilon)$ oracle calls.

\begin{theorem}
\label{thm}
Let $M$ be a (possibly non-deterministic) time-$T$, space-$S$ oracle-aided machine.  
We assume that the underlying primitives, $\mathsf{UpBARG}$, $\HT$, and (ordinary) $\IVC$, are sub-exponentially secure, even given oracle
access to $\calO$. Then there exists a privately verifiable $\IVC$ for
$\varepsilon$-robust oracle-aided computations.
\begin{itemize}
    \item Its public parameters, private verification key, and proof have length
$\poly(\lambda,S,\log T,1/\varepsilon)$.
\item Proof updates take time
$\poly(n,\lambda,S,\log T,1/\varepsilon)$.
\item Verification takes time
$\poly(n,\lambda,S,\log T,1/\varepsilon)$,
and makes $O(\lambda/\varepsilon)$ calls to $\calO$.  Setup makes no oracle
calls.
\end{itemize}

\end{theorem}

\paragraph{Proof of \Cref{thm}.}


For completeness, first consider the experiment in which
all outer reference strings are generated by
$\mathsf{UpBARG.Gen}(1^\kappa)$.
Completeness of $\mathsf{OrdIVC}$, incremental completeness
of $\mathsf{UpBARG}$, and correctness of the hash family
ensure that all public checks accept on an honest execution.
Replace these reference strings, one at a time, by outputs
of $\mathsf{UpBARG.TGen}(1^\kappa,j_{h,i})$.
In each hybrid, rerun the algorithm choosing the input and
honest execution, followed by the prescribed proof updates.
Public-check failure can be tested without any extraction
trapdoor, in time $\operatorname{poly}(T^*)$ with access
to $\mathcal O$, where $T^*$ is specified below.
Thus, the strengthened index-hiding guarantee changes its
probability by at most $\operatorname{negl}(T^*)$ per hybrid.
Since there are $r(H+1)\leq\operatorname{poly}(T^*)$
components, all public checks accept in the real experiment
except with probability $\operatorname{negl}(T^*)$.

In the real experiment, the somewhere argument-of-knowledge
property ensures that, except with negligible probability,
every unskipped extraction from an accepting component
provides a valid witness and hash openings.
By collision resistance with respect to opening, the
extracted records agree with the corresponding records
of the honest execution. Hence all private checks also
accept. This proves adaptive completeness.

We prove adaptive soundness. To this end, as in \Cref{sec:robust-oracle-construction}, let
\[
    T^*
    =
    2^{O(S)}
    \operatorname{poly}(n,T,\lambda,1/\varepsilon)
\]
be large enough to bound the reductions below.
We assume that the primitives, instantiated with the
chosen security parameter $\kappa$, satisfy the required
security properties against the corresponding
oracle-aided algorithms, with errors negligible in $T^*$.

Throughout the proof, $\delta=\operatorname{negl}(T^*)$
denotes a negligible error term, which may increase
to absorb polynomially many primitive-security errors.
The bounds are uniform over the indices used below.
In particular, the final accumulated error satisfies
\[
    Q(r-1)\delta=\operatorname{negl}(\lambda).
\]
Fix an oracle $\calO$ independently
of setup and an adversary with access to $\calO$. 
Let $(x,t,C,\pi)$ be the output of the adversary, and let $q$ be the claimed
number of oracle calls. 
If $t=0$, the verifier accepts only when
$C=C_{\mathrm{init}}(x)$ and $\pi=\bot$,
so soundness is immediate.
Thus, from now on we assume that $t>0$.
The verifier checks that
$q=\mathsf{HT.Size}(\val)=\operatorname{queries}(C)$ and that
$\operatorname{time}(C)=t$.

If $q=0$, an accepting certificate contains an ordinary-$\IVC$
proof from $C_{\mathrm{init}}(x)$ to $C$ in exactly $t$
ordinary steps. Soundness in this case follows directly
from soundness of $\mathsf{OrdIVC}$.

For $q\in[Q]$ and $s\in\{0,\ldots,r\}$, let $E_{q,s}$ be
the event that the adversary outputs an unreachable,
$\varepsilon$-robust endpoint with exactly $q$ oracle calls,
all public verification checks accept, including the
ordinary-$\IVC$ check, and the first $s$ private checks
at scale $h=\lceil\log_2 q\rceil$ accept. Set
\[
    p_{q,s}=\Pr[E_{q,s}]
\]
in the real experiment. We prove that, for every $s\in[r-1]$,
\begin{equation}
    p_{q,s}
    \le
    \left(1-\frac{\varepsilon}{2}\right)p_{q,s-1}
    +\delta.
    \label{eq:robust-ivc-contraction}
\end{equation}
Fix $q\in[Q]$ and $s\in[r-1]$, and let
\[
    E=E_{q,s-1},
    \qquad
    h=\lceil\log_2 q\rceil,
    \qquad
    K=2^h.
\]
Thus $q\le K<2q$. We use components $s$ and $r$ at this
scale in the hybrid argument. The event $E$ uses neither
of their trapdoors.

For a configuration $U$, let $d_x(U)$ be the minimum number
of incorrect oracle answers in a purported execution from
$C_{\mathrm{init}}(x)$ to $U$, at the time specified by the
step counter in $U$. Set $d_x(U)=+\infty$ if no such
execution exists, and define
\[
    \Phi_q^x(U)=\frac{\min\{d_x(U),q\}}{q}.
\]
This potential lies in $[0,1]$ and is zero at the initial
configuration. On $E$, endpoint robustness gives
\[
    \Phi_q^x(C)\ge\varepsilon.
\]
If a valid ordinary interval from $A$ to $B$ is followed
by a purported oracle transition from $B$ to $A'$ with
query-answer pair $(u,a)$, then
\begin{equation}
    \Phi_q^x(A')
    \le
    \Phi_q^x(A)
    +\frac{\mathbf{1}[a\ne \calO(u)]}{q}.
    \label{eq:robust-ivc-block-potential}
\end{equation}
These potentials and the event $E$ can be evaluated by
the bounded-space configuration-graph computation, together
with the indicated private checks.

Let $H_{a,b}$ denote the experiment in which components
$s$ and $r$ at scale $h$ have extraction indices $a$ and
$b$, respectively, with all other setup components
generated as in the real experiment. A star denotes an
independently uniform index in $[K]$.

For each $j\in[q]$, consider $H_{\star,j}$ and extract
record $j$ using component $r$. Write its configurations
as
\[
    \widehat D_j
    =
    (\widehat A_{j-1},\widehat B_j,\widehat A_j).
\]
An extraction is valid if the local relation, consistency
predicate, and required hash openings verify. It is aligned
if its common context is the expected $\theta$ and its
starting and ending oracle counters are $j-1$ and $j$.
Let
\[
    b_j=\mathbf{1}[a_j\ne \calO(u_j)]
\]
for the extracted query-answer pair, and define
\[
    F_j^-=
    \mathbb{E}_{H_{\star,j}}
    [\mathbf{1}_E\Phi_q^x(\widehat A_{j-1})],
    \qquad
    F_j^+=
    \mathbb{E}_{H_{\star,j}}
    [\mathbf{1}_E\Phi_q^x(\widehat A_j)],
\]
and
\[
    W_j=
    \mathbb{E}_{H_{\star,j}}
    [\mathbf{1}_E b_j].
\]
All contributions are set to zero when the extraction
is invalid or unaligned.

The extracted witness contains an accepting ordinary-$\IVC$
proof under the expected context. By adaptive soundness
of $\mathsf{OrdIVC}$, except with negligible probability
this proof certifies a valid ordinary interval on the
actual input $x$. Applying
\Cref{eq:robust-ivc-block-potential}, we obtain
\[
    F_j^+\le F_j^-+\frac{W_j}{q}+\delta.
\]

For every $j<q$, the two-component hybrid argument gives
\[
    F_{j+1}^-=F_j^+\pm\delta.
\]
Indeed, consider
\[
    H_{\star,j},\quad
    H_{j+1,j},\quad
    H_{j+1,j+1},\quad
    H_{\star,j+1}.
\]
Each change of an extraction index is justified by index
hiding while evaluating a quantity that uses only the
other component's trapdoor. In $H_{j+1,j}$, the two
extractions authenticate the same record at position $j$.
Collision resistance identifies these records, and the
consistency predicate identifies the shared boundary
configuration and propagates alignment. In
$H_{j+1,j+1}$, both components authenticate the same
record at position $j+1$, so we may switch back to
evaluating the potential with component $r$.

The first-record condition gives $F_1^-=0$. At the other
endpoint, last-statement collision resistance identifies
the record extracted at position $q$ with the claimed
last record, except with negligible probability. As in
\Cref{sec:IVC-signed}, this property is applied by changing the
unused component $s$ to ordinary generation while
evaluating the extraction from component $r$, and then
changing it back by index hiding.

The verifier's checks therefore establish alignment of
the last extracted record. Soundness of the trailing
ordinary-$\IVC$ proof gives
\[
    \Phi_q^x(\widehat A_q)\ge\Phi_q^x(C)
    \ge\varepsilon
\]
on $E$, except with negligible probability. Since $E$
does not use component $r$'s trapdoor, index hiding also
gives
\[
    \Pr_{H_{\star,q}}[E]=p_{q,s-1}\pm\delta.
\]
Consequently,
\[
    F_q^+\ge\varepsilon p_{q,s-1}-\delta.
\]
Telescoping the potential inequalities now yields
\[
    \sum_{j=1}^{q}W_j
    \ge
    \varepsilon q\,p_{q,s-1}-\delta,
\]
where $\delta$ absorbs the accumulated primitive-security
errors.

It remains to relate this expected cost to the $s$th
private check. For each $j\in[q]$, consider
\[
    H_{\star,j},\quad H_{j,j},\quad H_{j,\star}.
\]
Index hiding and collision resistance allow us to transfer
the statistic $\mathbf{1}_E b_j$ from the extraction by
component $r$ in the first experiment to the extraction
by component $s$ in the last. In the middle experiment,
both components authenticate the same record at position
$j$, and hence the same query-answer pair.

In the real experiment, component $s$ has a uniformly
sampled index in $[K]$. An extracted incorrect oracle
answer causes rejection. Therefore,
\[
 \Pr[E\land\text{the $s$th private check rejects}]
    \ge \frac{1}{K}\sum_{j=1}^{q}W_j-\delta
    \ge \frac{\varepsilon q}{K}p_{q,s-1}-\delta    \ge \frac{\varepsilon}{2}p_{q,s-1}-\delta.
\]
This proves \Cref{eq:robust-ivc-contraction}.

Finally, the events $E_{q,0}$ are disjoint, so
$\sum_{q=1}^{Q}p_{q,0}\le1$. Iterating the contraction
and summing over all possible query counts gives
\[
\begin{aligned}
    \Pr[\text{bad acceptance with }q>0]
    &\le \sum_{q=1}^{Q}p_{q,r-1}\\
    &\le
    \left(1-\frac{\varepsilon}{2}\right)^{r-1}
    \sum_{q=1}^{Q}p_{q,0}
    +Q(r-1)\delta\\
    &\le e^{-\lambda+1/2}
    +\operatorname{negl}(\lambda).
\end{aligned}
\]
Here we used $r=\lceil2\lambda/\varepsilon\rceil$ and
the stipulated quantitative security of the underlying
primitives. Together with the $q=0$ case, this proves
adaptive soundness.

\qed

\paragraph{Instantiation under $\LWE$.}
We instantiate $\mathsf{OrdIVC}$ using \Cref{sec:IVC-signed} restricted
to ordinary transitions, and use the $\LWE$-based
$\mathsf{UpBARG}$ from \Cref{sec:updatableBARG} for both the inner and
outer constructions.
The composed reductions have size and losses
$\operatorname{poly}(T^*)$: the ordinary-IVC analysis
enumerates configurations of the original space-$S$ machine,
not of the proof verifier.
Assuming sub-exponential hardness of $\LWE$, with
sub-exponentially small advantage, even given access to $\calO$,
we can therefore choose all underlying security parameters
as sufficiently large polynomials in $\lambda+\log T^*$
to obtain errors negligible in $T^*$, even after all
reductions and hybrids.

\paragraph{Intermediate configurations.}
For an arbitrary intermediate configuration we require the natural
endpoint version of robustness: either the configuration is genuinely
reachable, or every purported execution reaching it contains an
$\varepsilon$-fraction of incorrect oracle answers.  For an accepting
terminal configuration on an $\varepsilon$-robust no-instance, this
condition follows immediately from \Cref{def:oracle-robustness}.

\section*{AI Disclosure}
\addcontentsline{toc}{section}{AI Disclosure}
We used generative AI tools to edit the writing, proofread the paper to catch potential bugs, suggest writing improvements, and for a literature review. All AI-generated text was reviewed and verified by the authors. All the ideas, including the proof ideas, were developed solely by the authors. We take full responsibility for the content of this work.  

\section*{Acknowledgments}
\addcontentsline{toc}{section}{Acknowledgments}
This work is supported by the U.S.\ National Science Foundation under award No.~2534400, and by the UK AISI alignment project.

\ifnum\llncs=0
\bibliographystyle{alpha}
\else
\bibliographystyle{splncs04}
\fi

\bibliography{abbrev3,crypto,mybib}

\end{document}